\documentclass[12pt,preprint,eqsecnum,aps]{revtex4}%
\usepackage{amsfonts}
\usepackage{amsmath}
\usepackage{amssymb}
\usepackage{graphicx}%
\begin{document}
\preprint{ }
\title{Lattice simulations for few- and many-body systems}
\author{Dean Lee}
\affiliation{Department of Physics, North Carolina State University, Raleigh, NC 27695}
\keywords{lattice, effective field theory, nuclear lattice simulations, neutron matter,
nuclear matter, unitarity limit, BCS-BEC\ crossover, attractive Hubbard model}
\pacs{03.75.Ss, 21.30.-x, 21.45.-v, 21.60.Ka, 21.65.-f, 21.65.Cd, 71.10.Fd, 71.10.Hf}

\begin{abstract}
We review the recent literature on lattice simulations for few- and many-body
systems. \ We focus on methods that combine the framework of effective field
theory with computational lattice methods. \ Lattice effective field theory is
discussed for cold atoms as well as low-energy nucleons with and without
pions. \ A number of different lattice formulations and computational
algorithms are considered, and an effort is made to show common themes in
studies of cold atoms and low-energy nuclear physics as well as common themes
in work by different collaborations.

\end{abstract}
\maketitle
\tableofcontents

\section{Introduction}

In this article we review the literature on lattice simulations for few- and
many-body systems. \ We discuss methods which combine effective field theory
with lattice methods and which can be applied to both cold atomic systems and
low-energy nuclear physics. \ Several recent reviews have already been written
describing quantum Monte Carlo methods for a range of topics. \ These include
Monte Carlo calculations in continuous space for electronic orbitals in
chemistry \cite{Hammond:1994}, solid state materials \cite{Foulkes:2001},
superfluid helium \cite{Ceperley:1995RMP}, and few-nucleon systems
\cite{Carlson:qn}. \ There are also reviews of Monte Carlo lattice methods for
strongly-correlated lattice models \cite{vonderLinden:1992}, lattice quantum
chromodynamics at nonzero density \cite{Muroya:2003qs}, and a general
introduction to lattice quantum chromodynamics \cite{Davies:2002cx}.

Lattice simulations of quantum chromodynamics (QCD) are now able to accurately
describe the properties of many isolated hadrons. \ In addition to isolated
hadrons, it is also possible to calculate low-energy hadronic interactions
such as meson-meson scattering
\cite{Kuramashi:1993ka,Aoki:2002ny,Lin:2002aj,Beane:2005rj,Beane:2006gj,Beane:2007uh}%
.\ \ Other interactions such as baryon-baryon scattering are computationally
more difficult, but there has been promising work in this direction as well
\cite{Fukugita:1994ve,Beane:2003da,Beane:2006gf,Beane:2006mx,Ishii:2006ec,Aoki:2008hh,Nemura:2008sp}%
. \ A recent review of hadronic interaction results computed from lattice QCD
can be found in Ref.~\cite{Beane:2008dv}.

However for few- and many-body systems beyond two nucleons, lattice QCD
simulations are presently out of reach. \ Such simulations require pion masses
at or near the physical mass and lattices several times longer in each
dimension than used in current simulations. \ Another significant
computational challenge is to overcome the exponentially small signal-to-noise
ratio for simulations at large quark number. \ For few- and many-body systems
in low-energy nuclear physics one can make further progress by working
directly with hadronic degrees of freedom.

There are several choices one can make for the nuclear forces and the
calculational method used to describe interacting low-energy protons and
neutrons. \ For systems with four or fewer nucleons, a semi-analytic approach
is provided by the Faddeev-Yakubovsky integral equations. \ Using this method,
one study \cite{Nogga:2000uu} looked at three- and four-nucleon systems using
the Nijmegen potentials \cite{Stoks:1994wp}, CD-Bonn potential
\cite{Machleidt:1996km}, and AV18 potential \cite{Wiringa:1995wb}, together
with the Tucson-Melbourne \cite{Coon:1981TM} and Urbana-IX
\cite{Pudliner:1997ck} three-nucleon forces. \ A different investigation
considered the same observables using a two-nucleon potential derived from
chiral effective field theory \cite{Epelbaum:2000mx}. \ Another recent study
\cite{Nogga:2004ab}\ considered the low-momentum interaction potential
$V_{\text{low }k}$ \cite{Bogner:2001gq,Bogner:2003wn}. \ This method used the
renormalization group to derive effective interactions equivalent to potential
models but at low cutoff momentum.

For systems with more nucleons approaches such as Monte Carlo simulations or
basis-truncated eigenvector methods are needed. \ There is considerable
literature describing Green's Function Monte Carlo simulations of light nuclei
and neutron matter based on AV18 as well as other phenomenological potentials
\cite{Pudliner:1997ck,Wiringa:2000gb,Pieper:2001mp,Pieper:2001ap,Pieper:2002ne,Wiringa:2002ja,Pieper:2004qw,Nollett:2006su,Gezerlis:2007fs}%
. \ There is a review article detailing this method \cite{Carlson:qn} as well
as a more recent set of lecture notes \cite{Pieper:2007ax}. \ A related
technique known as auxiliary-field diffusion Monte Carlo simplifies the spin
structure of the same calculations by introducing auxiliary fields
\cite{Fantoni:2001ih,Sarsa:2003zu,Pederiva:2004iz,Chang:2004sj,Gandolfi:2007hs,Gandolfi:2008id}%
. \ The No-Core Shell Model (NCSM) is a different approach to light nuclei
which produces approximate eigenvectors in a reduced vector space. \ There
have been several NCSM calculations using various different phenomenological
potential models
\cite{Navratil:2000ww,Fayache:2001kq,Navratil:2003ef,Caurier:2005rb}. \ There
are also NCSM calculations which have used nuclear forces derived from chiral
effective field theory \cite{Forssen:2004dk,Nogga:2005hp,Navratil:2007we}.
\ Recently there has also been work in constructing a low-energy effective
theory within the framework of truncated basis states used in the NCSM
formalism \cite{Stetcu:2006ey}. \ A benchmark comparison of many of the
methods listed above as well as other techniques can be found in
Ref.~\cite{Kamada:2001tv}.

In this article we describe recent work by several different collaborations
which combine the framework of effective field theory with computational
lattice methods. \ The idea of lattice simulations using effective field
theory is rather new. \ The first quantum lattice study of nuclear
matter\ appears to be Ref.~\cite{Brockmann:1992in}, which used a momentum
lattice and the quantum hadrodynamics model of Walecka \cite{Walecka:1974qa}.
\ The first study combining lattice methods with an effective theory for
low-energy nuclear physics was Ref.~\cite{Muller:1999cp}. \ This study looked
at infinite nuclear and neutron matter at nonzero density and temperature.
\ After this there appeared a computational study of the attractive Hubbard
model in three dimensions \cite{Sewer:2002}, as well as a paper noting the
absence of sign oscillations for nonzero chemical potential and external
pairing field \cite{Chen:2003vy}. \ Another study looked at nonlinear
realizations of chiral symmetry with static nucleons on the lattice
\cite{Chandrasekharan:2003ub}, and there were also a number of investigations
of chiral perturbation theory with lattice regularization
\cite{Shushpanov:1998ms,Lewis:2000cc,Borasoy:2003pg}. \ This was followed by
the first many-body lattice calculation using chiral effective field theory
\cite{Lee:2004si}. \ From about this time forward there were a number of
lattice calculations for cold atoms and low-energy nuclear physics which we
discuss in this article.

The lattice effective field theory approach has some qualitative parallels
with digital media. \ In digital media input signals are compressed into
standard digital output that can be read by different devices. \ In our case
the input is low-energy scattering data, and the digital format is effective
field theory defined with lattice regularization. \ The process of sampling
and compression consists of matching low-energy scattering data using
effective interactions up to some chosen order in power counting. \ By
increasing the order, the accuracy in describing low-energy phenomena can be
systematically improved.

Just as standard digital format enables communication between different
devices, lattice effective field theory enables the study of many different
phenomena using the same lattice action. \ This includes few- and many-body
systems as well as ground state properties and thermodynamics at nonzero
temperature and density. \ Another attractive feature of lattice effective
field theory is the direct link with analytic calculations using effective
field theory. \ It is straightforward to derive lattice Feynman rules and
calculate diagrams using the same theory used in non-perturbative simulations.
\ At fixed lattice spacing all of the systematic error is introduced up front
when defining the low-energy lattice effective field theory and not determined
by the particular computational scheme used to calculate observables. \ This
allows for a wide degree of computational flexibility, and one can use a
number of efficient lattice methods already developed for lattice QCD and
condensed matter applications. \ This includes cluster algorithms,
auxiliary-field transformations, pseudofermion methods, and non-local
configuration updating schemes. \ We discuss all of these techniques in this
article. \ We also review the relevant principles of effective field theory as
well as different formalisms and algorithms used in lattice calculations.
\ Towards the end we discuss some recent results and compare with results
obtained using other methods.

\section{Effective field theory}

Effective field theory provides a systematic approach to studying low-energy
phenomena in few- and many-body systems. \ We give a brief overview of the
effective range expansion and the application of effective field theory to
cold atoms and low-energy nuclear physics. \ A more thorough review of
effective field theory methods applied to systems at nonzero density can be
found in Ref.~\cite{Furnstahl:2008df}.

\subsection{Effective range expansion}

At sufficiently low momentum the cross-section for two-body scattering is
dominated by the $S$-wave amplitude, and higher partial waves are suppressed
by powers of the relative momentum. \ The $S$-wave scattering amplitude for
two particles with mass $m$ and relative momentum $p$ is%
\begin{equation}
\mathcal{A}_{0}(p)=\frac{4\pi}{m}\frac{1}{p\cot\delta_{0}-ip},
\end{equation}
where $\delta_{0}$ is the $S$-wave phase shift. \ At low momentum the $S$-wave
phase shift for two-body scattering with short-range interactions can be
written in terms of the effective range expansion \cite{Bethe:1949yr},
\begin{equation}
p\cot\delta_{0}=-\frac{1}{a_{\text{scatt}}}+\frac{1}{2}r_{\text{eff}}%
p^{2}+\cdots. \label{swave}%
\end{equation}
Here $a_{\text{scatt}}$ is the $S$-wave scattering length, and $r_{\text{eff}%
}$ is the $S$-wave effective range. \ The radius of convergence of the
effective range expansion is controlled by the characteristic length scale of
the interaction. \ For example in low-energy nuclear physics the range of the
two-nucleon interaction is set by the Compton wavelength of the pion. \ The
generalization of the effective range expansion to partial wave $L$ has the
form
\begin{equation}
p^{2L+1}\cot\delta_{L}=-\frac{1}{a_{L}}+\frac{1}{2}r_{L}p^{2}+\cdots.
\label{Lwave}%
\end{equation}
The $\delta_{L}$ phase shift scales as $O(p^{2L+1}a_{L})$ in the low-momentum
limit, and higher-order terms are suppressed by further powers of $p^{2}$.
This establishes a hierarchy of low-energy two-body scattering parameters for
short-range interactions. \ For particles with intrinsic spin there is also
some mixing between partial waves carrying the same total angular momentum.

For many interacting systems we can characterize the low-energy phenomenology
according to exact and approximate symmetries and low-order interactions
according to some hierarchy of power counting. \ This universality is due to a
wide disparity between the long-distance\ scale of low-energy phenomena and
the short-distance scale of the underlying interaction. \ In some cases the
simple power counting of the effective range expansion must be rearranged or
resummed in order to accommodate non-perturbative effects. \ We discuss this
later in connection with singular potentials and three-body forces. \ A recent
review of universality in few-body systems at large scattering length can be
found in Ref.~\cite{Braaten:2004a}.

In many-body systems a prime example of universality is the unitarity limit.
\ The unitarity limit describes attractive two-component fermions in an
idealized limit where the range of the interaction is zero and the scattering
length is infinite. \ The name refers to the fact that the $S$-wave
cross-section saturates the limit imposed by unitarity, $\sigma_{0}(p)\leq
4\pi/p^{2}$, for low momenta $p$. \ While the unitarity limit has a
well-defined continuum limit and strong interactions, at zero temperature it
has no intrinsic physical scale other than the interparticle spacing.

Phenomenological interest in the unitarity limit extends across several
subfields of physics. \ The ground state of the unitarity limit is known to be
a superfluid with properties in between a Bardeen-Cooper-Schrieffer (BCS)
fermionic superfluid at weak attractive coupling and a Bose-Einstein
condensate (BEC) of bound dimers at strong attractive coupling
\cite{Eagles:1969PR,Leggett:1980pro,Nozieres:1985JLTP}. \ It has been
suggested that the crossover from fermionic to bosonic superfluid could be
qualitatively similar to pseudogap behavior in high-temperature
superconductors \cite{Chen:2005PhyRep}. \ In nuclear physics the unitarity
limit is relevant to the properties of cold dilute neutron matter. \ The
neutron scattering length is about $-18.5$ fm while the range of the
interaction is comparable to the Compton wavelength of the pion, $m_{\pi}%
^{-1}\approx1.4$ fm. \ Therefore the unitarity limit is approximately realized
when the interparticle spacing is about $5$ fm. \ Superfluid neutrons at
around this density may exist in the inner crust of neutron stars
\cite{Pethick:1995di,Lattimer:2004pg}.

\subsection{Effective field theory for cold atoms}

Physics near the unitarity limit has been experimentally observed in cold
degenerate gases of $^{6}$Li and $^{40}$K atoms. \ Alkali atoms are convenient
for evaporative cooling due to their predominantly elastic collisions. \ For
sufficiently dilute gases the effective range and higher partial wave effects
are negligible while the scattering length can be adjusted using a
magnetically-tuned Feshbach resonance
\cite{Tiesinga:1993PRA,Stwalley:1976PRL,Courteille:1998PRL,Inouye:1998Nat}.
\ Overviews of experiments using Feshbach resonances can be found in
Ref.~\cite{Koehler:2006A, Regal:2006thesis}, and there are a number of reviews
covering the theory of BCS-BEC\ crossover in cold atomic systems
\cite{Chen:2005PhyRep, Giorgini:2007a, Bloch:2007a}.

At long distances the interactions between alkali atoms are dominated by the
van der Waals $-C_{6}/r^{6}$ interaction. \ Power-law interactions complicate
the effective range expansion by producing a branch cut in each partial wave
at $p=0$. \ For the van der Waals interaction the expansion in $p^{2}$ is an
asymptotic expansion coinciding with the effective range expansions in
Eq.~(\ref{swave}) and (\ref{Lwave}) up through terms involving
$a_{\text{scatt}}$, $r_{\text{eff}}$, and $a_{1}$ \cite{Gao:1998A,Gao:1998B}.
\ Beyond this the asymptotic expansion involves powers of $p^{2}$ times $\ln
p^{2}$ or odd powers of $p$. \ All of the work discussed in this article
involves low-energy phenomena where these non-analytic terms can be neglected.

The low-energy effective field theory for the unitarity limit can be derived
from any theory of two-component fermions with infinite scattering length and
negligible higher-order scattering effects at the relevant low-momentum scale.
\ For example the two fermion components may correspond with dressed hyperfine
states $\left\vert f,m_{f}\right\rangle =\left\vert 9/2,-9/2\right\rangle $
and $\left\vert 9/2,-7/2\right\rangle $ of $^{40}$K with interactions given
either by a full multi-channel Hamiltonian or a simplified two-channel model
\cite{Goral:2004A,Szymanska:2005A,Nygaard:2007A}. The starting point does not
matter so long as the $S$-wave scattering length is tuned to infinity to
produce a zero-energy resonance.

In our notation $m$ is the atomic mass and $a_{i}$ and $a_{i}^{\dagger}$ are
annihilation and creation operators for two hyperfine states. \ We label these
as up and down spins, $i=\uparrow,\downarrow$, even though the connection with
actual intrinsic spin is not necessary. \ We enclose operator products with
the symbols $::$ to indicate normal ordering, where creation operators are on
the left and annihilation operators are on the right. \ The effective
Hamiltonian at leading order (LO) is%
\begin{equation}
H_{\text{LO}}=H_{\text{free}}+V_{\text{LO}}, \label{two_component_hamiltonian}%
\end{equation}
where%
\begin{equation}
H_{\text{free}}=\frac{1}{2m}\sum_{i=\uparrow,\downarrow}\int d^{3}\vec
{r}\;\vec{\nabla}a_{i}^{\dagger}(\vec{r})\cdot\vec{\nabla}a_{i}(\vec{r}),
\end{equation}%
\begin{equation}
V_{\text{LO}}=\frac{C}{2}\int d^{3}\vec{r}\;:\left[  \rho^{a^{\dagger},a}%
(\vec{r})\right]  ^{2}:,
\end{equation}
and $\rho^{a^{\dagger},a}(\vec{r})$ is the particle density operator,%
\begin{equation}
\rho^{a^{\dagger},a}(\vec{r})=\sum_{i=\uparrow,\downarrow}a_{i}^{\dagger}%
(\vec{r})a_{i}(\vec{r}).
\end{equation}
The coefficient $C$ depends on the cutoff scheme used to regulate ultraviolet
divergences in the effective theory. \ Higher-order effects may be introduced
systematically as higher-dimensional local operators with more derivatives
and/or more local fields.

\subsection{Pionless effective field theory}

For nucleons at momenta much smaller than the pion mass, all interactions
produced by the strong nuclear force can be treated as local interactions
among nucleons. \ The effective Hamiltonian in
Eq.~(\ref{two_component_hamiltonian}) also describes the interactions of
low-energy neutrons at leading order. \ For systems with both protons and
neutrons we label the nucleon annihilation operators with two subscripts,%
\begin{equation}
a_{0,0}=a_{\uparrow,p},\text{ \ }a_{0,1}=a_{\uparrow,n},
\end{equation}%
\begin{equation}
a_{1,0}=a_{\downarrow,p},\text{ \ }a_{1,1}=a_{\downarrow,n}.
\end{equation}
The first subscript is for spin $\uparrow,\downarrow$ and the second subscript
is for isospin $p,n$. \ We use $\sigma_{S}$ with $S=1,2,3$ to represent Pauli
matrices acting in spin space and $\tau_{I}$ with $I=1,2,3$ to represent Pauli
matrices acting in isospin space. \ The same letters $S$ and $I$ are also used
to indicate total spin and total isospin quantum numbers, but the intended
meaning will be clear from the context. \ If we neglect isospin breaking and
electromagnetic effects, the effective theory has exact SU$(2)$ spin and
SU$(2)$ isospin symmetries.

Let us define the total nucleon density%
\begin{equation}
\rho^{a^{\dagger},a}(\vec{r})=\sum_{i,j=0,1}a_{i,j}^{\dagger}(\vec{r}%
)a_{i,j}(\vec{r}). \label{density}%
\end{equation}
The total nucleon density is invariant under Wigner's SU(4) symmetry mixing
all spin and isospin degrees of freedom \cite{Wigner:1937}. \ Using
$\sigma_{S}$ and $\tau_{I}$, we also define the local spin density,%
\begin{equation}
\rho_{S}^{a^{\dagger},a}(\vec{r})=\sum_{i,j,i^{\prime}=0,1}a_{i,j}^{\dagger
}(\vec{r})\left[  \sigma_{S}\right]  _{ii^{\prime}}a_{i^{\prime},j}(\vec{r}),
\label{density_S}%
\end{equation}
isospin density$,$%
\begin{equation}
\rho_{I}^{a^{\dagger},a}(\vec{r})=\sum_{i,j,j^{\prime}=0,1}a_{i,j}^{\dagger
}(\vec{r})\left[  \tau_{I}\right]  _{jj^{\prime}}a_{i,j^{\prime}}(\vec{r}),
\label{density_I}%
\end{equation}
and spin-isospin density,%
\begin{equation}
\rho_{S,I}^{a^{\dagger},a}(\vec{r})=\sum_{i,j,i^{\prime},j^{\prime}%
=0,1}a_{i,j}^{\dagger}(\vec{r})\left[  \sigma_{S}\right]  _{ii^{\prime}%
}\left[  \tau_{I}\right]  _{jj^{\prime}}a_{i^{\prime},j^{\prime}}(\vec{r}).
\label{density_SI}%
\end{equation}

At leading order the effective Hamiltonian can be written as%
\begin{equation}
H_{\text{LO}}=H_{\text{free}}+V_{\text{LO}}, \label{pionless_hamiltonian}%
\end{equation}
where%
\begin{equation}
H_{\text{free}}=\frac{1}{2m}\sum_{i,j=0,1}\int d^{3}\vec{r}\;\vec{\nabla
}a_{i,j}^{\dagger}(\vec{r})\cdot\vec{\nabla}a_{i,j}(\vec{r}), \label{Hfree_4}%
\end{equation}%
\begin{equation}
V_{\text{LO}}=V+V_{I^{2}}+V^{(3N)},
\end{equation}%
\begin{equation}
V=\frac{C}{2}\int d^{3}\vec{r}\;:\left[  \rho^{a^{\dagger},a}(\vec{r})\right]
^{2}:, \label{V_4}%
\end{equation}%
\begin{equation}
V_{I^{2}}=\frac{C_{I^{2}}}{2}\sum_{I=1,2,3}\int d^{3}\vec{r}\;:\left[
\rho_{I}^{a^{\dagger},a}(\vec{r})\right]  ^{2}, \label{V_I2}%
\end{equation}%
\begin{equation}
V^{(3N)}=\frac{D}{6}\int d^{3}\vec{r}\;:\left[  \rho^{a^{\dagger},a}(\vec
{r})\right]  ^{3}:.
\end{equation}
Due to an instability in the limit of zero-range interactions
\cite{Thomas:1935},\ the SU(4)-symmetric three-nucleon force $V^{(3N)}$ is
needed for consistent renormalization at leading order
\cite{Bedaque:1998kg,Bedaque:1998km,Bedaque:1999ve}. \ With the constraint of
antisymmetry there are two independent $S$-wave nucleon-nucleon scattering
channels. \ These correspond with spin-isospin quantum numbers $S=1$, $I=0$
and $S=0$, $I=1$. \ Some analytic methods used in pionless effective field
theory are discussed in Ref.~\cite{vanKolck:1998bw,Chen:1999tn}. \ A general
overview of methods in pionless effective field theory can be found in recent
reviews \cite{vanKolck:1999mw,Bedaque:2002mn,Epelbaum:2005pn}. \ 

\subsection{Chiral effective field theory}

For nucleon momenta comparable to the pion mass, the contribution from pion
modes must be included in the effective theory. \ In the following $\vec{q}$
denotes the $t$-channel momentum transfer for nucleon-nucleon scattering while
$\vec{k}$ is the $u$-channel exchanged momentum transfer. At leading order in
the Weinberg power-counting scheme \cite{Weinberg:1990rz,Weinberg:1991um} the
nucleon-nucleon effective potential is%
\begin{equation}
H_{\text{LO}}=H_{\text{free}}+V_{\text{LO}},
\end{equation}%
\begin{equation}
V_{\text{LO}}=V+V_{I^{2}}+V^{\text{OPEP}}.
\end{equation}
$H_{\text{free}},$ $V$, $V_{I^{2}}$ are defined in the same manner as in
Eq.~(\ref{Hfree_4}), (\ref{V_4}), (\ref{V_I2}). $\ V^{\text{OPEP}}$ is the
instantaneous one-pion exchange potential,%
\begin{equation}
V^{\text{OPEP}}=\sum_{\substack{S_{1},S_{2},I=1,2,3}}\int d^{3}\vec{r}%
_{1}d^{3}\vec{r}_{2}G_{S_{1}S_{2}}(\vec{r}_{1}-\vec{r}_{2}):\rho_{S_{1}%
,I}^{a^{\dag},a}(\vec{r}_{1})\rho_{S_{2},I}^{a^{\dag},a}(\vec{r}_{2}):,
\end{equation}
where the spin-isospin density $\rho_{S,I}^{a^{\dag},a}$ is defined in
Eq.~(\ref{density_SI}) and%
\begin{equation}
G_{S_{1}S_{2}}(\vec{r}_{1}-\vec{r}_{2})=-\left(  \frac{g_{A}}{2f_{\pi}%
}\right)  ^{2}\int\frac{d^{3}\vec{q}}{\left(  2\pi\right)  ^{3}}\frac
{q_{S_{1}}q_{S_{2}}e^{i\vec{q}\cdot(\vec{r}_{1}-\vec{r}_{2})}}{q^{\,2}+m_{\pi
}^{2}}.
\end{equation}
For our physical constants we take $m=938.92$ MeV as the nucleon mass,
$m_{\pi}=138.08$ MeV as the pion mass, $f_{\pi}=93$ MeV as the pion decay
constant, and $g_{A}=1.26$ as the nucleon axial charge.

The terms in $V_{\text{LO}}$ can be written more compactly in terms of their
matrix elements with two-nucleon momentum states. \ The tree-level amplitude
for two-nucleon scattering consists of contributions from direct and exchange
diagrams. \ However for bookkeeping purposes we label the amplitude as though
the two interacting nucleons are distinguishable. \ We label one nucleon as
type $A$, the other nucleon as type $B$, and the interactions include
densities for both $A$ and $B$. \ For example the total nucleon density
becomes%
\begin{equation}
\rho^{a^{\dagger},a}\rightarrow\rho^{a_{A}^{\dagger},a_{A}}+\rho
^{a_{B}^{\dagger},a_{B}}.
\end{equation}
The amplitudes are then%
\begin{equation}
\mathcal{A}\left(  V\right)  =C,
\end{equation}%
\begin{equation}
\mathcal{A}\left(  V_{I^{2}}\right)  =C_{I^{2}}\sum_{I}\tau_{I}^{A}\tau
_{I}^{B},
\end{equation}%
\begin{equation}
\mathcal{A}\left(  V^{\text{OPEP}}\right)  =-\left(  \frac{g_{A}}{2f_{\pi}%
}\right)  ^{2}\frac{\sum_{I}\tau_{I}^{A}\tau_{I}^{B}\sum_{S}q_{S}\sigma
_{S}^{A}\sum_{S^{\prime}}q_{S^{\prime}}\sigma_{S^{\prime}}^{B}}{q^{\,2}%
+m_{\pi}^{2}}.
\end{equation}

At next-to-leading order (NLO) the effective potential introduces corrections
to the two LO\ contact terms, seven independent contact terms carrying two
powers of momentum, and instantaneous two-pion exchange (TPEP)
\cite{Ordonez:1992xp,Ordonez:1993tn,Ordonez:1996rz,Epelbaum:1998ka,Epelbaum:1999dj}%
. \ We write this as%
\begin{equation}
V_{\text{NLO}}=V_{\text{LO}}+\Delta V^{(0)}+V^{(2)}+V_{\text{NLO}%
}^{\text{TPEP}}.
\end{equation}
The tree-level amplitudes for the new contact interactions are%
\begin{equation}
\mathcal{A}\left(  \Delta V\right)  =\Delta C,
\end{equation}%
\begin{equation}
\mathcal{A}\left(  \Delta V_{I^{2}}\right)  =\Delta C_{I^{2}}\sum_{I}\tau
_{I}^{A}\tau_{I}^{B},
\end{equation}%
\begin{equation}
\mathcal{A}\left(  V_{q^{2}}\right)  =C_{q^{2}}q^{2},
\end{equation}%
\begin{equation}
\mathcal{A}\left(  V_{I^{2},q^{2}}\right)  =C_{I^{2},q^{2}}q^{2}\sum_{I}%
\tau_{I}^{A}\tau_{I}^{B},
\end{equation}%
\begin{equation}
\mathcal{A}\left(  V_{S^{2},q^{2}}\right)  =C_{S^{2},q^{2}}q^{2}\sum_{S}%
\sigma_{S}^{A}\sigma_{S}^{B},
\end{equation}%
\begin{equation}
\mathcal{A}\left(  V_{S^{2},I^{2},q^{2}}\right)  =C_{S^{2},I^{2},q^{2}}%
q^{2}\sum_{S}\sigma_{S}^{A}\sigma_{S}^{B}\sum_{I}\tau_{I}^{A}\tau_{I}^{B},
\end{equation}%
\begin{equation}
\mathcal{A}\left(  V_{(q\cdot S)^{2}}\right)  =C_{(q\cdot S)^{2}}\sum_{S}%
q_{S}\sigma_{S}^{A}\sum_{S^{\prime}}q_{S^{\prime}}\sigma_{S^{\prime}}^{B},
\end{equation}%
\begin{equation}
\mathcal{A}\left(  V_{I^{2},(q\cdot S)^{2}}\right)  =C_{I^{2},(q\cdot S)^{2}%
}\sum_{I}\tau_{I}^{A}\tau_{I}^{B}\sum_{S}q_{S}\sigma_{S}^{A}\sum_{S^{\prime}%
}q_{S^{\prime}}\sigma_{S^{\prime}}^{B},
\end{equation}%
\begin{equation}
\mathcal{A}\left(  V_{(iq\times S)\cdot k}\right)  =iC_{(iq\times S)\cdot
k}\sum_{l,S,l^{\prime}}\varepsilon_{lSl^{\prime}}q_{l}\left(  \sigma
^{A}+\sigma^{B}\right)  _{S}k_{l^{\prime}}.
\end{equation}
The amplitude for NLO two-pion exchange potential is
\cite{Friar:1994,Kaiser:1997mw}%
\begin{align}
\mathcal{A}\left(  V_{\text{NLO}}^{\text{TPEP}}\right)   &  =-\frac{\sum
_{I}\tau_{I}^{A}\tau_{I}^{B}}{384\pi^{2}f_{\pi}^{4}}L(q)\left[  4m_{\pi}%
^{2}\left(  5g_{A}^{4}-4g_{A}^{2}-1\right)  +q^{2}\left(  23g_{A}^{4}%
-10g_{A}^{2}-1\right)  +\frac{48g_{A}^{4}m_{\pi}^{4}}{4m_{\pi}^{2}+q^{2}%
}\right] \nonumber\\
&  -\frac{3g_{A}^{4}}{64\pi^{2}f_{\pi}^{4}}L(q)\left[  \sum_{S}q_{S}\sigma
_{S}^{A}\sum_{S^{\prime}}q_{S^{\prime}}\sigma_{S^{\prime}}^{B}-q^{2}\sum
_{S}\sigma_{S}^{A}\sigma_{S}^{B}\right]  ,
\end{align}
where%
\begin{equation}
L(q)=\frac{1}{2q}\sqrt{4m_{\pi}^{2}+q^{2}}\ln\frac{\sqrt{4m_{\pi}^{2}+q^{2}%
}+q}{\sqrt{4m_{\pi}^{2}+q^{2}}-q}.
\end{equation}
Recent reviews of chiral effective field theory can be found in
Ref.~\cite{vanKolck:1999mw,Bedaque:2002mn,Epelbaum:2005pn}.

\subsection{Three-nucleon forces}

The systematic framework provided by effective field theory becomes very
useful when discussing the form of the dominant three-nucleon interactions.
\ Few-nucleon forces in chiral effective field theory beyond two nucleons were
first discussed qualitatively in Ref.~\cite{Weinberg:1991um}. \ In
Ref.~\cite{vanKolck:1994yi} it was shown that the three-nucleon terms at NLO
cancelled, and the leading three-nucleon effects appeared at next-to-next-to
leading order (NNLO) in Weinberg power counting.

The NNLO three-nucleon effective potential arises from a pure contact
potential, $V_{\text{contact}}^{(3N)}$, one-pion exchange potential,
$V_{\text{OPE}}^{(3N)}$, and a two-pion exchange potential, $V_{\text{TPE}%
}^{(3N)}$. \ Parts of the NNLO three-nucleon potential are also contained in a
number of phenomenological three-nucleon potentials
\cite{Fujita:1957zz,Yang:1974zz,Coon:1978gr,Coon:1981TM,Carlson:1983kq,Pudliner:1997ck}%
. \ However there is clear value in identifying the full set of leading
interactions. \ Similar to our description above for two-nucleon scattering,
we write the tree-level amplitude for three-nucleon scattering where the first
nucleon is of type $A$, the second nucleon type $B$, and the three type $C$.
\ The amplitudes are \cite{Friar:1998zt,Epelbaum:2002vt}%
\begin{equation}
\mathcal{A}\left(  V_{\text{contact}}^{(3N)}\right)  =D_{\text{contact}},
\end{equation}%
\begin{equation}
\mathcal{A}\left(  V_{\text{OPE}}^{(3N)}\right)  =-D_{\text{OPE}}\frac{g_{A}%
}{2f_{\pi}}\sum_{\text{perm }A,B,C}\frac{\left(  \vec{q}_{A}\cdot\vec{\sigma
}_{A}\right)  \left(  \vec{q}_{A}\cdot\vec{\sigma}_{B}\right)  }{q_{A}%
^{2}+m_{\pi}^{2}}\left(  \vec{\tau}_{A}\cdot\vec{\tau}_{B}\right)  ,
\end{equation}%
\begin{align}
\mathcal{A}\left(  V_{\text{TPE}}^{(3N)}\right)   &  =c_{3}\frac{g_{A}^{2}%
}{4f_{\pi}^{4}}\sum_{\text{perm }A,B,C}\frac{\left(  \vec{q}_{A}\cdot
\vec{\sigma}_{A}\right)  \left(  \vec{q}_{B}\cdot\vec{\sigma}_{B}\right)
\left(  \vec{q}_{A}\cdot\vec{q}_{B}\right)  }{\left(  q_{A}^{2}+m_{\pi}%
^{2}\right)  \left(  q_{B}^{2}+m_{\pi}^{2}\right)  }\left(  \vec{\tau}%
_{A}\cdot\vec{\tau}_{B}\right) \nonumber\\
&  -c_{1}\frac{m_{\pi}^{2}g_{A}^{2}}{2f_{\pi}^{4}}\sum_{\text{perm }%
A,B,C}\frac{\left(  \vec{q}_{A}\cdot\vec{\sigma}_{A}\right)  \left(  \vec
{q}_{B}\cdot\vec{\sigma}_{B}\right)  }{\left(  q_{A}^{2}+m_{\pi}^{2}\right)
\left(  q_{B}^{2}+m_{\pi}^{2}\right)  }\left(  \vec{\tau}_{A}\cdot\vec{\tau
}_{B}\right) \nonumber\\
&  +c_{4}\frac{g_{A}^{2}}{8f_{\pi}^{4}}\sum_{\text{perm }A,B,C}\frac{\left(
\vec{q}_{A}\cdot\vec{\sigma}_{A}\right)  \left(  \vec{q}_{B}\cdot\vec{\sigma
}_{B}\right)  }{\left(  q_{A}^{2}+m_{\pi}^{2}\right)  \left(  q_{B}^{2}%
+m_{\pi}^{2}\right)  }\left[  \left(  \vec{q}_{A}\times\vec{q}_{B}\right)
\cdot\vec{\sigma}_{C}\right]  \left[  \left(  \vec{\tau}_{A}\times\vec{\tau
}_{B}\right)  \cdot\vec{\tau}_{C}\right]  .
\end{align}
In our notation $\vec{q}_{A}$, $\vec{q}_{B}$, $\vec{q}_{C}$ are the
differences between final and initial momenta for the respective nucleons.
\ The summations are over permutations of the bookkeeping labels $A,B,C$.

The coefficients $c_{1,3,4}$ are $\pi\pi NN$ interaction terms in the chiral
Lagrangian and are determined from fits to low-energy scattering data
\cite{Bernard:1995dp}. \ The remaining unknown coefficients $D_{\text{contact}%
}$ and $D_{\text{OPE}}$ are cutoff dependent. \ In Ref.~\cite{Epelbaum:2002vt}
these were fit to the triton binding energy and spin-doublet neutron-deuteron
scattering length. \ The resulting NNLO effective potential was shown to give
a prediction for the isospin-symmetric alpha binding energy accurate to within
a fraction of $1$ MeV.

\subsection{Non-perturbative physics and power counting}

When non-perturbative processes are involved, reaching the continuum limit and
power counting in effective field theory can sometimes become complicated.
\ The two-component effective Hamiltonian for cold atoms introduced in
Eq.~(\ref{two_component_hamiltonian}) has no such complications. \ Ultraviolet
divergences can be absorbed by renormalizing the interaction coefficient $C$,
and the cutoff momentum can be taken to infinity. \ Similarly the
leading-order pionless effective Hamiltonian in
Eq.~(\ref{pionless_hamiltonian}) has a well-defined continuum limit if we
neglect deeply-bound three-body states that decouple from the low-energy
effective theory. \ While these deeply-bound states generate instabilities in
numerical simulations they can be removed by hand in semi-analytic
calculations \cite{Bedaque:1998kg,Bedaque:1998km,Bedaque:1999ve}.

In chiral effective field theory there has been considerable study on the
consistency of the Weinberg power counting scheme at high momentum cutoff.
\ Complications arise from the singular behavior of the one-pion exchange
potential. In order to avoid unsubtracted ultraviolet divergences produced by
infinite iteration of the one-pion exchange potential, an alternative scheme
was proposed where pion exchange is treated perturbatively
\cite{Kaplan:1996xu,Kaplan:1998tg,Kaplan:1998we}. \ This approach, KSW power
counting, allows for systematic control of the ultraviolet divergence
structure of the effective theory. \ Unfortunately the convergence at higher
order is poor in some partial waves for momenta comparable to the pion mass
\cite{Fleming:1999ee}.

The most divergent short-distance part of the one-pion exchange potential is a
$f_{\pi}^{-2}r^{-3}$ singularity arising from the tensor force in the
spin-triplet channel. \ There are also subleading divergences at $r=0$ which
contain explicit factors of the pion mass. \ Based on this observation another
power counting scheme was proposed in Ref.~\cite{Beane:2001bc}. \ This new
scheme coincides with KSW power counting in the spin-singlet channel. \ But in
the spin-triplet channel the most singular piece of the one-pion exchange
potential is iterated non-perturbatively, while the rest is incorporated as a
perturbative expansion around $m_{\pi}=0$.

More recently a different power counting modification was proposed in
Ref.~\cite{Nogga:2005hy}. \ In this approach the one-pion exchange potential
is treated non-perturbatively in lower angular momentum channels along with
higher-derivative counterterms promoted to leading order. \ These counterterms
are used to cancel cutoff dependence in channels where the tensor force is
attractive and strong enough to overcome the centrifugal barrier. \ Advantages
over Weinberg power counting at leading order were shown for cutoff momenta
much greater than the pion mass. \ Further investigations of this approach in
higher partial waves and power counting with one-pion exchange were considered
in Ref.~\cite{Birse:2005um,Birse:2007sx}.

The choice of cutoff momentum and power counting scheme in lattice effective
field theory is shaped to a large extent by computational constraints. \ For
two-nucleon scattering in chiral effective field theory, small lattice
spacings corresponding with cutoff momenta many times greater than the pion
mass are no problem. \ However at small lattice spacing significant numerical
problems appear in simulations of few- and many-nucleon systems. \ In
attractive channels one must contend with spurious deeply-bound states that
spoil Euclidean time projection methods (a technique described later in this
review). \ In channels where the short-range interactions are repulsive a
different problem arises. \ In auxiliary-field and diagrammatic Monte Carlo
(methods we discuss later in this review), repulsive interactions produce sign
or complex phase oscillations that render the method ineffective. \ Due to
these practical computational issues one must settle for lattice simulations
where the cutoff momentum is only a few times the pion mass, and the
advantages of the improved scheme over Weinberg power counting are numerically
small \cite{Epelbaum:2006pt}.

\section{Lattice formulations for zero-range attractive two-component
fermions}

\begin{figure}
[ptb]
\begin{center}
\includegraphics[
height=1.8118in,
width=4.6959in
]%
{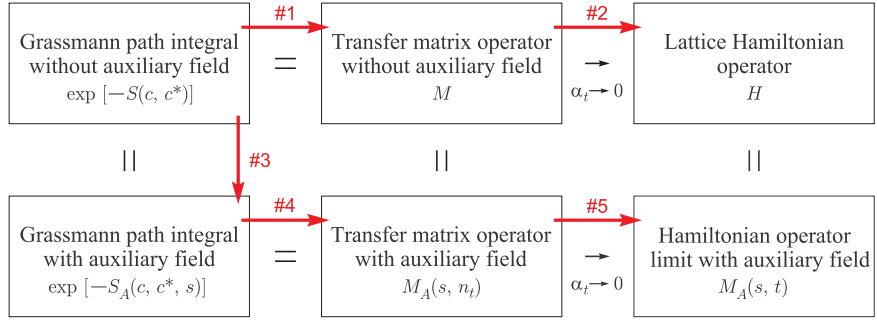}%
\caption{A schematic diagram of different lattice formulations. \ The numbered
arrows indicate the discussion order in the text.}%
\label{formulations}%
\end{center}
\end{figure}

In this section we introduce a number of different lattice formulations using
the example of zero-range attractive two-component fermions described by
$H_{\text{LO}}$ in Eq.~(\ref{two_component_hamiltonian}). \ In
Fig.~(\ref{formulations}) we show a schematic diagram of the different lattice
formulations. \ The numbered arrows indicate the discussion order in the text.%

%TCIMACRO{\FRAME{ftbpFU}{4.6959in}{1.8118in}{0pt}{\Qcb{A schematic diagram of
%different lattice formulations. \ The numbered arrows indicate the discussion
%order in the text.}}{\Qlb{formulations}}{formulations.eps}%
%width 4.6959in;  height 1.8118in;  depth 0pt;  original-width 7.7798in;
%original-height 2.9724in;  cropleft "0";  croptop "1";  cropright "1";
%cropbottom "0";  filename 'formulations.eps';file-properties "XNPEU";}} }%
%BeginExpansion
%EndExpansion
Throughout our discussion of the lattice formalism we use dimensionless
parameters and operators corresponding with physical values multiplied by the
appropriate power of the spatial lattice spacing $a$. \ In our notation the
three-component integer vector $\vec{n}$ labels the lattice sites of a
three-dimensional periodic lattice with dimensions $L^{3}$. \ The spatial
lattice unit vectors are denoted $\hat{l}=\hat{1}$, $\hat{2}$, $\hat{3}$. \ We
use $n_{t}$ to label lattice steps in the temporal direction, and $L_{t}$
denotes the total number of lattice time steps. \ The temporal lattice spacing
is given by $a_{t}$, and $\alpha_{t}=a_{t}/a$ is the ratio of the temporal to
spatial lattice spacing. \ We also define $h=\alpha_{t}/(2m)$, where $m$ is
the fermion mass in lattice units.

\subsection{Grassmann path integral without auxiliary field}

For two-component fermions with zero-range attractive interactions we start
with the lattice Grassmann path integral action without auxiliary fields. \ It
is the simplest formulation in which to derive the lattice Feynman rules.
\ Hence it is useful for both analytic lattice calculations and diagrammatic
lattice Monte Carlo simulations \cite{Burovski:2006a,Burovski:2006b}.

We let $c_{i}$ and $c_{i}^{\ast}$ be anticommuting Grassmann fields for spin
$i=\uparrow,\downarrow$. \ The Grassmann fields are periodic with respect to
the spatial lengths of the $L^{3}$ lattice,%
\begin{equation}
c_{i}(\vec{n}+L\hat{1},n_{t})=c_{i}(\vec{n}+L\hat{2},n_{t})=c_{i}(\vec
{n}+L\hat{3},n_{t})=c_{i}(\vec{n},n_{t}),
\end{equation}%
\begin{equation}
c_{i}^{\ast}(\vec{n}+L\hat{1},n_{t})=c_{i}^{\ast}(\vec{n}+L\hat{2}%
,n_{t})=c_{i}^{\ast}(\vec{n}+L\hat{3},n_{t})=c_{i}^{\ast}(\vec{n},n_{t}),
\end{equation}
and antiperiodic along the temporal direction,%
\begin{equation}
c_{i}(\vec{n},n_{t}+L_{t})=-c_{i}(\vec{n},n_{t}).
\end{equation}%
\begin{equation}
c_{i}^{\ast}(\vec{n},n_{t}+L_{t})=-c_{i}^{\ast}(\vec{n},n_{t}).
\end{equation}
We write $DcDc^{\ast}$ as shorthand for the integral measure,%
\begin{equation}
DcDc^{\ast}=\prod_{\vec{n},n_{t},i=\uparrow,\downarrow}dc_{i}(\vec{n}%
,n_{t})dc_{i}^{\ast}(\vec{n},n_{t}).
\end{equation}
We use the standard convention for Grassmann integration,%
\begin{equation}
\int dc_{i}(\vec{n},n_{t})=\int dc_{i}^{\ast}(\vec{n},n_{t})=0\text{,}%
\end{equation}%
\begin{equation}
\int dc_{i}(\vec{n},n_{t})c_{i}(\vec{n},n_{t})=\int dc_{i}^{\ast}(\vec
{n},n_{t})c_{i}^{\ast}(\vec{n},n_{t})=1\text{ \ (no sum on }i\text{)}.
\end{equation}

Local Grassmann densities $\rho_{\uparrow},\rho_{\downarrow},\rho$ are defined
in terms of bilinear products of the Grassmann fields,%
\begin{equation}
\rho_{\uparrow}(\vec{n},n_{t})=c_{\uparrow}^{\ast}(\vec{n},n_{t})c_{\uparrow
}(\vec{n},n_{t}),
\end{equation}%
\begin{equation}
\rho_{\downarrow}(\vec{n},n_{t})=c_{\downarrow}^{\ast}(\vec{n},n_{t}%
)c_{\downarrow}(\vec{n},n_{t}),
\end{equation}%
\begin{equation}
\rho(\vec{n},n_{t})=\rho_{\uparrow}(\vec{n},n_{t})+\rho_{\downarrow}(\vec
{n},n_{t}).
\end{equation}
We consider the Grassmann path integral%
\begin{equation}
\mathcal{Z}=\int DcDc^{\ast}\exp\left[  -S\left(  c,c^{\ast}\right)  \right]
, \label{defining_Z}%
\end{equation}
where%
\begin{equation}
S(c,c^{\ast})=S_{\text{free}}(c,c^{\ast})+C\alpha_{t}\sum_{\vec{n},n_{t}}%
\rho_{\uparrow}(\vec{n},n_{t})\rho_{\downarrow}(\vec{n},n_{t}).
\label{path_nonaux}%
\end{equation}
The action $S(c,c^{\ast})$ consists of the free nonrelativistic fermion action%
\begin{align}
S_{\text{free}}(c,c^{\ast})  &  =\sum_{\vec{n},n_{t},i=\uparrow,\downarrow
}\left[  c_{i}^{\ast}(\vec{n},n_{t})c_{i}(\vec{n},n_{t}+1)-(1-6h)c_{i}^{\ast
}(\vec{n},n_{t})c_{i}(\vec{n},n_{t})\right] \nonumber\\
&  -h\sum_{\vec{n},n_{t},i=\uparrow,\downarrow}\sum_{l=1,2,3}\left[
c_{i}^{\ast}(\vec{n},n_{t})c_{i}(\vec{n}+\hat{l},n_{t})+c_{i}^{\ast}(\vec
{n},n_{t})c_{i}(\vec{n}-\hat{l},n_{t})\right]  ,
\end{align}
and a contact interaction between up and down spins. \ We consider the case
where the coefficient $C$ is negative, corresponding with an attractive
interaction. \ Since we are considering nonrelativistic lattice fermions with
a quadratic dispersion relation, the lattice doubling problem associated with
relativistic fermions does not occur.

In the grand canonical ensemble a common chemical potential $\mu$ is added for
all spins. \ In this case the $\mu$-dependent path integral is%
\begin{equation}
\mathcal{Z}\text{(}\mu\text{)}=\int DcDc^{\ast}\exp\left[  -S(c,c^{\ast}%
,\mu)\right]  ,
\end{equation}
where%
\begin{equation}
S(c,c^{\ast},\mu)=S(e^{\mu\alpha_{t}}c,c^{\ast})+\sum_{\vec{n},n_{t}%
,i=\uparrow,\downarrow}\left[  \left(  1-e^{\mu\alpha_{t}}\right)  c_{i}%
^{\ast}(\vec{n},n_{t})c_{i}(\vec{n},n_{t}+1)\right]  ,
\end{equation}
and $S(e^{\mu\alpha_{t}}c,c^{\ast})$ is the same as $S(c,c^{\ast})$ defined in
Eq.~(\ref{path_nonaux}), but with $c$ replaced by $e^{\mu\alpha_{t}}c.$

\subsection{Transfer matrix operator without auxiliary field}

Let $a$ and $a^{\dagger}$ denote fermion annihilation and creation operators
satisfying the usual anticommutation relations%
\begin{equation}
\left\{  a,a\right\}  =\left\{  a^{\dagger},a^{\dagger}\right\}  =0,
\end{equation}%
\begin{equation}
\left\{  a,a^{\dagger}\right\}  =1.
\end{equation}
For any function $f\left(  a^{\dagger},a\right)  $ we note the identity
\cite{Creutz:1999zy}%
\begin{equation}
Tr\left[  \colon f\left(  a^{\dagger},a\right)  \colon\right]  =\int
dcdc^{\ast}e^{2c^{\ast}c}f(c^{\ast},c), \label{simple_correspondence}%
\end{equation}
where $c$ and $c^{\ast}$ are Grassmann variables. \ As before the $::$ symbols
in Eq.~(\ref{simple_correspondence}) indicate normal ordering, and the trace
is evaluated over all possible fermion states. \ This result can be checked
explicitly using the complete set of possible functions $\left\{
1,a,a^{\dagger},a^{\dag}a\right\}  $.

It is useful to write Eq.~(\ref{simple_correspondence}) in a form that
resembles a path integral over a short time interval with antiperiodic
boundary conditions,%
\begin{equation}
Tr\left[  \colon f\left(  a^{\dagger},a\right)  \colon\right]  =\int
dc(0)dc^{\ast}(0)e^{c^{\ast}(0)\left[  c(0)-c(1)\right]  }f\left[  c^{\ast
}(0),c(0)\right]  ,
\end{equation}%
\begin{equation}
c(1)=-c(0)\text{.}%
\end{equation}
This result can be generalized to products of normal-ordered functions of
several creation and annihilation operators. \ Let $a_{i}(\vec{n})$ and
$a_{i}^{\dagger}(\vec{n})$ denote fermion annihilation and creation operators
for spin $i$ at lattice site $\vec{n}$. \ We can write any Grassmann path
integral with instantaneous interactions as the trace of a product of
operators using the identity \cite{Creutz:1988wv,Creutz:1999zy}%
\begin{align}
&  Tr\left\{  \colon F_{L_{t}-1}\left[  a_{i^{\prime}}^{\dagger}(\vec
{n}^{\prime}),a_{i}(\vec{n})\right]  \colon\times\cdots\times\colon
F_{0}\left[  a_{i^{\prime}}^{\dagger}(\vec{n}^{\prime}),a_{i}(\vec{n})\right]
\colon\right\} \nonumber\\
&  =\int DcDc^{\ast}\exp\left\{  \sum_{n_{t}=0}^{L_{t}-1}\sum_{\vec{n},i}%
c_{i}^{\ast}(\vec{n},n_{t})\left[  c_{i}(\vec{n},n_{t})-c_{i}(\vec{n}%
,n_{t}+1)\right]  \right\} \nonumber\\
&  \qquad\qquad\qquad\times\prod_{n_{t}=0}^{L_{t}-1}F_{n_{t}}\left[
c_{i^{\prime}}^{\ast}(\vec{n}^{\prime},n_{t}),c_{i}(\vec{n},n_{t})\right]  ,
\label{correspondence}%
\end{align}
where $c_{i}(\vec{n},L_{t})=-c_{i}(\vec{n},0)$.

Let us define the free nonrelativistic lattice Hamiltonian
\begin{equation}
H_{\text{free}}=\frac{3}{m}\sum_{\vec{n},i=\uparrow,\downarrow}a_{i}^{\dagger
}(\vec{n})a_{i}(\vec{n})-\frac{1}{2m}\sum_{\vec{n},i=\uparrow,\downarrow}%
\sum_{l=1,2,3}\left[  a_{i}^{\dagger}(\vec{n})a_{i}(\vec{n}+\hat{l}%
)+a_{i}^{\dagger}(\vec{n})a_{i}(\vec{n}-\hat{l})\right]  ,
\end{equation}
as well as the lattice density operators%
\begin{equation}
\rho_{\uparrow}^{a^{\dagger}a}(\vec{n})=a_{\uparrow}^{\dagger}(\vec
{n})a_{\uparrow}(\vec{n}),
\end{equation}%
\begin{equation}
\rho_{\downarrow}^{a^{\dagger}a}(\vec{n})=a_{\downarrow}^{\dagger}(\vec
{n})a_{\downarrow}(\vec{n}),
\end{equation}%
\begin{equation}
\rho^{a^{\dagger}a}(\vec{n})=\rho_{\uparrow}^{a^{\dagger}a}(\vec{n}%
)+\rho_{\downarrow}^{a^{\dagger}a}(\vec{n}).
\end{equation}
Using the correspondence Eq.~(\ref{correspondence}), we can rewrite the path
integral $\mathcal{Z}$ defined in Eq.~(\ref{defining_Z}) as a transfer matrix
partition function,%
\begin{equation}
\mathcal{Z}=Tr\left(  M^{L_{t}}\right)  ,
\end{equation}
where $M$ is the normal-ordered transfer matrix operator%
\begin{equation}
M=:\exp\left[  -H_{\text{free}}\alpha_{t}-C\alpha_{t}\sum_{\vec{n}}%
\rho_{\uparrow}^{a^{\dagger}a}(\vec{n})\rho_{\downarrow}^{a^{\dagger}a}%
(\vec{n})\right]  :. \label{transfer_noaux}%
\end{equation}
Roughly speaking the transfer matrix operator is the exponential of the
Hamiltonian operator over one Euclidean lattice time step, $e^{-H\alpha_{t}}$.
\ In order to satisfy the identity Eq.~(\ref{correspondence}), we work with
normal-ordered transfer matrix operators. \ In the limit of zero temporal
lattice spacing, $\alpha_{t}\rightarrow0$, we obtain the Hamiltonian lattice
formulation with Hamiltonian
\begin{equation}
H=H_{\text{free}}+C\sum_{\vec{n}}\rho_{\uparrow}^{a^{\dagger}a}(\vec{n}%
)\rho_{\downarrow}^{a^{\dagger}a}(\vec{n}).
\end{equation}
This is also the defining Hamiltonian for the attractive Hubbard model in
three dimensions.

In the grand canonical ensemble the effect of the chemical potential is
equivalent to replacing $M$ by%
\begin{equation}
M(\mu)=M\exp\left\{  \mu\alpha_{t}\sum_{\vec{n}}\rho^{a^{\dagger}a}(\vec
{n})\right\}  . \label{M_mu_as_product}%
\end{equation}
For the Hamiltonian lattice formulation the effect of the chemical potential
has the familiar form%
\begin{equation}
H(\mu)=H_{\text{free}}+C\sum_{\vec{n}}\rho_{\uparrow}^{a^{\dagger}a}(\vec
{n})\rho_{\downarrow}^{a^{\dagger}a}(\vec{n})-\mu\sum_{\vec{n}}\rho
^{a^{\dagger}a}(\vec{n}).
\end{equation}

\subsection{Grassmann path integral with auxiliary field}

We can re-express the Grassmann path integral using an auxiliary field coupled
to the particle density. \ This lattice formulation has been used in several
lattice studies at nonzero temperature
\cite{Chen:2003vy,Lee:2004qd,Lee:2004si,Wingate:2005xy,Lee:2005is,Lee:2005it,Abe:2007fe,Abe:2007ff}%
. \ Due to the simple contact interaction $\rho_{\uparrow}(\vec{n},n_{t}%
)\rho_{\downarrow}(\vec{n},n_{t})$ and the anticommutation of Grassmann
variables, there is a large class of auxiliary-field transformations which
reproduce the same action.

Let us write the Grassmann path integral using the auxiliary field $s,$%
\begin{equation}
\mathcal{Z}=\prod\limits_{\vec{n},n_{t}}\left[  \int d_{A}s(\vec{n}%
,n_{t})\right]  \int DcDc^{\ast}\exp\left[  -S_{A}\left(  c,c^{\ast},s\right)
\right]  , \label{Z_A}%
\end{equation}
where%
\begin{equation}
S_{A}\left(  c,c^{\ast},s\right)  =S_{\text{free}}(c,c^{\ast})-\sum_{\vec
{n},n_{t}}A\left[  s(\vec{n},n_{t})\right]  \rho(\vec{n},n_{t}). \label{Aj}%
\end{equation}
One possible example is a Gaussian-integral transformation similar to the
original Hubbard-Stratonovich transformation
\cite{Stratonovich:1958,Hubbard:1959ub} where%
\begin{equation}
\int d_{A}s(\vec{n},n_{t})=\frac{1}{\sqrt{2\pi}}\int_{-\infty}^{+\infty
}ds(\vec{n},n_{t})e^{-\frac{1}{2}s^{2}(\vec{n},n_{t})},
\end{equation}%
\begin{equation}
A\left[  s(\vec{n},n_{t})\right]  =\sqrt{-C\alpha_{t}}\,s(\vec{n},n_{t}).
\end{equation}
Another possibility is a discrete auxiliary-field transformation similar to
that used in Ref.~\cite{Hirsch:1983}. \ In our notation this can be written
as
\begin{equation}%
%TCIMACRO{\dint }%
%BeginExpansion
{\displaystyle\int}
%EndExpansion
d_{A}s(\vec{n},n_{t})=%
%TCIMACRO{\dint _{-1/2}^{1/2}}%
%BeginExpansion
{\displaystyle\int_{-1/2}^{1/2}}
%EndExpansion
ds(\vec{n},n_{t}),
\end{equation}%
\begin{equation}
A\left[  s(\vec{n},n_{t})\right]  =\sqrt{-C\alpha_{t}}\,\text{sgn}\left[
s(\vec{n},n_{t})\right]  ,
\end{equation}
where sgn equals $+1$ for positive values and $-1$ for negative values. \ In
Ref.~\cite{Lee:2008xs} the performance of four different auxiliary-field
transformations were compared.

We intentionally leave the forms for $d_{A}s(\vec{n},n_{t})$ and $A\left[
s(\vec{n},n_{t})\right]  $ unspecified, except for a number of conditions
needed to recover Eq.~(\ref{defining_Z}) upon integrating out the auxiliary
field $s$. \ The first two conditions we set are
\begin{equation}
\int d_{A}s(\vec{n},n_{t})1=1,
\end{equation}%
\begin{equation}
\int d_{A}s(\vec{n},n_{t})\,A\left[  s(\vec{n},n_{t})\right]  =0.
\end{equation}
Since all even products of Grassmann variables commute, we can factor out the
term in Eq.~(\ref{Z_A}) involving the auxiliary field $s$ at $\vec{n},n_{t}$.
\ To shorten the notation we temporarily omit writing $\vec{n},n_{t}$
explicitly. \ We find%
\begin{align}
\int d_{A}s\exp\left[  \,A\left(  s\right)  \left(  \rho_{\uparrow}%
+\rho_{\downarrow}\right)  \right]   &  =\int d_{A}s\left[  1+\,A\left(
s\right)  \left(  \rho_{\uparrow}+\rho_{\downarrow}\right)  +A^{2}\left(
s\right)  \rho_{\uparrow}\rho_{\downarrow}\right] \nonumber\\
&  =1+\int d_{A}s\,A^{2}\left(  s\right)  \rho_{\uparrow}\rho_{\downarrow
}=\exp\left[  \int d_{A}s\,A^{2}\left(  s\right)  \rho_{\uparrow}%
\rho_{\downarrow}\right]  .
\end{align}
Therefore the last condition needed to recover Eq.~(\ref{defining_Z}) is
\begin{equation}
-C\alpha_{t}=\int d_{A}s\,A^{2}\left(  s\right)  .
\end{equation}
In the grand canonical ensemble, the auxiliary-field path integral at chemical
potential $\mu$ is%
\begin{equation}
\mathcal{Z}\text{(}\mu\text{)}=\prod\limits_{\vec{n},n_{t}}\left[  \int
d_{A}s(\vec{n},n_{t})\right]  \int DcDc^{\ast}\exp\left[  -S_{A}\left(
c,c^{\ast},s,\mu\right)  \right]  , \label{Z_A_mu_path}%
\end{equation}
where%
\begin{equation}
S_{A}\left(  c,c^{\ast},s,\mu\right)  =S_{A}(e^{\mu\alpha_{t}}c,c^{\ast
},s)+\sum_{\vec{n},n_{t},i=\uparrow,\downarrow}\left[  \left(  1-e^{\mu
\alpha_{t}}\right)  c_{i}^{\ast}(\vec{n},n_{t})c_{i}(\vec{n},n_{t}+1)\right]
. \label{S_A_mu}%
\end{equation}

\subsection{Transfer matrix operator with auxiliary field}

Using Eq.~(\ref{correspondence}) and (\ref{Z_A}) we can write $\mathcal{Z}$ as
a product of transfer matrix operators which depend on the auxiliary field,%
\begin{equation}
\mathcal{Z}=\prod\limits_{\vec{n},n_{t}}\left[  \int d_{A}s(\vec{n}%
,n_{t})\right]  Tr\left\{  M_{A}(s,L_{t}-1)\cdot\cdots\cdot M_{A}%
(s,0)\right\}  ,
\end{equation}
where%
\begin{equation}
M_{A}(s,n_{t})=\colon\exp\left\{  -H_{\text{free}}\alpha_{t}+\sum_{\vec{n}%
}A\left[  s(\vec{n},n_{t})\right]  \rho^{a^{\dagger}a}(\vec{n})\right\}
\colon. \label{transfer_aux}%
\end{equation}
This form has been used in a number of lattice simulations
\cite{Muller:1999cp,Bulgac:2005a,Lee:2005fk,Lee:2005xy,Lee:2006hr,Borasoy:2006qn,Juillet:2007a,Abe:2007fe,Abe:2007ff,Borasoy:2007vi,Borasoy:2007vk,Bulgac:2008b,Lee:2008xs,Bulgac:2008c}%
. \ In some of these studies the Hamiltonian limit $\alpha_{t}\rightarrow0$ is
also taken.

In the grand canonical ensemble at chemical potential $\mu$ the partition
function is
\begin{equation}
\mathcal{Z}(\mu)=\prod\limits_{\vec{n},n_{t}}\left[  \int d_{A}s(\vec{n}%
,n_{t})\right]  Tr\left\{  M_{A}(s,L_{t}-1,\mu)\cdot\cdots\cdot M_{A}%
(s,0,\mu)\right\}  , \label{Z_mu_aux}%
\end{equation}
where $M_{A}(s,n_{t},\mu)$ is defined as%
\begin{equation}
M_{A}(s,n_{t},\mu)=M_{A}(s,n_{t})\exp\left\{  \mu\alpha_{t}\sum_{\vec{n}}%
\rho^{a^{\dagger}a}(\vec{n})\right\}  . \label{MAmu}%
\end{equation}

\subsection{Improved lattice dispersion relations}

In Ref.~\cite{Bulgac:2005a,Bulgac:2008b,Bulgac:2008c} the transfer matrix
operator at chemical potential $\mu$ was written as%
\begin{equation}
\exp\left[  -\frac{\alpha_{t}}{2}\left(  H_{\text{free}}-\mu\hat{N}\right)
\right]  \exp\left[  -C\alpha_{t}\sum_{\vec{n}}\rho_{\uparrow}^{a^{\dagger}%
a}(\vec{n})\rho_{\downarrow}^{a^{\dagger}a}(\vec{n})\right]  \exp\left[
-\frac{\alpha_{t}}{2}\left(  H_{\text{free}}-\mu\hat{N}\right)  \right]
\label{Bulgac_transfer}%
\end{equation}
in the Hamiltonian limit, $\alpha_{t}\rightarrow0,$ with
\begin{equation}
\hat{N}=\sum_{\vec{n}}\rho^{a^{\dagger}a}(\vec{n}).
\end{equation}
This is different from $M(\mu)$ in Eq.~(\ref{M_mu_as_product}), but the two
are the same in the Hamiltonian limit. \ The exponential interaction term in
Eq.$~$(\ref{Bulgac_transfer}) was treated using a discrete auxiliary field.
\ Also the matrix elements of $H_{\text{free}}$ were computed by Fast Fourier
Transform in momentum space using the quadratic dispersion relation%
\begin{equation}
\omega^{(\text{quad})}(\vec{p})=\frac{1}{2m}\sum_{l=1,2,3}p_{l}^{2},
\label{omega_quad}%
\end{equation}
with $p_{l}$ defined in the first Brillouin zone, $\left\vert p_{l}\right\vert
\leq\pi$. \ The motivation for this approach was to remove errors associated
with the standard lattice dispersion relation%
\begin{equation}
\omega(\vec{p})=\frac{1}{m}\sum_{l=1,2,3}\left(  1-\cos p_{l}\right)  .
\end{equation}

In Ref.~\cite{Lee:2005is,Lee:2005it} lattice calculations at nonzero
temperature and large scattering length found significant errors due to
lattice artifacts. \ A detailed analysis in Ref.~\cite{Lee:2007jd} showed that
the large errors were produced by broken Galilean invariance on the lattice.
\ As an alternative to the momentum space approach in Eq.$~$(\ref{omega_quad}%
), improved lattice dispersions were investigated that could be derived from
local lattice actions.

A class of improved single-particle dispersion relations can be defined on the
lattice,%
\begin{equation}
\omega^{(n)}(\vec{p})=\frac{1}{m}\sum_{j=0,1,2,\cdots}\sum_{l=1,2,3}%
(-1)^{j}v_{j}^{(n)}\cos\left(  jp_{l}\right)  .
\end{equation}
$\omega^{(0)}(\vec{p})$ corresponds with the standard action, $\omega
^{(1)}(\vec{p})$ is the $O(a^{2})$-improved action, and so on.\ \ The improved
actions eliminate lattice artifacts in the Taylor expansion of $\omega
^{(n)}(\vec{p})$ about $\vec{p}=0,$%
\begin{equation}
\omega^{(n)}(\vec{p})=\frac{1}{2m}\sum_{l=1,2,3}p_{l}^{2}\times\left[
1+O(a^{2n+2})\right]  .
\end{equation}
The lattice action corresponding with $\omega^{(n)}$ contains hopping terms in
each spatial direction that extend $n$ lattice steps beyond the nearest
neighbor. \ The hopping coefficients $v_{j}^{(n)}$ for actions up to
$O(a^{4})$ are shown in Table \ref{hopping_coeff}.

\begin{table}[tbh]
\caption{Hopping coefficients for lattice actions up to $O(a^{4}).$}%
\label{hopping_coeff}%
$%
\begin{tabular}
[c]{|c|c|c|c|}\hline
& standard & $O(a^{2})$-improved & $O(a^{4})$-improved\\\hline
$v_{0}$ & $1$ & $\frac{5}{4}$ & $\frac{49}{36}$\\\hline
$v_{1}$ & $1$ & $\frac{4}{3}$ & $\frac{3}{2}$\\\hline
$v_{2}$ & $0$ & $\frac{1}{12}$ & $\frac{3}{20}$\\\hline
$v_{3}$ & $0$ & $0$ & $\frac{1}{90}$\\\hline
\end{tabular}
$\end{table}In addition to these improved actions, new lattice actions called
well-tempered actions were also introduced. \ These were defined implicitly in
terms of their dispersion relation,%
\begin{equation}
\omega^{(\text{wt}n)}(\vec{p})=\omega^{(n-1)}(\vec{p})+c\left[  \omega
^{(n)}(\vec{p})-\omega^{(n-1)}(\vec{p})\right]  ,
\end{equation}
where the unknown constant $c$ was determined by the integral constraint,%
\begin{equation}%
%TCIMACRO{\dint \limits_{-\pi}^{\pi}}%
%BeginExpansion
{\displaystyle\int\limits_{-\pi}^{\pi}}
%EndExpansion%
%TCIMACRO{\dint \limits_{-\pi}^{\pi}}%
%BeginExpansion
{\displaystyle\int\limits_{-\pi}^{\pi}}
%EndExpansion%
%TCIMACRO{\dint \limits_{-\pi}^{\pi}}%
%BeginExpansion
{\displaystyle\int\limits_{-\pi}^{\pi}}
%EndExpansion
dp_{1}dp_{2}dp_{3}\left[  \omega^{(\text{wt}n)}(\vec{p})-\frac{1}{2m}%
\sum_{l=1,2,3}p_{l}^{2}\right]  =0.
\end{equation}
At nonzero temperature and large scattering length the local well-tempered
action corresponding with $\omega^{(\text{wt}1)}$ was shown to be comparable
in accuracy to the nonlocal action defined by $\omega^{(\text{quad})}$
\cite{Lee:2007jd}.

\section{Lattice formulations for low-energy nucleons}

\subsection{Pionless effective field theory}

Analogous with the continuum densities in Eq.~(\ref{density}),
(\ref{density_S}), (\ref{density_I}), and (\ref{density_SI}), we define the
lattice operators%
\begin{equation}
\rho^{a^{\dagger},a}(\vec{n})=\sum_{i,j=0,1}a_{i,j}^{\dagger}(\vec{n}%
)a_{i,j}(\vec{n}),
\end{equation}%
\begin{equation}
\rho_{S}^{a^{\dagger},a}(\vec{n})=\sum_{i,j,i^{\prime}=0,1}a_{i,j}^{\dagger
}(\vec{n})\left[  \sigma_{S}\right]  _{ii^{\prime}}a_{i^{\prime},j}(\vec{n}),
\end{equation}%
\begin{equation}
\rho_{I}^{a^{\dagger},a}(\vec{n})=\sum_{i,j,j^{\prime}=0,1}a_{i,j}^{\dagger
}(\vec{n})\left[  \tau_{I}\right]  _{jj^{\prime}}a_{i,j^{\prime}}(\vec{n}),
\end{equation}%
\begin{equation}
\rho_{S,I}^{a^{\dagger},a}(\vec{n})=\sum_{i,j,i^{\prime},j^{\prime}%
=0,1}a_{i,j}^{\dagger}(\vec{n})\left[  \sigma_{S}\right]  _{ii^{\prime}%
}\left[  \tau_{I}\right]  _{jj^{\prime}}a_{i^{\prime},j^{\prime}}(\vec{n}).
\end{equation}
At leading order in pionless effective field theory,
\begin{equation}
\mathcal{Z}=Tr\left(  M^{L_{t}}\right)  ,
\end{equation}
where%
\begin{align}
M  &  =\colon\exp\left\{  -H_{\text{free}}\alpha_{t}-\frac{1}{2}C\alpha
_{t}\sum_{\vec{n}}\left[  \rho^{a^{\dag},a}(\vec{n})\right]  ^{2}\right.
\nonumber\\
&  \qquad\left.  -\frac{1}{2}C_{I^{2}}\alpha_{t}\sum_{\vec{n},I}\left[
\rho_{I}^{a^{\dag},a}(\vec{n})\right]  ^{2}-\frac{1}{6}D\alpha_{t}\sum
_{\vec{n}}\left[  \rho^{a^{\dag},a}(\vec{n})\right]  ^{3}\right\}  \colon.
\label{pionless_transfer}%
\end{align}
This formalism was used to study the triton and three-body forces on the
lattice \cite{Borasoy:2005yc}. \ The triton can be regarded as an approximate
example of the Efimov effect, which in the limit of zero range and infinite
scattering length predicts a geometric sequence of trimer bound states
\cite{Efimov:1971a,Efimov:1993a,Bedaque:1998kg,Bedaque:1998km,Bedaque:1999ve,Braaten:2004a}%
. \ The Efimov effect is not possible for two-component fermions due to Pauli
exclusion but is allowed for more than two components. \ Once the binding
energy of the trimer system is fixed, the binding energy of the four-body
system is also determined \cite{Platter:2004pra,Platter:2004zs,Hammer:2006ct}.
\ This is in analogy with the Tjon line relating the nuclear binding energies
of $^{3}$H and $^{4}$He. \ In two dimensions a different geometric sequence
has been predicted for zero-range attractive interactions. \ In this case the
geometric sequence describes the binding energy of $N$-body clusters as a
function of $N$ in the large $N$ limit \cite{Hammer:2004x, Platter:2004x,
Blume:2004}. \ These two-dimensional clusters have been studied using lattice
effective field theory for up to $10$ particles and the geometric scaling has
been confirmed \cite{Lee:2005xy}.

\subsection{Pionless effective field theory with auxiliary fields}

In terms of auxiliary fields
\begin{align}
\mathcal{Z}  &  =\prod\limits_{\vec{n},n_{t}}\left[  \int d_{A}s(\vec{n}%
,n_{t})\right]  \prod\limits_{\vec{n},n_{t},I}\left[  \frac{1}{\sqrt{2\pi}%
}\int_{-\infty}^{\infty}ds_{I}(\vec{n},n_{t})e^{-\frac{1}{2}s_{I}^{2}(\vec
{n},n_{t})}\right] \nonumber\\
&  \qquad\times Tr\left\{  M_{A}(s,L_{t}-1)\cdot\cdots\cdot M_{A}%
(s,0)\right\}  ,
\end{align}
where the auxiliary-field transfer matrix is%
\begin{align}
M_{A}(s,n_{t})  &  =\colon\exp\left\{  -H_{\text{free}}\alpha_{t}+\sum
_{\vec{n}}A\left[  s(\vec{n},n_{t})\right]  \rho^{a^{\dagger}a}(\vec
{n})\right. \nonumber\\
&  +\left.  i\sqrt{C_{I}\alpha_{t}}\sum_{\vec{n},I}s_{I}(\vec{n},n_{t}%
)\rho_{I}^{a^{\dagger}a}(\vec{n})\right\}  \colon. \label{MA_pionless}%
\end{align}
Let $\left\langle A^{k}\right\rangle $ be the expectation value of the
$k^{\text{th}}$ power of $A$ with respect to the measure $d_{A}s,$%
\begin{equation}
\left\langle A^{k}\right\rangle =\int d_{A}s(\vec{n},n_{t})\,\left\{  A\left[
s(\vec{n},n_{t})\right]  \right\}  ^{k},\text{ \ }k=0,1,2,3,4.
\end{equation}
In order to reproduce the interactions in Eq.~(\ref{pionless_transfer}) we
require that
\begin{equation}
\left\langle A^{0}\right\rangle =1,\qquad\left\langle A^{1}\right\rangle
=0,\qquad\left\langle A^{2}\right\rangle =-C\alpha_{t},\qquad\left\langle
A^{3}\right\rangle =-D\alpha_{t},\qquad\left\langle A^{4}\right\rangle
=3C^{2}\alpha_{t}^{2}.
\end{equation}

The existence of a positive definite measure $d_{A}s$ and real-valued $A$ is
essential for Monte Carlo simulations without sign and phase oscillations.
\ Sufficient and necessary conditions for the existence of a positive definite
$d_{A}s$ and real-valued $A$ is known in the mathematics literature as the
truncated Hamburger moment problem. \ This problem has been solved
\cite{Curto:1991,Adamyan:2003,Chen:2004rq}, and the conditions are satisfied
if and only if the block-Hankel matrix,%
\begin{equation}%
\begin{bmatrix}
\left\langle A^{0}\right\rangle  & \left\langle A^{1}\right\rangle  &
\left\langle A^{2}\right\rangle \\
\left\langle A^{1}\right\rangle  & \left\langle A^{2}\right\rangle  &
\left\langle A^{3}\right\rangle \\
\left\langle A^{2}\right\rangle  & \left\langle A^{3}\right\rangle  &
\left\langle A^{4}\right\rangle
\end{bmatrix}
=%
\begin{bmatrix}
1 & 0 & -C\alpha_{t}\\
0 & -C\alpha_{t} & -D\alpha_{t}\\
-C\alpha_{t} & -D\alpha_{t} & 3C^{2}\alpha_{t}^{2}%
\end{bmatrix}
,
\end{equation}
is positive semi-definite. \ The determinant of this matrix is $-2C^{3}%
\alpha_{t}^{3}-D^{2}\alpha_{t}^{2}$. \ With an attractive two-nucleon force
where $C<0$ the conditions are satisfied provided that the three-body
interaction coefficient $D$ is not too large. \ We note that the positivity
condition is spoiled more easily in the Hamiltonian limit where $\alpha
_{t}\rightarrow0$.

\subsection{Instantaneous free pion action}

Before discussing lattice actions for chiral effective field theory, we first
consider the lattice action for free pions with mass $m_{\pi}$ and purely
instantaneous propagation,%
\begin{equation}
S_{\pi\pi}(\pi_{I})=\alpha_{t}(\tfrac{m_{\pi}^{2}}{2}+3)\sum_{\vec{n},n_{t}%
,I}\pi_{I}(\vec{n},n_{t})\pi_{I}(\vec{n},n_{t})-\alpha_{t}\sum_{\vec{n}%
,n_{t},I,l}\pi_{I}(\vec{n},n_{t})\pi_{I}(\vec{n}+\hat{l},n_{t}).
\label{S_pipi}%
\end{equation}
The pion field $\pi_{I}$ is labelled with isospin index $I$. \ Pion fields at
different time steps $n_{t}$ and $n_{t}^{\prime}$ are not coupled due to the
omission of time derivatives. \ This generates instantaneous propagation\ at
each time step when computing one-pion exchange diagrams. \ It also eliminates
unwanted pion couplings contributing to nucleon self-energy diagrams found in
earlier work \cite{Lee:2004si}. \ Though we call it a pion field, it is more
accurate to regard $\pi_{I}$ as an auxiliary field which is used to reproduce
the one-pion exchange potential on the lattice. \ If for example we wish to
consider low-energy physical pions within the framework of chiral effective
field theory\ \cite{Weinberg:1992yk}, these scattering processes can be
introduced perturbatively using external pion fields and additional auxiliary
fields to reproduce the corresponding Feynman diagrams at each order.

Following the notation in Ref.~\cite{Borasoy:2006qn}, it is useful to define a
rescaled pion field, $\pi_{I}^{\prime}$,%
\begin{equation}
\pi_{I}^{\prime}(\vec{n},n_{t})=\sqrt{q_{\pi}}\pi_{I}(\vec{n},n_{t}),
\end{equation}
where%
\begin{equation}
q_{\pi}=\alpha_{t}(m_{\pi}^{2}+6).
\end{equation}
In terms of $\pi_{I}^{\prime}$,%
\begin{equation}
S_{\pi\pi}(\pi_{I}^{\prime})=\frac{1}{2}\sum_{\vec{n},n_{t},I}\pi_{I}^{\prime
}(\vec{n},n_{t})\pi_{I}^{\prime}(\vec{n},n_{t})-\frac{\alpha_{t}}{q_{\pi}}%
\sum_{\vec{n},n_{t},I,l}\pi_{I}^{\prime}(\vec{n},n_{t})\pi_{I}^{\prime}%
(\vec{n}+\hat{l},n_{t}),
\end{equation}
and in momentum space we have%
\begin{equation}
S_{\pi\pi}(\pi_{I}^{\prime})=\frac{1}{L^{3}}\sum_{I,\vec{k}}\pi_{I}^{\prime
}(-\vec{k},n_{t})\pi_{I}^{\prime}(\vec{k},n_{t})\left[  \frac{1}{2}%
-\frac{\alpha_{t}}{q_{\pi}}\sum_{l}\cos k_{l}\right]  .
\end{equation}
The instantaneous pion correlation function at spatial separation $\vec{n}$ is%
\begin{equation}
\left\langle \pi_{I}^{\prime}(\vec{n},n_{t})\pi_{I}^{\prime}(\vec{0}%
,n_{t})\right\rangle =\frac{1}{L^{3}}\sum_{\vec{k}}e^{-i\vec{k}\cdot\vec{n}%
}D_{\pi}(\vec{k}),
\end{equation}
where%
\begin{equation}
D_{\pi}(\vec{k})=\frac{1}{1-\tfrac{2\alpha_{t}}{q_{\pi}}\sum_{l}\cos k_{l}}.
\end{equation}

\subsection{Chiral effective field theory on the lattice}

We define some lattice derivative notation which will be useful later. \ There
are various ways to introduce spatial derivatives of the pion field on the
lattice. \ The simplest definition for the gradient of $\pi_{I}^{\prime}$ is
to define a forward-backward lattice derivative. \ For example we can write%
\begin{equation}
\partial_{1}\pi_{I}^{\prime}(\vec{n})=\frac{1}{2}\left[  \pi_{I}^{\prime}%
(\vec{n}+\hat{1})-\pi_{I}^{\prime}(\vec{n}-\hat{1})\right]  .
\end{equation}
This is the method used in Ref.~\cite{Lee:2004si}. \ One disadvantage is that
it is a coarse derivative involving a separation distance of two lattice
units. \ We can avoid this if we think of the pion lattice points as being
shifted by $-1/2$ lattice unit from the nucleon lattice points in each of the
three spatial directions. \ For each nucleon lattice point $\vec
{n}_{\text{nucleon}}$ we associate a pion lattice point $\vec{n}_{\text{pion}%
}$,%
\begin{equation}
\vec{n}_{\text{pion}}=\vec{n}_{\text{nucleon}}-\frac{1}{2}\hat{1}-\frac{1}%
{2}\hat{2}-\frac{1}{2}\hat{3}.
\end{equation}
Then we have eight pion lattice points forming a cube centered at $\vec
{n}_{\text{nucleon}}$,%
\begin{align}
&  \vec{n}_{\text{pion}},\quad\vec{n}_{\text{pion}}+\hat{1},\quad\vec
{n}_{\text{pion}}+\hat{2},\quad\vec{n}_{\text{pion}}+\hat{3},\nonumber\\
&  \vec{n}_{\text{pion}}+\hat{1}+\hat{2},\quad\vec{n}_{\text{pion}}+\hat
{2}+\hat{3},\quad\vec{n}_{\text{pion}}+\hat{3}+\hat{1},\quad\vec
{n}_{\text{pion}}+\hat{1}+\hat{2}+\hat{3}. \label{pionlattice}%
\end{align}
For derivatives of the pion field we use the eight vertices of this unit cube
on the lattice to define spatial derivatives. \ For each spatial direction
$l=1,2,3$ and any lattice function $f(\vec{n})$ we define%
\begin{equation}
\Delta_{l}f(\vec{n})=\frac{1}{4}\sum_{\substack{\nu_{1},\nu_{2},\nu_{3}%
=0,1}}(-1)^{\nu_{l}+1}f(\vec{n}+\vec{\nu}),\qquad\vec{\nu}=\nu_{1}\hat{1}%
+\nu_{2}\hat{2}+\nu_{3}\hat{3}. \label{derivative}%
\end{equation}
For double spatial derivatives of nucleon fields along direction $l$ we use
the simpler definition,%
\begin{equation}
\triangledown_{l}^{2}f(\vec{n})=f(\vec{n}+\hat{l})+f(\vec{n}-\hat{l}%
)-2f(\vec{n}).
\end{equation}

At leading order in chiral effective field theory, the first partition
function and transfer matrix operator considered in Ref.~\cite{Borasoy:2006qn}
was
\begin{equation}
\mathcal{Z}_{\text{LO}_{1}}=Tr\left[  \left(  M_{\text{LO}_{1}}\right)
^{L_{t}}\right]  ,
\end{equation}
where%
\begin{align}
M_{\text{LO}_{1}}  &  =\colon\exp\left\{  -H_{\text{free}}\alpha_{t}-\frac
{1}{2}C\alpha_{t}\sum_{\vec{n}}\left[  \rho^{a^{\dag},a}(\vec{n})\right]
^{2}-\frac{1}{2}C_{I^{2}}\alpha_{t}\sum_{\vec{n},I}\left[  \rho_{I}^{a^{\dag
},a}(\vec{n})\right]  ^{2}\right. \nonumber\\
&  +\left.  \frac{g_{A}^{2}\alpha_{t}^{2}}{8f_{\pi}^{2}q_{\pi}}\sum
_{\substack{S_{1},S_{2},I}}\sum_{\vec{n}_{1},\vec{n}_{2}}G_{S_{1}S_{2}}%
(\vec{n}_{1}-\vec{n}_{2})\rho_{S_{1},I}^{a^{\dag},a}(\vec{n}_{1})\rho
_{S_{2},I}^{a^{\dag},a}(\vec{n}_{2})\right\}  \colon,
\end{align}
and
\begin{align}
G_{S_{1}S_{2}}(\vec{n})  &  =\left\langle \Delta_{S_{1}}\pi_{I}^{\prime}%
(\vec{n},n_{t})\Delta_{S_{2}}\pi_{I}^{\prime}(\vec{0},n_{t})\right\rangle
\text{ \ (no sum on }I\text{)}\nonumber\\
&  =\frac{1}{16}\sum_{\nu_{1},\nu_{2},\nu_{3}=0,1}\sum_{\nu_{1}^{\prime}%
,\nu_{2}^{\prime},\nu_{3}^{\prime}=0,1}(-1)^{\nu_{S_{1}}}(-1)^{\nu_{S_{2}%
}^{\prime}}\left\langle \pi_{I}^{\prime}(\vec{n}+\vec{\nu}-\vec{\nu}^{\prime
},n_{t})\pi_{I}^{\prime}(\vec{0},n_{t})\right\rangle .
\end{align}
This leading-order transfer matrix, labelled $M_{\text{LO}_{1}}$, has
zero-range contact interactions analogous to the pionless transfer matrix in
Eq.~(\ref{pionless_transfer}). \ The $O(a^{4})$-improved action was used for
$H_{\text{free}}$.

A second leading-order partition function and transfer matrix was also
considered,%
\begin{equation}
\mathcal{Z}_{\text{LO}_{2}}=Tr\left[  \left(  M_{\text{LO}_{2}}\right)
^{L_{t}}\right]  ,
\end{equation}
where%

\begin{align}
M_{\text{LO}_{2}}  &  =\colon\exp\left\{  -H_{\text{free}}\alpha_{t}%
-\frac{\alpha_{t}}{2L^{3}}\sum_{\vec{q}}f(\vec{q})\left[  C\rho^{a^{\dag}%
,a}(\vec{q})\rho^{a^{\dag},a}(-\vec{q})+C_{I^{2}}\sum_{I}\rho_{I}^{a^{\dag}%
,a}(\vec{q})\rho_{I}^{a^{\dag},a}(-\vec{q})\right]  \right. \nonumber\\
&  +\left.  \frac{g_{A}^{2}\alpha_{t}^{2}}{8f_{\pi}^{2}q_{\pi}}\sum
_{\substack{S_{1},S_{2},I}}\sum_{\vec{n}_{1},\vec{n}_{2}}G_{S_{1}S_{2}}%
(\vec{n}_{1}-\vec{n}_{2})\rho_{S_{1},I}^{a^{\dag},a}(\vec{n}_{1})\rho
_{S_{2},I}^{a^{\dag},a}(\vec{n}_{2})\right\}  \colon.
\end{align}
The momentum-dependent coefficient function $f(\vec{q})$ has the form
\begin{equation}
f(\vec{q})=f_{0}^{-1}\exp\left[  -b%
%TCIMACRO{\dsum \limits_{l=1,2,3}}%
%BeginExpansion
{\displaystyle\sum\limits_{l=1,2,3}}
%EndExpansion
\left(  1-\cos q_{l}\right)  \right]  ,
\end{equation}
and the normalization factor $f_{0}$ is determined by the condition%
\begin{equation}
f_{0}=\frac{1}{L^{3}}\sum_{\vec{q}}\exp\left[  -b%
%TCIMACRO{\dsum \limits_{l=1,2,3}}%
%BeginExpansion
{\displaystyle\sum\limits_{l=1,2,3}}
%EndExpansion
\left(  1-\cos q_{l}\right)  \right]  .
\end{equation}
The coefficient $b$ was determined by fitting to reproduce the correct average
effective range for the two $S$-wave channels. \ For small $\vec{q}$ the
function $f(\vec{q})$ reduces to a Gaussian function,%
\begin{equation}
f(\vec{q})\approx f_{0}^{-1}\exp\left(  -\frac{b}{2}q^{2}\right)  .
\end{equation}
This Gaussian smearing of the contact interactions in $M_{\text{LO}_{2}}$ was
found to remove four-nucleon clustering instabilities at lattice spacing
$a=(100$ MeV$)^{-1}$ \cite{Borasoy:2006qn}.

\subsection{Chiral effective field theory with auxiliary fields}

Let us define the auxiliary-field action%
\begin{equation}
S_{ss}^{\text{LO}_{1}}=\frac{1}{2}\sum_{\vec{n},n_{t}}s^{2}(\vec{n}%
,n_{t})+\frac{1}{2}\sum_{\vec{n},n_{t},I}s_{I}^{2}(\vec{n},n_{t}).
\end{equation}
In terms of auxiliary and pion fields, the partition function for LO$_{1}$ is
\begin{align}
\mathcal{Z}_{\text{LO}_{1}}  &  =%
%TCIMACRO{\dint }%
%BeginExpansion
{\displaystyle\int}
%EndExpansion
D\pi_{I}^{\prime}DsDs_{I}\;\exp\left[  -S_{\pi\pi}-S_{ss}^{\text{LO}_{1}%
}\right] \nonumber\\
&  \times Tr\left\{  M_{\text{LO}_{1}}(\pi_{I}^{\prime},s,s_{I},L_{t}%
-1)\times\cdots\times M_{\text{LO}_{1}}(\pi_{I}^{\prime},s,s_{I},0)\right\}  ,
\end{align}
where%
\begin{align}
M_{\text{LO}_{1}}(\pi_{I}^{\prime},s,s_{I},n_{t})  &  =\colon\exp\left[
-H_{\text{free}}\alpha_{t}+\sqrt{-C\alpha_{t}}\sum_{\vec{n}}s(\vec{n}%
,n_{t})\rho^{a^{\dag},a}(\vec{n})\right. \nonumber\\
&  +\left.  i\sqrt{C_{I}\alpha_{t}}\sum_{\vec{n},I}s_{I}(\vec{n},n_{t}%
)\rho_{I}^{a^{\dag},a}(\vec{n})-\tfrac{g_{A}\alpha_{t}}{2f_{\pi}\sqrt{q_{\pi}%
}}%
%TCIMACRO{\dsum _{\vec{n},S,I}}%
%BeginExpansion
{\displaystyle\sum_{\vec{n},S,I}}
%EndExpansion
\Delta_{S}\pi_{I}^{\prime}(\vec{n},n_{t})\rho_{S,I}^{a^{\dag},a}(\vec
{n})\right]  \colon,
\end{align}
and $D\pi_{I}^{\prime}DsDs_{I}$ is the functional measure,%
\begin{equation}
D\pi_{I}^{\prime}DsDs_{I}=\prod\limits_{\vec{n},n_{t}}\left[  \frac{ds(\vec
{n},n_{t})}{\sqrt{2\pi}}\right]  \prod\limits_{\vec{n},n_{t},I}\left[
\frac{d\pi_{I}^{\prime}(\vec{n},n_{t})ds_{I}(\vec{n},n_{t})}{2\pi}\right]  .
\end{equation}
The instantaneous free pion action $S_{\pi\pi}$ was already defined in
Eq.~(\ref{S_pipi}). \ 

For the LO$_{2}$ action we have
\begin{align}
\mathcal{Z}_{\text{LO}_{2}}  &  =%
%TCIMACRO{\dprod \limits_{\vec{q}}}%
%BeginExpansion
{\displaystyle\prod\limits_{\vec{q}}}
%EndExpansion
\frac{1}{f^{2}(\vec{q})}\times%
%TCIMACRO{\dint }%
%BeginExpansion
{\displaystyle\int}
%EndExpansion
D\pi_{I}^{\prime}DsDs_{I}\;\exp\left[  -S_{\pi\pi}-S_{ss}^{\text{LO}_{2}%
}\right] \nonumber\\
&  \times Tr\left\{  M_{\text{LO}_{2}}(\pi_{I}^{\prime},s,s_{I},L_{t}%
-1)\times\cdots\times M_{\text{LO}_{2}}(\pi_{I}^{\prime},s,s_{I},0)\right\}  .
\end{align}
The functional form of the transfer matrices are the same,%
\begin{equation}
M_{\text{LO}_{2}}(\pi_{I}^{\prime},s,s_{I},n_{t})=M_{\text{LO}_{1}}(\pi
_{I}^{\prime},s,s_{I},n_{t})\text{,}%
\end{equation}
but for LO$_{2}$ the auxiliary-field action has the non-local form%
\begin{align}
S_{ss}^{\text{LO}_{2}}  &  =\frac{1}{2}\sum_{\vec{n},\vec{n}^{\prime},n_{t}%
}s(\vec{n},n_{t})f^{-1}\left(  \vec{n}-\vec{n}^{\prime}\right)  s(\vec
{n}^{\prime},n_{t})\nonumber\\
&  +\frac{1}{2}\sum_{I}\sum_{\vec{n},\vec{n}^{\prime},n_{t}}s_{I}(\vec
{n},n_{t})f^{-1}(\vec{n}-\vec{n}^{\prime})s_{I}(\vec{n}^{\prime},n_{t}),
\end{align}
where the inverse function $f^{-1}$ is defined as%
\begin{equation}
f^{-1}(\vec{n}-\vec{n}^{\prime})=\frac{1}{L^{3}}\sum_{\vec{q}}\frac{1}%
{f(\vec{q})}e^{-i\vec{q}\cdot(\vec{n}-\vec{n}^{\prime})}.
\end{equation}

\subsection{Next-to-leading-order interactions on the lattice}

The lattice studies in Ref.~\cite{Borasoy:2007vi,Borasoy:2007vk} considered
low-energy nucleon-nucleon scattering at momenta less than or equal to the
pion mass, $m_{\pi}$. \ On the lattice the ultraviolet cutoff momentum,
$\Lambda$, equals $\pi$ divided by the lattice spacing, $a$. \ As noted
earlier, serious numerical difficulties appear at large $\Lambda$ in Monte
Carlo simulations of few- and many-nucleon systems. \ In attractive channels
unphysical deeply-bound states appear at large $\Lambda$. \ In other channels
short-range repulsion becomes prominent, producing destructive sign or complex
phase oscillations. \ The severity of the problem scales exponentially with
system size and strength of the repulsive interaction.

In order to avoid these difficulties the approach advocated in
Ref.~\cite{Borasoy:2007vi,Borasoy:2007vk} was to set the cutoff momentum
$\Lambda$ as low as possible for describing physical momenta up to $m_{\pi}$.
\ In most of the published work so far the value chosen was $\Lambda=314$ MeV
$\approx2.3m_{\pi}$, corresponding with $a=(100$ MeV$)^{-1}$. \ \ This coarse
lattice approach is similar in motivation to the continuum low-momentum
renormalization group approach using $V_{\text{low }k}$
\cite{Bogner:2001gq,Bogner:2003wn}.

For nearly all $\left\vert q\right\vert <\Lambda$ the two-pion exchange
potential can be expanded in powers of $q^{2}/(4m_{\pi}^{2}),$%
\begin{equation}
L(q)=1+\frac{1}{3}\frac{q^{2}}{4m_{\pi}^{2}}+\cdots,
\end{equation}%
\begin{equation}
\frac{4m_{\pi}^{2}}{4m_{\pi}^{2}+q^{2}}L(q)=1-\frac{2}{3}\frac{q^{2}}{4m_{\pi
}^{2}}+\cdots,
\end{equation}%
\begin{align}
\mathcal{A}\left(  V_{\text{NLO}}^{\text{TPEP}}\right)   &  =-\frac
{\boldsymbol\tau_{1}\cdot\boldsymbol\tau_{2}}{384\pi^{2}f_{\pi}^{4}}\left[
4m_{\pi}^{2}\left(  8g_{A}^{4}-4g_{A}^{2}-1\right)  +\frac{2}{3}q^{2}\left(
34g_{A}^{4}-17g_{A}^{2}-2\right)  +O\left(  \left(  \tfrac{q^{2}}{4m_{\pi}%
^{2}}\right)  ^{2}\right)  \right] \nonumber\\
&  -\frac{3g_{A}^{4}}{64\pi^{2}f_{\pi}^{4}}\left[  \left(  \vec{q}\cdot
\vec{\sigma}_{1}\right)  \left(  \vec{q}\cdot\vec{\sigma}_{2}\right)
-q^{2}\left(  \vec{\sigma}_{1}\cdot\vec{\sigma}_{2}\right)  \right]  \left[
1+O\left(  \tfrac{q^{2}}{4m_{\pi}^{2}}\right)  \right]  . \label{localTPEP}%
\end{align}
This expansion fails to converge only for $q$ near the cutoff scale $\Lambda$
$\approx2.3m_{\pi}$, and so there is no practical advantage in keeping the
full non-local structure of $V_{\text{NLO}}^{\text{TPEP}}$ at this lattice
spacing. \ Instead we simply use%
\begin{equation}
V_{\text{LO}}=V^{(0)}+V^{\text{OPEP}},
\end{equation}%
\begin{equation}
V_{\text{NLO}}=V_{\text{LO}}+\Delta V^{(0)}+V^{(2)},
\end{equation}
where the terms in Eq.~(\ref{localTPEP}) with up to two powers of $q$ are
absorbed in the definition of the coefficients for $\Delta V^{(0)}$ and
$V^{(2)}$.

Before describing the NLO lattice interactions in $\Delta V^{(0)}$ and
$V^{(2)}$, we first define lattice current densities for total nucleon number,
spin, isospin, and spin-isospin. \ Similar to the definition of $\Delta_{l}$
in Eq.~(\ref{derivative}), we use the eight vertices of a unit cube,
\begin{equation}
\vec{\nu}=\nu_{1}\hat{1}+\nu_{2}\hat{2}+\nu_{3}\hat{3},
\end{equation}
for $\nu_{1},\nu_{2},\nu_{3}=0,1$. \ Let $\vec{\nu}(-l)$ for $l=1,2,3$ be the
reflection of the $l^{\text{th}}$-component of $\vec{\nu}$ about the center of
the cube,%
\begin{equation}
\vec{\nu}(-l)=\vec{\nu}+(1-2\nu_{l})\hat{l}.
\end{equation}
The $l^{\text{th}}$-component of the SU(4)-invariant current density is
defined as%
\begin{equation}
\Pi_{l}^{a^{\dagger},a}(\vec{n})=\frac{1}{4}\sum_{\substack{\nu_{1},\nu
_{2},\nu_{3}=0,1}}\sum_{i,j=0,1}(-1)^{\nu_{l}+1}a_{i,j}^{\dagger}(\vec{n}%
+\vec{\nu}(-l))a_{i,j}(\vec{n}+\vec{\nu}).
\end{equation}
Similarly for spin current density,%
\begin{equation}
\Pi_{l,S}^{a^{\dagger},a}(\vec{n})=\frac{1}{4}\sum_{\substack{\nu_{1},\nu
_{2},\nu_{3}=0,1}}\sum_{i,j,i^{\prime}=0,1}(-1)^{\nu_{l}+1}a_{i,j}^{\dagger
}(\vec{n}+\vec{\nu}(-l))\left[  \sigma_{S}\right]  _{ii^{\prime}}a_{i^{\prime
},j}(\vec{n}+\vec{\nu}),
\end{equation}
isospin current density,%
\begin{equation}
\Pi_{l,I}^{a^{\dagger},a}(\vec{n})=\frac{1}{4}\sum_{\substack{\nu_{1},\nu
_{2},\nu_{3}=0,1}}\sum_{i,j,j^{\prime}=0,1}(-1)^{\nu_{l}+1}a_{i,j}^{\dagger
}(\vec{n}+\vec{\nu}(-l))\left[  \tau_{I}\right]  _{jj^{\prime}}a_{i,j^{\prime
}}(\vec{n}+\vec{\nu}),
\end{equation}
and spin-isospin current density,%
\begin{equation}
\Pi_{l,S,I}^{a^{\dagger},a}(\vec{n})=\frac{1}{4}\sum_{\substack{\nu_{1}%
,\nu_{2},\nu_{3}=0,1}}\sum_{i,j,i^{\prime},j^{\prime}=0,1}(-1)^{\nu_{l}%
+1}a_{i,j}^{\dagger}(\vec{n}+\vec{\nu}(-l))\left[  \sigma_{S}\right]
_{ii^{\prime}}\left[  \tau_{I}\right]  _{jj^{\prime}}a_{i^{\prime},j^{\prime}%
}(\vec{n}+\vec{\nu}).
\end{equation}

In Ref.~\cite{Borasoy:2007vi} the next-to-leading-order transfer matrices
$M_{\text{NLO}_{1}}$ and $M_{\text{NLO}_{2}}$ were defined by adding the
following nine local interactions to the leading-order transfer matrices
$M_{\text{LO}_{1}}$ and $M_{\text{LO}_{2}}$. \ The two corrections to the
leading-order contact interactions are%
\begin{equation}
\Delta V=\frac{1}{2}\Delta C:\sum\limits_{\vec{n}}\rho^{a^{\dagger},a}(\vec
{n})\rho^{a^{\dagger},a}(\vec{n}):,
\end{equation}%
\begin{equation}
\Delta V_{I^{2}}=\frac{1}{2}\Delta C_{I^{2}}:\sum\limits_{\vec{n},I}\rho
_{I}^{a^{\dagger},a}(\vec{n})\rho_{I}^{a^{\dagger},a}(\vec{n}):.
\end{equation}
At next-to-leading order there are seven independent contact interactions with
two derivatives. \ These are%
\begin{equation}
V_{q^{2}}=-\frac{1}{2}C_{q^{2}}:\sum\limits_{\vec{n},l}\rho^{a^{\dagger}%
,a}(\vec{n})\triangledown_{l}^{2}\rho^{a^{\dagger},a}(\vec{n}):,
\end{equation}%
\begin{equation}
V_{I^{2},q^{2}}=-\frac{1}{2}C_{I^{2},q^{2}}:\sum\limits_{\vec{n},I,l}\rho
_{I}^{a^{\dagger},a}(\vec{n})\triangledown_{l}^{2}\rho_{I}^{a^{\dagger}%
,a}(\vec{n}):,
\end{equation}%
\begin{equation}
V_{S^{2},q^{2}}=-\frac{1}{2}C_{S^{2},q^{2}}:\sum\limits_{\vec{n},S,l}\rho
_{S}^{a^{\dagger},a}(\vec{n})\triangledown_{l}^{2}\rho_{S}^{a^{\dagger}%
,a}(\vec{n}):,
\end{equation}%
\begin{equation}
V_{S^{2},I^{2},q^{2}}=-\frac{1}{2}C_{S^{2},I^{2},q^{2}}:\sum\limits_{\vec
{n},S,I,l}\rho_{S,I}^{a^{\dagger},a}(\vec{n})\triangledown_{l}^{2}\rho
_{S,I}^{a^{\dagger},a}(\vec{n}):,
\end{equation}%
\begin{equation}
V_{(q\cdot S)^{2}}=\frac{1}{2}C_{(q\cdot S)^{2}}:\sum\limits_{\vec{n}}%
\sum\limits_{S}\Delta_{S}\rho_{S}^{a^{\dagger},a}(\vec{n})\sum
\limits_{S^{\prime}}\Delta_{S^{\prime}}\rho_{S^{\prime}}^{a^{\dagger},a}%
(\vec{n}):,
\end{equation}%
\begin{equation}
V_{I^{2},(q\cdot S)^{2}}=\frac{1}{2}C_{I^{2},(q\cdot S)^{2}}:\sum
\limits_{\vec{n},I}\sum\limits_{S}\Delta_{S}\rho_{S,I}^{a^{\dagger},a}(\vec
{n})\sum\limits_{S^{\prime}}\Delta_{S^{\prime}}\rho_{S^{\prime},I}%
^{a^{\dagger},a}(\vec{n}):,
\end{equation}%
\begin{equation}
V_{(iq\times S)\cdot k}=-\frac{i}{2}C_{(iq\times S)\cdot k}:\sum
\limits_{\vec{n},l,S,l^{\prime}}\varepsilon_{l,S,l^{\prime}}\left[  \Pi
_{l}^{a^{\dagger},a}(\vec{n})\Delta_{l^{\prime}}\rho_{S}^{a^{\dagger},a}%
(\vec{n})+\Pi_{l,S}^{a^{\dagger},a}(\vec{n})\Delta_{l^{\prime}}\rho
^{a^{\dagger},a}(\vec{n})\right]  :.
\end{equation}

\subsection{Model independence at fixed lattice spacing}

In effective field theory calculations model independence is often tested by
checking sensitivity on the cutoff scale $\Lambda$. \ At a given order the
difference between calculations for two different cutoff scales $\Lambda_{1}$
and $\Lambda_{2}$ should be no larger than the omitted corrections at the next
order. \ On the lattice this test is problematic since the lattice spacing
cannot be changed by a large amount due to computational constraints.
\ Instead a different approach was introduced in Ref.~\cite{Borasoy:2007vi} to
test model independence at fixed lattice spacing which we summarize in the following.

The notation $V^{Q^{n}/\Lambda^{n}}$ is used to denote two-nucleon operators
with the following properties. \ $V^{Q^{n}/\Lambda^{n}}$ is a sum of local
two-nucleon interactions that is an analytic function of momenta below the
cutoff scale $\Lambda$ and scales as $n$ or more powers of momenta in the
asymptotic low-momentum limit. \ The term \textquotedblleft
quasi-local\textquotedblright\ is used to describe $V^{Q^{n}/\Lambda^{n}}$
since the interactions are short-ranged. \ At fixed lattice spacing we may
consider two different lowest-order actions with interactions of the form%
\begin{align}
V_{\text{LO}_{1}}  &  =V_{1}^{(0)}+V^{\text{OPEP}}+V_{1}^{Q^{2}/\Lambda^{2}%
},\label{LO1}\\
V_{\text{LO}_{2}}  &  =V_{2}^{(0)}+V^{\text{OPEP}}+V_{2}^{Q^{2}/\Lambda^{2}},
\label{LO2}%
\end{align}
where $V_{1}^{Q^{2}/\Lambda^{2}}$ and $V_{2}^{Q^{2}/\Lambda^{2}}$ are
different quasi-local operators with at least two powers of momenta. \ Since
the leading-order interactions are iterated non-perturbatively the contact
terms $V_{1}^{(0)}$ and $V_{2}^{(0)}$ in general have different coefficients.
\ However low-energy physical observables should agree up to differences the
same size as the omitted contributions at next-to-leading-order.

Similarly at next-to-leading order we may consider two different actions of
the form%
\begin{align}
V_{\text{NLO}_{1}}  &  =V_{\text{LO}_{1}}+\Delta V_{1}^{(0)}+V_{1}^{(2)}%
+V_{1}^{Q^{4}/\Lambda^{4}},\\
V_{\text{NLO}_{2}}  &  =V_{\text{LO}_{2}}+\Delta V_{2}^{(0)}+V_{2}^{(2)}%
+V_{2}^{Q^{4}/\Lambda^{4}},
\end{align}
where $V_{1}^{Q^{4}/\Lambda^{4}}$ and $V_{2}^{Q^{4}/\Lambda^{4}}$ are
different quasi-local operators with at least four powers of momenta.
\ Low-energy physical observables should again agree up to differences the
same size as the omitted contributions at the next order.

This technique provides a method for testing model independence of the
low-energy lattice effective theory without changing the lattice spacing. \ In
principle however it is good to check model independence in multiple ways,
including different variations for $V^{Q^{n}/\Lambda^{n}}$ as well as changing
the lattice spacing as much as allowed by computational constraints.

\section{Two-particle scattering on the lattice}

\subsection{Cubic rotation group}

Lattice regularization reduces the SO$(3)$ rotational symmetry of continuous
space to the cubic rotational group SO$(3,Z)$. \ This group is also known as
the proper octahedral group and abbreviated as O. \ This lack of exact
rotational symmetry complicates the extraction of partial wave amplitudes.
\ SO$(3,Z)$ consists of $24$ group elements generated by products of $\pi/2$
rotations about the $x$, $y$, $z$\ axes. \ Since SO$(3,Z)$ is discrete,
angular momentum operators $J_{x}$, $J_{y}$, $J_{z}$ cannot be defined in the
usual sense. \ Let $R_{\hat{z}}\left(  \pi/2\right)  $ be the group element
for a $\pi/2$ rotation about the $z$ axis. \ The SO$(3)$ relation%
\begin{equation}
R_{\hat{z}}\left(  \pi/2\right)  =\exp\left[  -i\frac{\pi}{2}J_{z}\right]
\label{Jz}%
\end{equation}
can be used to define $J_{z}$. \ The eigenvalues of $J_{z}$ are integers
specified modulo 4. \ $J_{x}$ and $J_{y}$ may be defined in the same way using
$R_{\hat{x}}\left(  \pi/2\right)  $ and $R_{\hat{y}}\left(  \pi/2\right)  $.

There are five irreducible representations of the cubic rotational group.
\ These are usually called $A_{1}$, $T_{1}$, $E$, $T_{2}$, and $A_{2}$. \ Some
of their properties and examples using low-order spherical harmonics
$Y_{L,L_{z}}(\theta,\phi)$ are listed in Table \ref{reps}. \ The $2J+1$
elements of the total angular momentum $J$ representation of SO$(3)$ break up
into smaller pieces associated with the five irreducible representations.
\ Examples for $J\leq5$ are shown in Table \ref{decomp} \cite{Johnson:1982yq}.
\ \setcounter{table}{0}\begin{table}[tbh]
\caption{Irreducible SO$(3,Z)$ representations.}%
\label{reps}%
$%
\begin{tabular}
[c]{|c|c|c|}\hline
Re$\text{presentation}$ & $J_{z}$ & Ex$\text{ample}$\\\hline
$A_{1}$ & $0\operatorname{mod}4$ & $Y_{0,0}$\\\hline
$T_{1}$ & $0,1,3\operatorname{mod}4$ & $\left\{  Y_{1,0},Y_{1,1}%
,Y_{1,-1}\right\}  $\\\hline
$E$ & $0,2\operatorname{mod}4$ & $\left\{  Y_{2,0},\frac{Y_{2,-2}+Y_{2,2}%
}{\sqrt{2}}\right\}  $\\\hline
$T_{2}$ & $1,2,3\operatorname{mod}4$ & $\left\{  Y_{2,1},\frac{Y_{2,-2}%
-Y_{2,2}}{\sqrt{2}},Y_{2,-1}\right\}  $\\\hline
$A_{2}$ & $2\operatorname{mod}4$ & $\frac{Y_{3,2}-Y_{3,-2}}{\sqrt{2}}$\\\hline
\end{tabular}
\ $\end{table}\begin{table}[tbhtbh]
\caption{SO$(3,Z)$ decompositions for $J\leq5.$}%
\label{decomp}%
\begin{tabular}
[c]{|c|c|}\hline
$\text{SO}(3)$ & $\text{SO}(3,Z)$\\\hline
$J=0$ & $A_{1}$\\\hline
$J=1$ & $T_{1}$\\\hline
$J=2$ & $E\oplus T_{2}$\\\hline
$J=3$ & $T_{1}\oplus T_{2}\oplus A_{2}$\\\hline
$J=4$ & $A_{1}\oplus T_{1}\oplus E\oplus T_{2}$\\\hline
$J=5$ & $T_{1}\oplus T_{1}\oplus E\oplus T_{2}$\\\hline
\end{tabular}
\end{table}In lattice QCD these irreducible representations have been used to
classify glueball states \cite{Morningstar:1999rf} as well as predict the
spectrum and properties of baryon resonances \cite{Basak:2005aq, Basak:2005ir}.

\subsection{L\"{u}scher's finite volume formula}

L\"{u}scher's finite volume formula
\cite{Luscher:1985dn,Luscher:1986pf,Luscher:1991ux} relates the energy levels
of two-body states in a finite volume cubic box with periodic boundaries to
the infinite volume scattering matrix. \ Recently L\"{u}scher's method has
been studied and extended in a number of different ways. \ Several
investigations have looked at asymmetric boxes\ \cite{Li:2003jn,Feng:2004ua},
while another considered small volumes where the lattice length $L$ is smaller
than the scattering length \cite{Beane:2003da}. \ There have also been studies
of moving frames \cite{Rummukainen:1995vs, Kim:2005gf}, Yukawa interactions
\cite{deSoto:2006pe}, pion-exchange windings around the periodic boundary
\cite{Sato:2007ms}, modifications at nonzero lattice spacing
\cite{Seki:2005ns}, and techniques to distinguish shallow bound states from
scattering states using Levinson's theorem \cite{Sasaki:2006jn}. \ Several
recent studies derived finite volume formulas for systems of $n$ bosons with
short-range interactions \cite{Beane:2007qr,Detmold:2008gh}.

L\"{u}scher's method can be summarized as follows. \ We consider one up-spin
and one down-spin in a periodic cube of length $L$. \ The two-particle energy
levels in the center-of-mass frame are related to the $S$-wave phase shift,%
\begin{equation}
p\cot\delta_{0}(p)=\frac{1}{\pi L}S\left(  \eta\right)  ,\qquad\eta=\left(
\frac{Lp}{2\pi}\right)  ^{2}, \label{lusch}%
\end{equation}
where $S(\eta)$ is the three-dimensional zeta function,%
\begin{equation}
S(\eta)=\lim_{\Lambda\rightarrow\infty}\left[  \sum_{\vec{n}}\frac
{\theta(\Lambda^{2}-\vec{n}^{2})}{\vec{n}^{2}-\eta}-4\pi\Lambda\right]  .
\label{S}%
\end{equation}
The $S$-wave effective range expansion gives another expression for the
left-hand side of Eq.~(\ref{lusch}),%
\begin{equation}
p\cot\delta_{0}(p)\approx-\frac{1}{a_{\text{scatt}}}+\frac{1}{2}r_{0}%
p^{2}+\cdots\text{.} \label{effrange}%
\end{equation}

\begin{figure}
[ptb]
\begin{center}
\includegraphics[
height=0.9617in,
width=4.0041in
]%
{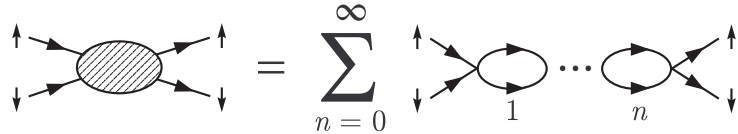}%
\caption{Sum of bubble diagrams contributing to two-particle scattering.}%
\label{twotwo}%
\end{center}
\end{figure}

In terms of $\eta$, the energy of the two-particle scattering state is%
\begin{equation}
E_{\text{pole}}=\frac{p^{2}}{m}=\frac{\eta}{m}\left(  \frac{2\pi}{L}\right)
^{2}. \label{Epole}%
\end{equation}
For the case of zero-range interactions, the location of the two-particle
scattering pole is calculated by summing the bubble diagrams shown in
Fig.~\ref{twotwo}.%
%TCIMACRO{\FRAME{ftbpFU}{4.0041in}{0.9617in}{0pt}{\Qcb{Sum of bubble diagrams
%contributing to two-particle scattering.}}{\Qlb{twotwo}}{twotwo.eps}%
%width 4.0041in;  height 0.9617in;  depth 0pt;  original-width 5.0004in;
%original-height 1.1831in;  cropleft "0";  croptop "1";  cropright "1";
%cropbottom "0";  filename 'twotwo.eps';file-properties "XNPEU";}} }%
%BeginExpansion
%EndExpansion
The relation between $C$ and $E_{\text{pole}}$ is \cite{Lee:2004qd}%
\begin{equation}
-\frac{1}{C\alpha_{t}}=\lim_{L\rightarrow\infty}\frac{1}{L^{3}}\sum_{\vec
{k}\text{ }\operatorname{integer}}\frac{1}{e^{-E_{\text{pole}}\alpha_{t}%
}-1+2\alpha_{t}\omega(2\pi\vec{k}/L)-\alpha_{t}^{2}\omega^{2}(2\pi\vec{k}/L)},
\label{bubble}%
\end{equation}
where%
\begin{equation}
\omega(\vec{p})=\frac{1}{m}\sum_{l=1,2,3}\left(  1-\cos p_{l}\right)
\end{equation}
for the standard lattice action. \ In this manner the coefficient $C$ can be
tuned to produce the desired scattering length $a_{\text{scatt}}$ at infinite
volume. \ Higher-order scattering parameters can also be extracted in this
way. \ However for zero-range interactions the characteristic scale of these
higher-order parameters is the lattice spacing, and so higher-order scattering
corrections are the same size as lattice discretization errors produced by
broken Galilean invariance and other lattice effects.

\subsection{Spherical wall method}

While L\"{u}scher's method is very useful at low momenta, it is not so useful
for determining phase shifts on the lattice at higher energies and higher
orbital angular momenta. \ Furthermore spin-orbit coupling and partial-wave
mixing are difficult to measure accurately using L\"{u}scher's method due to
multiple-scattering artifacts produced by the periodic cubic boundary. \ A
more robust approach was proposed in Ref.~\cite{Borasoy:2007vy} to measure
phase shifts for nonrelativistic point particles on the lattice using a
spherical wall boundary. \ Similar techniques have long been used in nuclear
physics (see for example Problem 5-7 in Ref.~\cite{Preston:1975}) dating back
to early work on $R$-matrix methods \cite{Wigner:1947a}. \ We summarize the
method as follows.

\begin{figure}
[ptb]
\begin{center}
\includegraphics[
height=1.6259in,
width=1.6259in
]%
{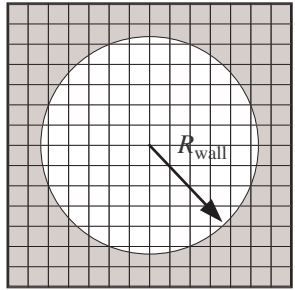}%
\caption{Spherical wall imposed in the center-of-mass frame.}%
\label{spherical_wall}%
\end{center}
\end{figure}

A hard spherical wall boundary is imposed on the relative separation between
the two particles at some chosen radius $R_{\text{wall}}$. \ This boundary
condition removes copies of the interactions produced by the periodic lattice.
\ Viewed in the center-of-mass frame we solve the Schr\"{o}dinger equation for
spherical standing waves which vanish at $r=R_{\text{wall}}$ as indicated in
Fig.~\ref{spherical_wall}.%
%TCIMACRO{\FRAME{ftbpFU}{1.6259in}{1.6259in}{0pt}{\Qcb{Spherical wall imposed
%in the center-of-mass frame.}}{\Qlb{spherical_wall}}{spherical_wall.eps}%
%width 1.6259in;  height 1.6259in;  depth 0pt;  original-width 3.1955in;
%original-height 3.1955in;  cropleft "0";  croptop "1";  cropright "1";
%cropbottom "0";  filename 'spherical_wall.eps';file-properties "XNPEU";}} }%
%BeginExpansion
%EndExpansion

When the combined intrinsic spin of the two interacting particles is zero
there is no mixing between partial waves. \ At values of $r$ beyond the range
of the interaction, the spherical standing wave can be decomposed as a
superposition of products of spherical harmonics and spherical Bessel
functions. \ Explicitly we have%
\begin{equation}
\left[  \cos\delta_{L}\cdot j_{L}(kr)-\sin\delta_{L}\cdot y_{L}(kr)\right]
Y_{L,L_{z}}(\theta,\phi), \label{wavefunction}%
\end{equation}
where the center-of-mass energy of the spherical wave is%
\begin{equation}
E=2\frac{k^{2}}{2m}=\frac{k^{2}}{m},
\end{equation}
and the phase shift for partial wave $L$ is $\delta_{L}$. \ We can determine
$k$ from the energy $E$ of the standing wave, and the phase shift $\delta_{L}$
is calculated by setting the wavefunction in Eq.~(\ref{wavefunction}) equal to
zero at the wall boundary,%
\begin{equation}
\cos\delta_{L}\cdot j_{L}(kR_{\text{wall}})=\sin\delta_{L}\cdot y_{L}%
(kR_{\text{wall}}),
\end{equation}%
\begin{equation}
\delta_{L}=\tan^{-1}\left[  \frac{j_{L}(kR_{\text{wall}})}{y_{L}%
(kR_{\text{wall}})}\right]  . \label{simple_phaseshift}%
\end{equation}
On the lattice there is some ambiguity on the value of $R_{\text{wall}}$ since
the components of $\vec{r}$ must be integer multiples of the lattice spacing.
\ The ambiguity is resolved by fine-tuning the value of $R_{\text{wall}}$ for
each standing wave so that $\delta_{L}$ equals zero when the particles are non-interacting.

When the combined intrinsic spin of the two interacting particles is nonzero,
spin-orbit coupling generates mixing between partial waves. $\ $For nucleons
the interesting case is $S=1$ where there is mixing between $L=J-1$ and
$L=J+1$. \ We discuss this case here using the two-component notation,
\begin{equation}
\left[
\begin{array}
[c]{c}%
R_{J-1}(r)\\
R_{J+1}(r)
\end{array}
\right]  , \label{twocomponent}%
\end{equation}
for the radial part of the wavefunction. \ Since we are considering a
two-channel system, there are two independent standing wave solutions of the
form%
\begin{equation}
\Psi^{I}\propto\frac{1}{k^{I}r}\left[
\begin{array}
[c]{c}%
A_{J-1}^{I}\sin\left(  k^{I}r-\frac{J-1}{2}\pi+\Delta_{J-1}^{I}\right) \\
A_{J+1}^{I}\sin\left(  k^{I}r-\frac{J+1}{2}\pi+\Delta_{J+1}^{I}\right)
\end{array}
\right]
\end{equation}
at energy $E^{I}=(k^{I})^{2}/m$ and%
\begin{equation}
\Psi^{II}\propto\frac{1}{k^{II}r}\left[
\begin{array}
[c]{c}%
A_{J-1}^{II}\sin\left(  k^{II}r-\frac{J-1}{2}\pi+\Delta_{J-1}^{II}\right) \\
A_{J+1}^{II}\sin\left(  k^{II}r-\frac{J+1}{2}\pi+\Delta_{J+1}^{II}\right)
\end{array}
\right]
\end{equation}
at $E^{II}=(k^{II})^{2}/m$. \ These can be used to derive the phase shifts
$\delta_{J-1}$ and $\delta_{J+1}$ and mixing angle $\varepsilon_{J}$ using
\cite{Borasoy:2007vy}%
\begin{equation}
\tan\left(  -\Delta_{J-1}^{I}+\delta_{J-1}\right)  \tan\left(  -\Delta
_{J+1}^{I}+\delta_{J+1}\right)  =\tan^{2}\varepsilon_{J},
\label{newconstraint1}%
\end{equation}%
\begin{equation}
\tan\left(  -\Delta_{J-1}^{II}+\delta_{J-1}\right)  \tan\left(  -\Delta
_{J+1}^{II}+\delta_{J+1}\right)  =\tan^{2}\varepsilon_{J},
\label{newconstraint2}%
\end{equation}%
\begin{equation}
A_{J-1}^{I}\tan\varepsilon_{J}=-A_{J+1}^{I}\frac{\sin\left(  -\Delta_{J+1}%
^{I}+\delta_{J+1}\right)  }{\cos\left(  -\Delta_{J-1}^{I}+\delta_{J-1}\right)
}, \label{newconstraint3}%
\end{equation}%
\begin{equation}
A_{J-1}^{II}\tan\varepsilon_{J}=-A_{J+1}^{II}\frac{\sin\left(  -\Delta
_{J+1}^{II}+\delta_{J+1}\right)  }{\cos\left(  -\Delta_{J-1}^{II}+\delta
_{J-1}\right)  }. \label{newconstraint4}%
\end{equation}

The phase shifts and mixing angle in Eq.~(\ref{newconstraint1}) and
(\ref{newconstraint3}) are at momentum $k^{I}$ while the phase shifts and
mixing angle in Eq.~(\ref{newconstraint2}) and (\ref{newconstraint4}) are at
momentum $k^{II}$. \ Nearly equal pairs $k^{I}\approx k^{II}$ are used in
solving the coupled constraints Eq.~(\ref{newconstraint1}%
)-(\ref{newconstraint4}). \ In practice this amounts to considering the
$(n+1)^{\text{st}}$-radial excitation of $L=J-1$ together with the
$n^{\text{th}}$-radial excitation of $L=J+1$. \ Then we use%
\begin{equation}
\tan\left(  -\Delta_{J-1}^{I}+\delta_{J-1}(k^{I})\right)  \tan\left(
-\Delta_{J+1}^{I}+\delta_{J+1}(k^{I})\right)  =\tan^{2}\left[  \varepsilon
_{J}(k^{I})\right]  , \label{kI_1}%
\end{equation}%
\begin{equation}
\tan\left(  -\Delta_{J-1}^{II}+\delta_{J-1}(k^{I})\right)  \tan\left(
-\Delta_{J+1}^{II}+\delta_{J+1}(k^{I})\right)  \approx\tan^{2}\left[
\varepsilon_{J}(k^{I})\right]  , \label{kI_2}%
\end{equation}%
\begin{equation}
A_{J-1}^{I}\tan\left[  \varepsilon_{J}(k^{I})\right]  =-A_{J+1}^{I}\frac
{\sin\left(  -\Delta_{J+1}^{I}+\delta_{J+1}(k^{I})\right)  }{\cos\left(
-\Delta_{J-1}^{I}+\delta_{J-1}(k^{I})\right)  }, \label{kI_3}%
\end{equation}
for the phase shifts and mixing angle at $k=k^{I}$, and%
\begin{equation}
\tan\left(  -\Delta_{J-1}^{I}+\delta_{J-1}(k^{II})\right)  \tan\left(
-\Delta_{J+1}^{I}+\delta_{J+1}(k^{II})\right)  \approx\tan^{2}\left[
\varepsilon_{J}(k^{II})\right]  , \label{kII_1}%
\end{equation}%
\begin{equation}
\tan\left(  -\Delta_{J-1}^{II}+\delta_{J-1}(k^{II})\right)  \tan\left(
-\Delta_{J+1}^{II}+\delta_{J+1}(k^{II})\right)  =\tan^{2}\left[
\varepsilon_{J}(k^{II})\right]  , \label{kII_2}%
\end{equation}%
\begin{equation}
A_{J-1}^{II}\tan\left[  \varepsilon_{J}(k^{II})\right]  =-A_{J+1}^{II}%
\frac{\sin\left(  -\Delta_{J+1}^{II}+\delta_{J+1}(k^{II})\right)  }%
{\cos\left(  -\Delta_{J-1}^{II}+\delta_{J-1}(k^{II})\right)  }, \label{kII_3}%
\end{equation}
for the phase shifts and mixing angle at $k=k^{II}$.

\subsection{Scattering at NLO in chiral effective field theory}

\begin{figure}
[ptb]
\begin{center}
\includegraphics[
height=3.7749in,
width=4.6959in
]%
{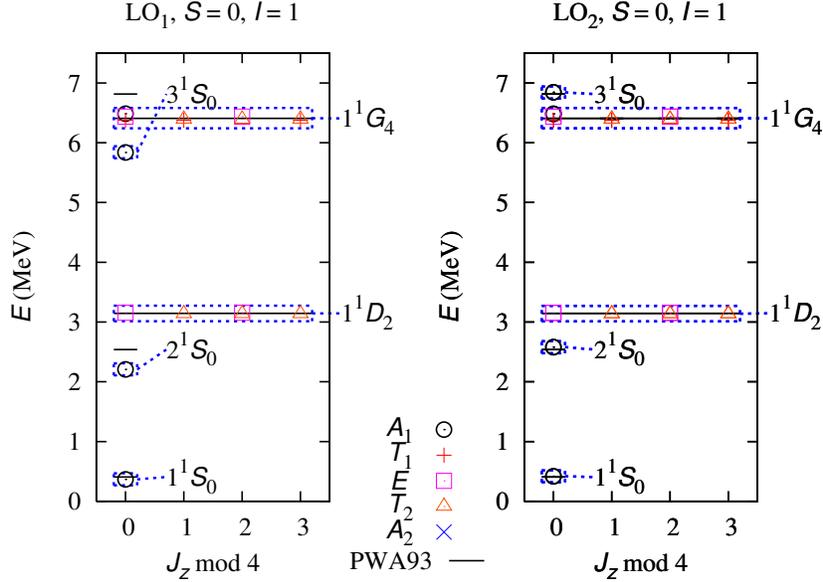}%
\caption{Energy levels for $S=0$, $I=1$ using lattice actions LO$_{1}$ and
LO$_{2}$ and a spherical wall at radius $R_{\text{wall}}=10+\epsilon$ lattice
units \cite{Borasoy:2007vi}. \ The solid line indicates the exact energy
levels which reproduce data from the partial wave analysis of
\cite{Stoks:1993tb}.}%
\label{s0_i1_r10}%
\end{center}
\end{figure}

Lattice phase shifts and mixing angles at leading order and next-to-leading
order were calculated in Ref.~\cite{Borasoy:2007vi} using the spherical wall
method at lattice spacings $a=(100$ MeV$)^{-1}$, $a_{t}=(70$ MeV$)^{-1}$. \ We
summarize the results here. \ Fig.~\ref{s0_i1_r10} shows energy levels for
spin $S=0$ and isospin $I=1$ using lattice actions LO$_{1}$ and LO$_{2}$.
\ The spherical wall is at radius $R_{\text{wall}}=10+\epsilon$ lattice units
where $\epsilon$ is a small positive number. \ The $\epsilon$ notation makes
explicit that $\left\vert \vec{r}\right\vert =10$ lattice units\ is inside the
spherical wall but all lattice sites with $\left\vert \vec{r}\right\vert >10$
lattice units lie outside. \ The solid lines indicate the exact energy levels
which would reproduce data from the partial wave analysis in
\cite{Stoks:1993tb}. \
%TCIMACRO{\FRAME{ftbpFU}{4.6959in}{3.7749in}{0pt}{\Qcb{Energy levels for $S=0$,
%$I=1$ using lattice actions LO$_{1}$ and LO$_{2}$ and a spherical wall at
%radius $R_{\text{wall}}=10+\epsilon$ lattice units \cite{Borasoy:2007vi}.
%\ The solid line indicates the exact energy levels which reproduce data from
%the partial wave analysis of \cite{Stoks:1993tb}.}}{\Qlb{s0_i1_r10}%
%}{s0_i1_r10.eps}{\special{ language "Scientific Word";  type "GRAPHIC";
%width 4.6959in;  height 3.7749in;  depth 0pt;  original-width 7.7798in;
%original-height 6.2457in;  cropleft "0";  croptop "1";  cropright "1";
%cropbottom "0";  filename 'S0_I1_R10.eps';file-properties "XNPEU";}} }%
%BeginExpansion
%EndExpansion
The energy levels for the standard action LO$_{1}$ are $10\%$ to $15\%$ too
low for the $^{1}S_{0}$ states, while the improved action LO$_{2}$ is correct
to a few of percent for all $^{1}S_{0}$ states$.$ \ Deviations for higher
partial waves are smaller than one percent for both LO$_{1}$ and LO$_{2}$.

\begin{figure}
[ptb]
\begin{center}
\includegraphics[
height=3.7749in,
width=4.6959in
]%
{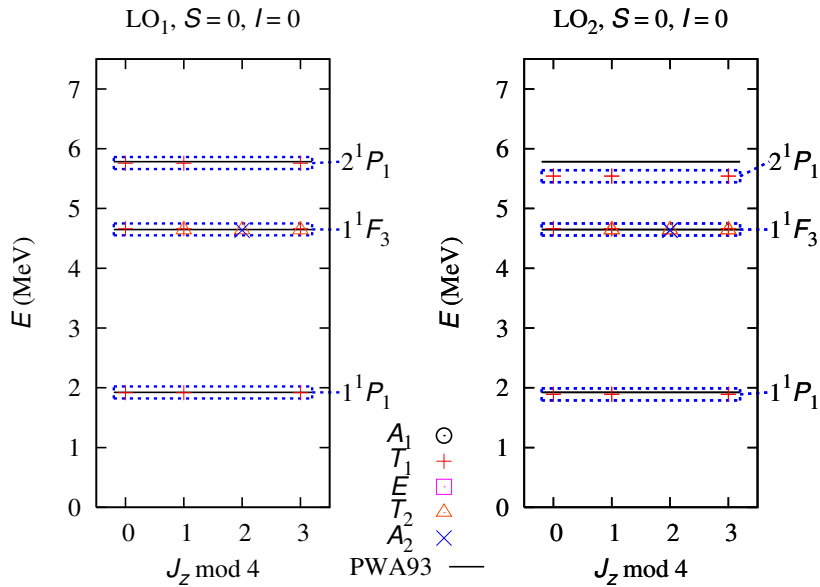}%
\caption{Energy levels for $S=0$, $I=0$ using lattice actions LO$_{1}$ and
LO$_{2}$ and a spherical wall at radius $R_{\text{wall}}=10+\epsilon$ lattice
units \cite{Borasoy:2007vi}. \ The solid line indicates the exact energy
levels which reproduce data from the partial wave analysis of
\cite{Stoks:1993tb}.}%
\label{s0_i0_r10}%
\end{center}
\end{figure}

The energy levels for spin $S=0$, isospin $I=0$, and $R_{\text{wall}%
}=10+\epsilon$ lattice units are shown in Fig.~\ref{s0_i0_r10}. \
%TCIMACRO{\FRAME{ftbpFU}{4.6959in}{3.7749in}{0pt}{\Qcb{Energy levels for $S=0$,
%$I=0$ using lattice actions LO$_{1}$ and LO$_{2}$ and a spherical wall at
%radius $R_{\text{wall}}=10+\epsilon$ lattice units \cite{Borasoy:2007vi}.
%\ The solid line indicates the exact energy levels which reproduce data from
%the partial wave analysis of \cite{Stoks:1993tb}.}}{\Qlb{s0_i0_r10}%
%}{s0_i0_r10.eps}{\special{ language "Scientific Word";  type "GRAPHIC";
%width 4.6959in;  height 3.7749in;  depth 0pt;  original-width 6.992in;
%original-height 10.0024in;  cropleft "0";  croptop "1";  cropright "1";
%cropbottom "0";  filename 'S0_I0_R10.eps';file-properties "XNPEU";}} }%
%BeginExpansion
%EndExpansion
In this case LO$_{1}$ is better for the $^{1}P_{1}$ states and is within one
percent of the exact values. \ The LO$_{2}$ energy levels are further away,
though still within a few percent for the $^{1}P_{1}$ states.

\begin{figure}
[ptb]
\begin{center}
\includegraphics[
height=2.6057in,
width=5.0289in
]%
{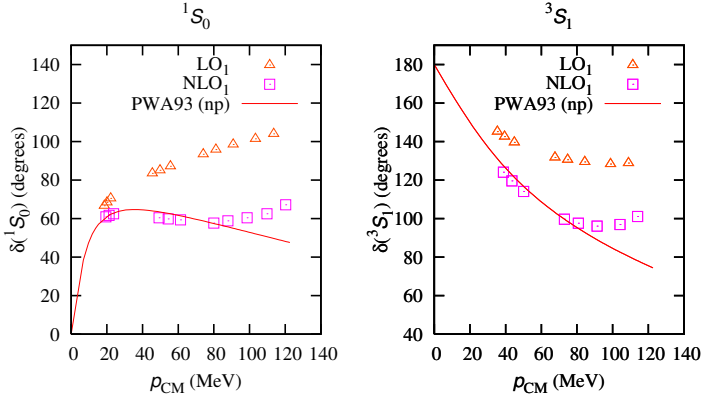}%
\caption{$S$-wave phase shifts versus center-of-mass momentum for LO$_{1}$ and
NLO$_{1}$ \cite{Borasoy:2007vi}.}%
\label{swave_b0}%
\end{center}
\end{figure}

\begin{figure}
[ptbptb]
\begin{center}
\includegraphics[
height=2.6048in,
width=5.0298in
]%
{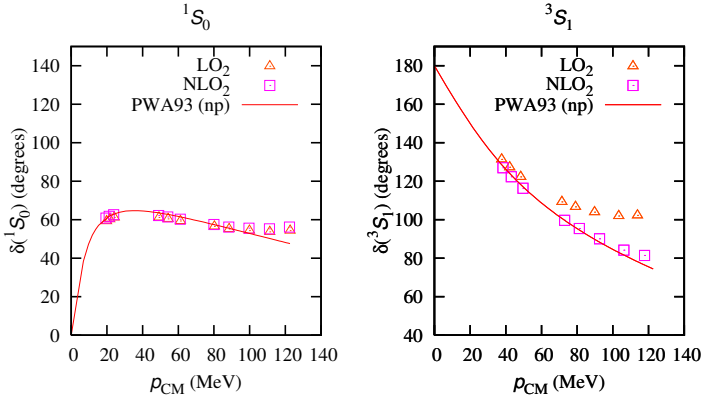}%
\caption{$S$-wave phase shifts versus center-of-mass momentum for LO$_{2}$ and
NLO$_{2}$ \cite{Borasoy:2007vi}.}%
\label{swave_b6}%
\end{center}
\end{figure}

In Ref.~\cite{Borasoy:2007vi} the nine unknown operator coefficients at
next-to-leading order were determined by matching three $S$-wave scattering
data points, four $P$-wave scattering data points, as well as the deuteron
binding energy and quadrupole moment. \ Each of the next-to-leading-order
corrections were computed perturbatively. \ The $S$-wave phase shifts for
LO$_{1}$ and NLO$_{1}$ versus center-of-mass momentum $p_{\text{CM}}$ are
shown in Fig.~\ref{swave_b0}, and the $S$-wave phase shifts for LO$_{2}$ and
NLO$_{2}$ are shown in Fig.~\ref{swave_b6}. \ The NLO$_{1}$ and NLO$_{2}$
results are both in good agreement with partial wave results from
\cite{Stoks:1993tb}. \ Systematic errors can be seen at momenta greater than
about $80$ MeV and are larger for NLO$_{1}$. \ But in both cases the
deviations are at larger momenta and consistent with higher-order effects.%
%TCIMACRO{\FRAME{ftbpFU}{5.0289in}{2.6057in}{0pt}{\Qcb{$S$-wave phase shifts
%versus center-of-mass momentum for LO$_{1}$ and NLO$_{1}$
%\cite{Borasoy:2007vi}.}}{\Qlb{swave_b0}}{swave_b0.eps}%
%width 5.0289in;  height 2.6057in;  depth 0pt;  original-width 2.4414in;
%original-height 3in;  cropleft "0";  croptop "1";  cropright "1";
%cropbottom "0";  filename 'Swave_b0.eps';file-properties "XNPEU";}} }%
%BeginExpansion
%EndExpansion
%TCIMACRO{\FRAME{ftbpFU}{5.0298in}{2.6048in}{0pt}{\Qcb{$S$-wave phase shifts
%versus center-of-mass momentum for LO$_{2}$ and NLO$_{2}$
%\cite{Borasoy:2007vi}.}}{\Qlb{swave_b6}}{swave_b6.eps}%
%width 5.0298in;  height 2.6048in;  depth 0pt;  original-width 8.3359in;
%original-height 4.2964in;  cropleft "0";  croptop "1";  cropright "1";
%cropbottom "0";  filename 'Swave_b6.eps';file-properties "XNPEU";}} }%
%BeginExpansion
%EndExpansion

\begin{figure}
[ptb]
\begin{center}
\includegraphics[
height=5.1145in,
width=5.028in
]%
{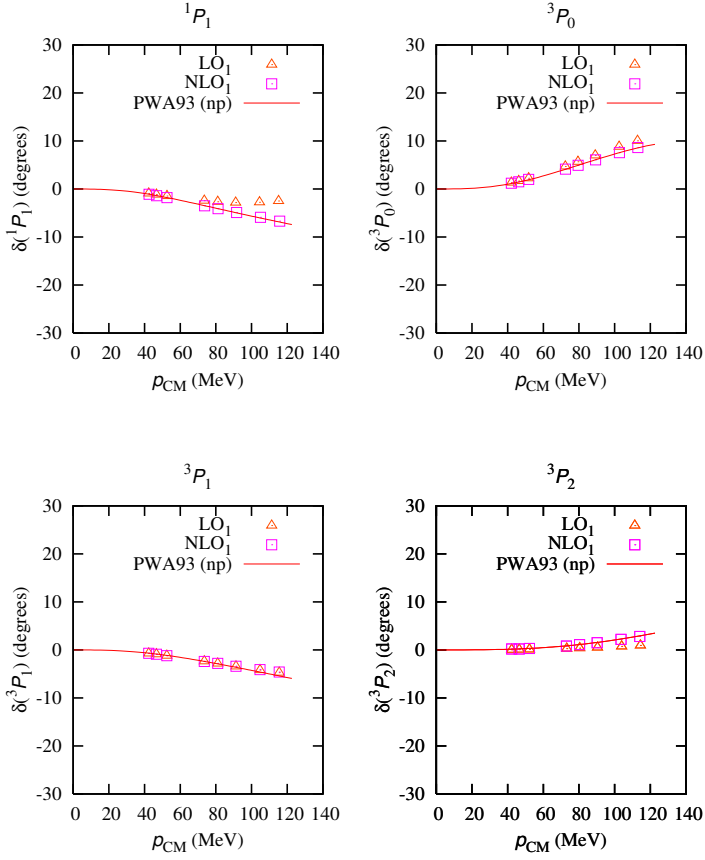}%
\caption{$P$-wave phase shifts versus center-of-mass momentum for LO$_{1}$ and
NLO$_{1}$ \cite{Borasoy:2007vi}.}%
\label{pwave_b0}%
\end{center}
\end{figure}

\begin{figure}
[ptbptb]
\begin{center}
\includegraphics[
height=5.1145in,
width=5.028in
]%
{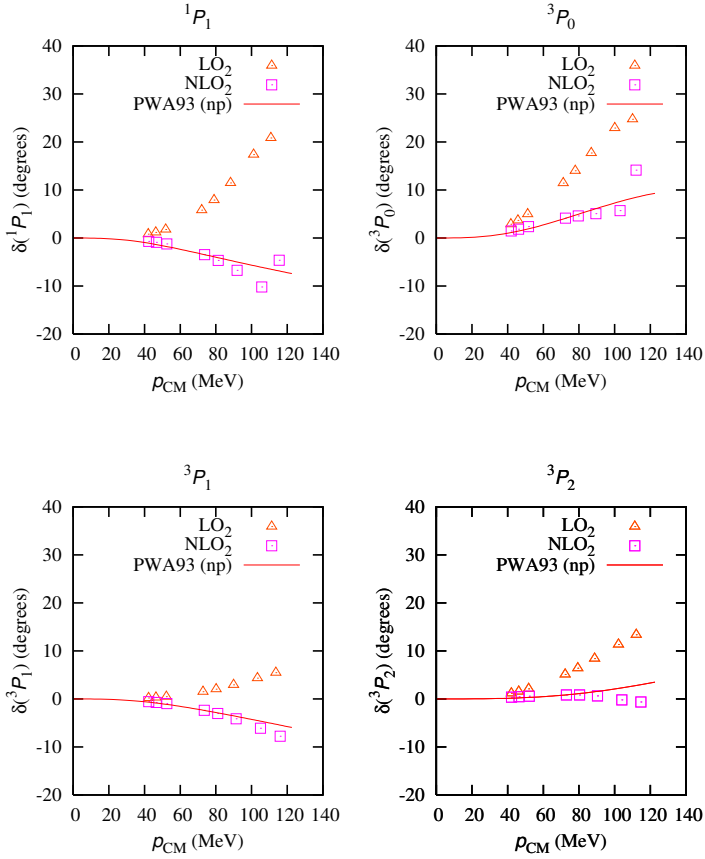}%
\caption{$P$-wave phase shifts versus center-of-mass momentum for LO$_{2}$ and
NLO$_{2}$ \cite{Borasoy:2007vi}.}%
\label{pwave_b6}%
\end{center}
\end{figure}

$P$-wave phase shifts are shown in Fig.~\ref{pwave_b0} and \ref{pwave_b6}
\cite{Borasoy:2007vi}. \ In this case the phase shifts are already close for
LO$_{1}$ and quite accurate for NLO$_{1}$. \ This suggests that only a small
correction is needed on top of $P$-wave interactions produced by one-pion
exchange. \ The results for LO$_{2}$ and NLO$_{2}$ are not quite as good.
\ The Gaussian smearing introduced in LO$_{2}$ produces attractive forces in
each $P$-wave channel that must be cancelled by next-to-leading-order
corrections. \ However the residual deviations in the NLO$_{2}$ results appear
consistent with effects that can be cancelled by higher-order terms.%
%TCIMACRO{\FRAME{ftbpFU}{5.028in}{5.1145in}{0pt}{\Qcb{$P$-wave phase shifts
%versus center-of-mass momentum for LO$_{1}$ and NLO$_{1}$
%\cite{Borasoy:2007vi}.}}{\Qlb{pwave_b0}}{pwave_b0.eps}%
%width 5.028in;  height 5.1145in;  depth 0pt;  original-width 2.4414in;
%original-height 3in;  cropleft "0";  croptop "1";  cropright "1";
%cropbottom "0";  filename 'Pwave_b0.eps';file-properties "XNPEU";}} }%
%BeginExpansion
%EndExpansion
%TCIMACRO{\FRAME{ftbpFU}{5.028in}{5.1145in}{0pt}{\Qcb{$P$-wave phase shifts
%versus center-of-mass momentum for LO$_{2}$ and NLO$_{2}$
%\cite{Borasoy:2007vi}.}}{\Qlb{pwave_b6}}{pwave_b6.eps}%
%width 5.028in;  height 5.1145in;  depth 0pt;  original-width 8.3333in;
%original-height 8.4769in;  cropleft "0";  croptop "1";  cropright "1";
%cropbottom "0";  filename 'Pwave_b6.eps';file-properties "XNPEU";}} }%
%BeginExpansion
%EndExpansion

\begin{figure}
[ptb]
\begin{center}
\includegraphics[
height=2.6048in,
width=5.0289in
]%
{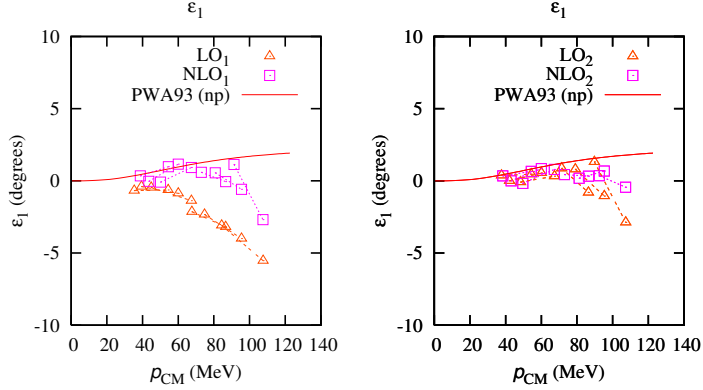}%
\caption{$\varepsilon_{1}$ mixing angle for LO$_{1}$ and NLO$_{1}$ on the
left, LO$_{2}$ and NLO$_{2}$ on the right \cite{Borasoy:2007vi}.}%
\label{eps1}%
\end{center}
\end{figure}

The mixing parameter $\varepsilon_{1}$ for $J=1$ is shown in Fig.~\ref{eps1}
\cite{Borasoy:2007vi}. \ The mixing angle is defined according to the Stapp
parameterization \cite{Stapp:1956mz}. \ Results for LO$_{1}$ and NLO$_{1}$ are
on the left, and results for LO$_{2}$ and NLO$_{2}$ are on the right. \ The
pairs of points connected by dotted lines indicate pairs of solutions at
$k=k^{I}$ and $k=k^{II}$ for the coupled $^{3}S_{1}$-$^{3}D_{1}$
channels.\ \ For LO$_{1}$ we note that $\varepsilon_{1}$ has the wrong sign.
\ This suggests that the mixing angle may be more sensitive to lattice
discretization errors than other scattering parameters. \ However for both
NLO$_{1}$ and NLO$_{2}$ results the remaining deviations appear consistent
with effects produced by higher-order interactions.%
%TCIMACRO{\FRAME{ftbpFU}{5.0289in}{2.6048in}{0pt}{\Qcb{$\varepsilon_{1}$ mixing
%angle for LO$_{1}$ and NLO$_{1}$ on the left, LO$_{2}$ and NLO$_{2}$ on the
%right \cite{Borasoy:2007vi}.}}{\Qlb{eps1}}{eps1.eps}%
%width 5.0289in;  height 2.6048in;  depth 0pt;  original-width 2.4414in;
%original-height 3in;  cropleft "0";  croptop "1";  cropright "1";
%cropbottom "0";  filename 'eps1.eps';file-properties "XNPEU";}} }%
%BeginExpansion
%EndExpansion

\section{Monte Carlo algorithms}

\subsection{Worldline methods}

\begin{figure}
[ptb]
\begin{center}
\includegraphics[
height=2.7103in,
width=2.5002in
]%
{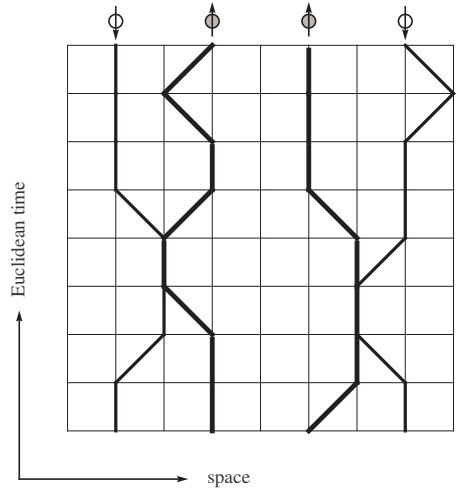}%
\caption{Example of a worldline configuration for two-component fermions.}%
\label{worldlines}%
\end{center}
\end{figure}

In bosonic systems or few-body systems where the problem of fermion sign
cancellation is not severe, lattice simulations can be performed by directly
sampling particle worldline configurations. \ We sketch an example of a
lattice worldline configuration for two-component fermions in one spatial
dimension in Fig.~(\ref{worldlines}). \
%TCIMACRO{\FRAME{ftbpFU}{2.5002in}{2.7103in}{0pt}{\Qcb{Example of a worldline
%configuration for two-component fermions.}}{\Qlb{worldlines}}{worldlines.eps}%
%width 2.5002in;  height 2.7103in;  depth 0pt;  original-width 4.1208in;
%original-height 4.4711in;  cropleft "0";  croptop "1";  cropright "1";
%cropbottom "0";  filename 'worldlines.eps';file-properties "XNPEU";}} }%
%BeginExpansion
%EndExpansion
This technique was used in the simulation of the triton using pionless
effective field theory \cite{Borasoy:2005yc}. \ A number of efficient cluster
algorithms have been developed for condensed matter applications to generate
new worldline configurations based on loop and worm updates
\cite{Kawashima:1994,Brower:1998,Chandrasekharan:1999,Evertz:2003,Kawashima:2004,Boninsegni:2006}%
.

While there are techniques which address the sign problem in certain cases
\cite{Chandrasekharan:1999}, there is no general method known for eliminating
sign oscillations in fermionic systems due to identical particle permutations.
\ For Monte Carlo simulations extending over Euclidean time $t$, the sign of
the configuration, sgn$(C)$, averaged over all configurations $C$ scales as%
\begin{equation}
\left\langle \text{sgn}(C)\right\rangle \sim\exp\left[  \left(  E_{0}%
^{\text{bosonic}}-E_{0}^{\text{fermionic}}\right)  t\right]  ,
\end{equation}
where $E_{0}^{\text{fermionic}}$ is the physical ground state energy and
$E_{0}^{\text{bosonic}}$ is the fictitious ground state energy for bosons with
the same interactions. \ The severity of the problem scales exponentially with
the size of the system and inverse temperature. \ In nuclear physics the same
issue arises in continuous-space worldline methods such as Green's Function
Monte Carlo and auxiliary-field diffusion Monte Carlo. \ In each case some
supplementary condition is used to fix fermion nodal boundaries or constrain
the domain of path integration \cite{zhang:1995a,zhang:1997a}.

\subsection{Determinantal diagrammatic methods}

Determinantal diagrammatic Monte Carlo was used in
\cite{Burovski:2006a,Burovski:2006b} to study two-component fermions in the
unitarity limit near the critical point. \ This method is structurally similar
to loop and worm updates of worldlines, however each configuration involves a
complete summation of diagrams for a given set of vertices in Euclidean space.
\ We discuss the method briefly here.

\begin{figure}
[ptb]
\begin{center}
\includegraphics[
height=2.1404in,
width=2.8444in
]%
{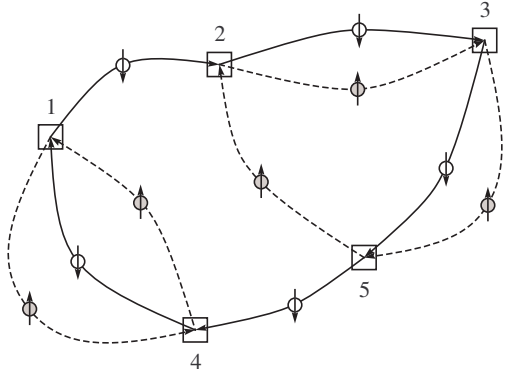}%
\caption{One diagram contributing to the sum of diagrams with vertices located
at the coordinates of $\mathcal{S}_{5}.$}%
\label{loops}%
\end{center}
\end{figure}

Let $G^{(0)}$ be the free-particle propagator in Euclidean space. \ We note
that $G^{(0)}$ is real-valued. \ We define a set of $n$ vertex locations
\begin{equation}
\mathcal{S}_{n}=\left\{  (\vec{r}_{j},t_{j})\right\}  _{j=1,\cdots,n},
\end{equation}
where $\vec{r}_{j}$ is the spatial location and $t_{j}$ is the Euclidean time
for the $j^{\text{th}}$ vertex. \ We also define a matrix of vertex-to-vertex
propagators $\mathbf{A}[\mathcal{S}_{n}]$, where%
\begin{equation}
A_{ij}[\mathcal{S}_{n}]=G^{(0)}(\vec{r}_{i}-\vec{r}_{j},t_{i}-t_{j}).
\end{equation}
As an example we choose a set of five points $\mathcal{S}_{5}$, and in
Fig.~(\ref{loops}) we draw a Feynman diagram with vertices located at the
coordinates of $\mathcal{S}_{5}$. \
%TCIMACRO{\FRAME{ftbpFU}{2.8444in}{2.1404in}{0pt}{\Qcb{One diagram contributing
%to the sum of diagrams with vertices located at the coordinates of
%$\mathcal{S}_{5}.$}}{\Qlb{loops}}{loops.eps}%
%width 2.8444in;  height 2.1404in;  depth 0pt;  original-width 4.6942in;
%original-height 3.5215in;  cropleft "0";  croptop "1";  cropright "1";
%cropbottom "0";  filename 'loops.eps';file-properties "XNPEU";}} }%
%BeginExpansion
%EndExpansion
The propagators for the down spins in Fig.~(\ref{loops}) give one term in the
expansion of $\det\mathbf{A}[\mathcal{S}_{5}]$,%
\begin{equation}
\det\mathbf{A}[\mathcal{S}_{5}]=\cdots+A_{14}A_{45}A_{53}A_{32}A_{21}+\cdots.
\end{equation}
The same is true for the up spins in Fig.~(\ref{loops}),%
\begin{equation}
\det\mathbf{A}[\mathcal{S}_{5}]=\cdots-A_{25}A_{53}A_{32}A_{14}A_{41}+\cdots,
\end{equation}
and the determinant expansion shows that there is a relative minus sign
between the up and down contributions. \ From this example it is clear that
the total contribution of all Feynman diagrams with vertices given by
$\mathcal{S}_{n}$ is%
\begin{equation}
dP[\mathcal{S}_{n}]=\left(  -C\alpha_{t}\right)  ^{n}\left\{  \det
\mathbf{A}[\mathcal{S}_{n}]\right\}  ^{2}.
\end{equation}
We note that $dP[\mathcal{S}_{n}]$ is positive definite when the interaction
is attractive, $C<0$. \ Convergence of the series in powers of $C$ is
guaranteed by finiteness of the Grassmann path integral at finite volume. \ 

\begin{figure}
[ptb]
\begin{center}
\includegraphics[
height=2.1404in,
width=2.8444in
]%
{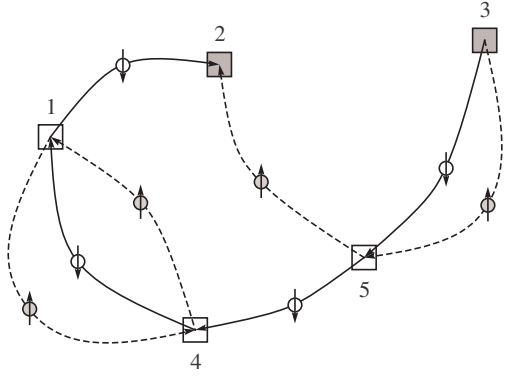}%
\caption{Single worm diagram with pairs of fermion lines created at vertex $3$
and annihilated at vertex $2$.}%
\label{worm}%
\end{center}
\end{figure}

In order to compute the full path integral%
\begin{equation}
\mathcal{Z}=\sum_{n}\int_{\mathcal{S}_{n}}dP[\mathcal{S}_{n}],
\end{equation}
the sampling of vertex configurations can be generated using a worm algorithm
that produces closed loop diagrams such as Fig.~(\ref{loops}) as well as
single worm diagrams such as the example shown in Fig.~(\ref{worm}). \ In this
diagram pairs of fermion lines are created at vertex $3$ and annihilated at
vertex $2$. \ The sum of all diagrams of the type shown in Fig.~(\ref{worm})
can be written in terms of the derivative of $\det\mathbf{A}[\mathcal{S}_{5}]$
with respect to $A_{32}$,%
\begin{equation}
\left(  -C\alpha_{t}\right)  ^{5}\left\{  \frac{\partial}{\partial A_{32}}%
\det\mathbf{A}[\mathcal{S}_{5}]\right\}  ^{2}.
\end{equation}
From this we see that the contribution from worm diagrams is also positive.
\ These diagrams are used to calculate the expectation value of the pair
correlation function,%
\begin{equation}
\left\langle c_{\downarrow}(\vec{r}_{2},t_{2})c_{\uparrow}(\vec{r}_{2}%
,t_{2})c_{\uparrow}^{\ast}(\vec{r}_{3},t_{3})c_{\downarrow}^{\ast}(\vec{r}%
_{3},t_{3})\right\rangle .
\end{equation}
Further details of the worm updating algorithm and determinantal diagrammatic
Monte Carlo can be found in
\cite{Burovski:2006a,Burovski:2006b,vanhoucke:2008}.%

%TCIMACRO{\FRAME{ftbpFU}{2.8444in}{2.1404in}{0pt}{\Qcb{Single worm diagram with
%pairs of fermion lines created at vertex $3$ and annihilated at vertex $2$.}%
%}{\Qlb{worm}}{worm.eps}{\special{ language "Scientific Word";
%type "GRAPHIC";  maintain-aspect-ratio TRUE;  display "USEDEF";
%valid_file "F";  width 2.8444in;  height 2.1404in;  depth 0pt;
%original-width 4.6942in;  original-height 3.5215in;  cropleft "0";
%filename 'worm.eps';file-properties "XNPEU";}} }%
%BeginExpansion
%EndExpansion

\subsection{Projection Monte Carlo with auxiliary field}

Projection Monte Carlo was used to compute the ground state energy of
two-components fermions at unitarity
\cite{Lee:2005fk,Lee:2006hr,Juillet:2007a,Lee:2008xs}. \ It was also used to
study light nuclei and dilute neutrons in chiral effective field theory
\cite{Borasoy:2006qn,Borasoy:2007vk}. \ We briefly describe the method here
using first the example of zero-range attractive two-component fermions.

Let $E_{N,N}^{0}$ be the ground state for the interacting system of $N$ up
spins and $N$ down spins. \ Let $\left\vert \Psi_{N,N}^{0,\text{free}%
}\right\rangle $ be the normalized Slater-determinant ground state for a
non-interacting system of $N$ up spins and $N$ down spins. \ We use the
auxiliary-field transfer matrix defined in Eq.~(\ref{transfer_aux}) to
construct the Euclidean-time projection amplitude%
\begin{equation}
Z_{N,N}(t)=\prod\limits_{\vec{n},n_{t}}\left[  \int d_{A}s(\vec{n}%
,n_{t})\right]  \left\langle \Psi_{N,N}^{0,\text{free}}\right\vert
M_{A}(s,L_{t}-1)\cdot\cdots\cdot M_{A}(s,0)\left\vert \Psi_{N,N}%
^{0,\text{free}}\right\rangle ,
\end{equation}
where $t=L_{t}\alpha_{t}$. \ We define $E_{N,N}(t)$ as the transient energy
measured at time $t$,%
\begin{equation}
E_{N,N}(t)=\frac{1}{\alpha_{t}}\ln\frac{Z_{N,N}(t-\alpha_{t})}{Z_{N,N}(t)}.
\end{equation}
So long as the overlap between $\left\vert \Psi_{N,N}^{0,\text{free}%
}\right\rangle $ and the ground state of the interacting system is nonzero,
the ground state $E_{N,N}^{0}$ is given by the limit%
\begin{equation}
E_{N,N}^{0}=\lim_{t\rightarrow\infty}E_{N,N}(t).
\end{equation}

As a result of normal ordering, $M_{A}(s,n_{t})$ consists of single-particle
operators interacting with the background auxiliary field and contains no
direct interactions between particles. \ Therefore we can write%
\begin{equation}
\left\langle \Psi_{N,N}^{0,\text{free}}\right\vert M_{A}(s,L_{t}-1)\cdot
\cdots\cdot M_{A}(s,0)\left\vert \Psi_{N,N}^{0,\text{free}}\right\rangle
=\left[  \det\mathbf{M}_{A}(s,t)\right]  ^{2}, \label{detsquare}%
\end{equation}
where%
\begin{equation}
\left[  \mathbf{M}_{A}(s,t)\right]  _{k^{\prime}k}=\left\langle \vec
{p}_{k^{\prime}}\right\vert M_{A}(s,L_{t}-1)\cdot\cdots\cdot M_{A}%
(s,0)\left\vert \vec{p}_{k}\right\rangle , \label{one_particle}%
\end{equation}
for matrix indices $k,k^{\prime}=1,\cdots,N$. $\ \left\vert \vec{p}%
_{k}\right\rangle ,\left\vert \vec{p}_{k^{\prime}}\right\rangle $ are
single-particle momentum states comprising the Slater-determinant state
$\left\vert \Psi_{N,N}^{0,\text{free}}\right\rangle $. \ The single-particle
interactions in $M_{A}(s,n_{t})$ are the same for both up and down spins.
\ Since the matrix is real-valued, the square of the determinant is
nonnegative and there is no problem with sign oscillations. \ New
configurations are accepted or rejected according to the probability weight%
\begin{equation}
dP(s)=\left[  \det\mathbf{M}_{A}(s,L_{t}\alpha_{t})\right]  ^{2}d_{A}s.
\end{equation}
We note that $\left[  \mathbf{M}_{A}(s,t)\right]  _{k^{\prime}k}$ is only an
$N\times N$ matrix. \ This is considerably smaller than matrices encountered
in most other determinantal methods and contributes to the relative efficiency
of projection Monte Carlo. \ For the case when the auxiliary field $s$ is
continuous, new configurations can be generated using a non-local updating
algorithm called hybrid Monte Carlo
\cite{Scalettar:1986uy,Gottlieb:1987mq,Duane:1987de}. \ This scheme is widely
used in lattice QCD simulations.

\subsection{Hybrid Monte Carlo}

We describe the hybrid Monte Carlo algorithm in general terms for probability
weight
\begin{equation}
P(s)\propto\exp\left[  -V(s)\right]  ,
\end{equation}
which depends on the lattice field $s(\vec{n},n_{t})$ and some function $V(s)$
which may be a non-local function of $s$. \ The method proposes new
configurations by means of molecular dynamics trajectories for%
\begin{equation}
H(s,p)=\frac{1}{2}\sum_{\vec{n},n_{t}}\left[  p(\vec{n},n_{t})\right]
^{2}+V(s),
\end{equation}
where $p(\vec{n},n_{t})$ is the conjugate momentum for $s(\vec{n},n_{t})$.
\ The steps of the algorithm are as follows.

\begin{itemize}
\item[Step 1:] Select an arbitrary initial configuration $s^{0}$.

\item[Step 2:] Select a configuration $p^{0}$ according to the Gaussian random
distribution%
\begin{equation}
P\left[  p^{0}(\vec{n},n_{t})\right]  \propto\exp\left\{  -\frac{1}{2}\left[
p^{0}(\vec{n},n_{t})\right]  ^{2}\right\}  .
\end{equation}

\item[Step 3:] For each $\vec{n},n_{t}$ let%
\begin{equation}
\tilde{p}^{0}(\vec{n},n_{t})=p^{0}(\vec{n},n_{t})-\frac{\varepsilon
_{\text{step}}}{2}\left[  \frac{\partial V(s)}{\partial s(\vec{n},n_{t}%
)}\right]  _{s=s^{0}} \label{step3}%
\end{equation}
for some small positive $\varepsilon_{\text{step}}$.

\item[Step 4:] For steps $i=0,1,...,N_{\text{step}}-1$, let%
\begin{equation}
s^{i+1}(\vec{n},n_{t})=s^{i}(\vec{n},n_{t})+\varepsilon_{\text{step}}\tilde
{p}^{i}(\vec{n},n_{t}),
\end{equation}%
\begin{equation}
\tilde{p}^{i+1}(\vec{n},n_{t})=\tilde{p}^{i}(\vec{n},n_{t})-\varepsilon
_{\text{step}}\left[  \frac{\partial V_{j}(s)}{\partial s(\vec{n},n_{t}%
)}\right]  _{s=s^{i+1}},
\end{equation}
for each $\vec{n},n_{t}.$

\item[Step 5:] For each $\vec{n},n_{t}$ let%
\begin{equation}
p^{N_{\text{step}}}(\vec{n},n_{t})=\tilde{p}^{N_{\text{step}}}(\vec{n}%
,n_{t})+\frac{\varepsilon_{\text{step}}}{2}\left[  \frac{\partial
V(s)}{\partial s(\vec{n},n_{t})}\right]  _{s=s^{N_{\text{step}}}}.
\end{equation}

\item[Step 6:] Select a random number $r\in$ $[0,1).$ \ If
\begin{equation}
r<\exp\left[  -H(s^{N_{\text{step}}},p^{N_{\text{step}}})+H(s^{0}%
,p^{0})\right]
\end{equation}
then set $s^{0}=s^{N_{\text{step}}}$. \ Otherwise leave $s^{0}$ as is. \ In
either case go back to Step 2 to start a new trajectory.
\end{itemize}

\subsection{Grand canonical simulations with auxiliary field}

In Eq.~(\ref{Z_mu_aux}) we introduced the partition function for zero-range
attractive two-component fermions at chemical potential $\mu$,%
\begin{equation}
\mathcal{Z}(\mu)=\prod\limits_{\vec{n},n_{t}}\left[  \int d_{A}s(\vec{n}%
,n_{t})\right]  Tr\left\{  M_{A}(s,L_{t}-1,\mu)\cdot\cdots\cdot M_{A}%
(s,0,\mu)\right\}  , \label{Z_mu_aux_again}%
\end{equation}
with auxiliary-field transfer matrix%
\begin{equation}
M_{A}(s,n_{t},\mu)=M_{A}(s,n_{t})\exp\left\{  \mu\alpha_{t}\sum_{\vec{n}}%
\rho^{a^{\dagger}a}(\vec{n})\right\}  .
\end{equation}
Let $\left\vert \vec{n}\right\rangle $ be the quantum state with one particle
at lattice site $\vec{n}$ and no other particles. \ As in
Eq.~(\ref{one_particle}), we define the one-particle matrix amplitudes%
\begin{equation}
\left[  \mathbf{M}_{A}(s,t)\right]  _{\vec{n}^{\prime}\vec{n}}=\left\langle
\vec{n}^{\prime}\right\vert M_{A}(s,L_{t}-1)\cdot\cdots\cdot M_{A}%
(s,0)\left\vert \vec{n}\right\rangle .
\end{equation}
However in this case the matrix $\left[  \mathbf{M}_{A}(s,t)\right]  _{\vec
{n}^{\prime}\vec{n}}$ has dimensions $L^{3}\times L^{3}$.

The trace over states in Eq.~(\ref{Z_mu_aux_again}) can now be written as%
\begin{equation}
Tr\left\{  M_{A}(s,L_{t}-1,\mu)\cdot\cdots\cdot M_{A}(s,0,\mu)\right\}
=\left\{  \det\left[  1+e^{\mu\alpha_{t}}\mathbf{M}_{A}(s,L_{t}\alpha
_{t})\right]  \right\}  ^{2}.
\end{equation}
New configurations for $s$ can be updated locally using the Metropolis
algorithm. \ This method has been used in lattice calculations to study the
thermodynamics of two-component fermions near unitarity
\cite{Abe:2007fe,Abe:2007ff,Bulgac:2005a,Bulgac:2008b,Bulgac:2008c}\ and, more
generally, the attractive Hubbard model and repulsive Hubbard model near
half-filling in various spatial dimensions \cite{Sewer:2002,Hirsch:1983}. \ A
review of\ numerical aspects of this method can be found in
\cite{Gubernatis:1992a}.

\subsection{Pseudofermion methods}

The same grand canonical partition function $\mathcal{Z}(\mu)$ in
Eq.~(\ref{Z_mu_aux_again}) can be evaluated in the Grassmann path integral
formulation with auxiliary fields,%
\begin{equation}
\mathcal{Z}\text{(}\mu\text{)}=\prod\limits_{\vec{n},n_{t}}\left[  \int
d_{A}s(\vec{n},n_{t})\right]  \int DcDc^{\ast}\exp\left[  -S_{A}\left(
c,c^{\ast},s,\mu\right)  \right]  ,
\end{equation}
where%
\begin{equation}
S_{A}\left(  c,c^{\ast},s,\mu\right)  =S_{A}(e^{\mu\alpha_{t}}c,c^{\ast
},s)+\sum_{\vec{n},n_{t},i=\uparrow,\downarrow}\left[  \left(  1-e^{\mu
\alpha_{t}}\right)  c_{i}^{\ast}(\vec{n},n_{t})c_{i}(\vec{n},n_{t}+1)\right]
,
\end{equation}
and%
\begin{equation}
S_{A}\left(  c,c^{\ast},s\right)  =S_{\text{free}}(c,c^{\ast})-\sum_{\vec
{n},n_{t}}A\left[  s(\vec{n},n_{t})\right]  \rho(\vec{n},n_{t}).
\end{equation}

We note that $S_{A}\left(  c,c^{\ast},s,\mu\right)  $ is a bilinear form
coupling $c$ and $c^{\ast}$ with a block-diagonal spin structure which is the
same for up and down spins,%
\begin{align}
S_{A}\left(  c,c^{\ast},s,\mu\right)   &  =\sum_{\vec{n},n_{t}}\sum_{\vec
{n}^{\prime},n_{t}^{\prime}}c_{\uparrow}^{\ast}(\vec{n},n_{t})\left[
\mathbf{S}_{A}\left(  s,\mu\right)  \right]  _{\vec{n},n_{t};\vec{n}^{\prime
},n_{t}^{\prime}}c_{\uparrow}(\vec{n}^{\prime},n_{t}^{\prime})\nonumber\\
&  +\sum_{\vec{n},n_{t}}\sum_{\vec{n}^{\prime},n_{t}^{\prime}}c_{\downarrow
}^{\ast}(\vec{n},n_{t})\left[  \mathbf{S}_{A}\left(  s,\mu\right)  \right]
_{\vec{n},n_{t};\vec{n}^{\prime},n_{t}^{\prime}}c_{\downarrow}(\vec{n}%
^{\prime},n_{t}^{\prime}).
\end{align}
Therefore the integration over Grassmann variables gives the square of the
determinant of $\mathbf{S}_{A}\left(  s,\mu\right)  $,%
\begin{equation}
\mathcal{Z}\text{(}\mu\text{)}=\prod\limits_{\vec{n},n_{t}}\left[  \int
d_{A}s(\vec{n},n_{t})\right]  \left\{  \det\mathbf{S}_{A}\left(  s,\mu\right)
\right\}  ^{2}.
\end{equation}
This result can also be written as a path integral over a complex bosonic
field $\phi(\vec{n},n_{t})$,%
\begin{equation}
\mathcal{Z}\text{(}\mu\text{)}=\prod\limits_{\vec{n},n_{t}}\left[  \int
d_{A}s(\vec{n},n_{t})\frac{d\phi_{\text{real}}(\vec{n},n_{t})d\phi
_{\text{imag}}(\vec{n},n_{t})}{2\pi}\right]  \exp\left[  -T_{A}\left(
\phi,s,\mu\right)  \right]  ,
\end{equation}
where%
\begin{equation}
T_{A}\left(  \phi,s,\mu\right)  =\sum_{\vec{n},n_{t}}\sum_{\vec{n}^{\prime
},n_{t}^{\prime}}\phi^{\ast}(\vec{n},n_{t})\left[  \mathbf{S}_{A}^{-1\dagger
}\left(  s,\mu\right)  \mathbf{S}_{A}^{-1}\left(  s,\mu\right)  \right]
_{\vec{n},n_{t};\vec{n}^{\prime},n_{t}^{\prime}}\phi(\vec{n}^{\prime}%
,n_{t}^{\prime}). \label{T}%
\end{equation}
The bosonic field $\phi(\vec{n},n_{t})$ is called a pseudofermion field.
\ This technique was first implemented for fermions in lattice QCD
\cite{Weingarten:1980hx}. \ The non-local action in Eq.~(\ref{T}) can be
updated using a non-local algorithm such as hybrid Monte Carlo. \ Typically an
iterative sparse matrix solver is used such as the conjugate gradient method. \ 

Pseudofermion methods have been used to study the thermodynamics of
two-component fermions near unitarity
\cite{Lee:2004qd,Wingate:2005xy,Lee:2005is,Lee:2005it,Abe:2007fe,Abe:2007ff}.
\ For the case when an external field $J$ is coupled to the difermion pair,%
\begin{equation}
\sum_{\vec{n},n_{t}}\left[  J^{\ast}(\vec{n},n_{t})c_{\uparrow}^{\ast}(\vec
{n},n_{t})c_{\downarrow}^{\ast}(\vec{n},n_{t})+J(\vec{n},n_{t})c_{\downarrow
}(\vec{n},n_{t})c_{\uparrow}(\vec{n},n_{t})\right]  ,
\end{equation}
the block structure of the Grassmann action is more complicated. \ However the
analysis in Ref.~\cite{Chen:2003vy} shows that the path integral can still be
written in terms of a positive-definite Pfaffian. \ Lattice simulations using
this formalism were carried out using pseudofermion methods and hybrid\ Monte
Carlo \cite{Wingate:2005xy}.

\subsection{Applications to low-energy nucleons}

\begin{figure}
[ptb]
\begin{center}
\includegraphics[
height=1.6786in,
width=5.1378in
]%
{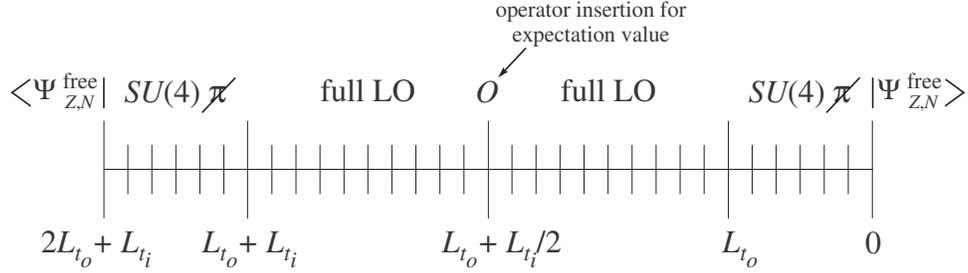}%
\caption{Schematic overview of the projection Monte Carlo calculation for
nucleons in chiral effective field theory.}%
\label{time_steps}%
\end{center}
\end{figure}

The projection Monte Carlo method with auxiliary fields has been used to study
low-energy nucleons in chiral effective field theory
\cite{Borasoy:2006qn,Borasoy:2007vi,Borasoy:2007vk}. \ A two-step approach was
used where a pionless SU(4)-symmetric transfer matrix acts as an approximate
and inexpensive low-energy filter at the beginning and end time steps. \ For
time steps in the midsection, the full leading-order transfer matrix was used
and next-to-leading-order operators were evaluated perturbatively by insertion
at the middle time step. \ A schematic overview of the transfer matrix
calculation is shown in Fig. \ref{time_steps}. \
%TCIMACRO{\FRAME{ftbpFU}{5.1378in}{1.6786in}{0pt}{\Qcb{Schematic overview of
%the projection Monte Carlo calculation for nucleons in chiral effective field
%theory.}}{\Qlb{time_steps}}{time_steps.eps}%
%width 5.1378in;  height 1.6786in;  depth 0pt;  original-width 8.5167in;
%original-height 2.7518in;  cropleft "0";  croptop "1";  cropright "1";
%cropbottom "0";  filename 'time_steps.eps';file-properties "XNPEU";}} }%
%BeginExpansion
%EndExpansion

The pionless SU(4)-symmetric transfer matrix is computationally inexpensive
because the path integral in the SU(4) limit is strictly positive for any even
number of nucleons with either spin-singlet or isospin-singlet quantum numbers
\cite{Chen:2004rq}. \ Although there is no positivity theorem for odd numbers
of nucleons, sign oscillations are relatively mild in odd systems which are
only one particle or one hole away from an even system with no sign
oscillations. \ Some general results on positivity of the path integral and
spectral inequalities in pionless effective theory have been discussed in
\cite{Lee:2004ze,Lee:2004hc,Chen:2004rq,Wu:2005PRB}.

\begin{figure}
[ptb]
\begin{center}
\includegraphics[
height=3.4714in,
width=3.6236in
]%
{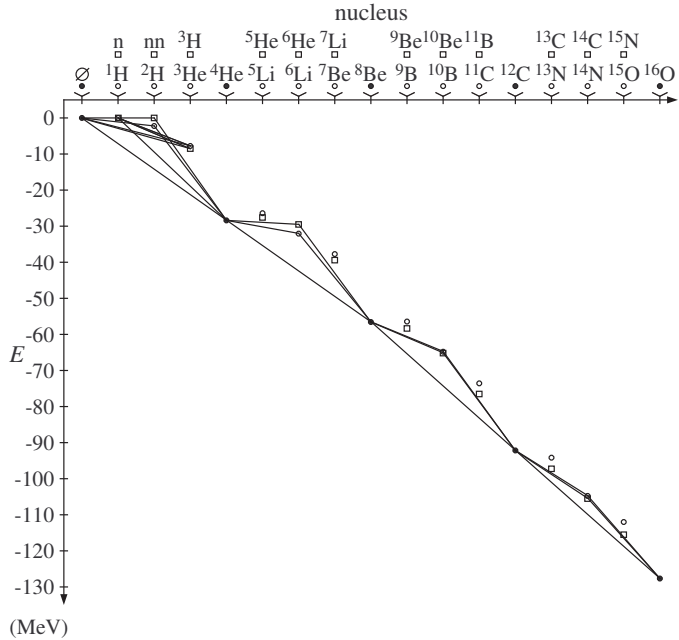}%
\caption{Plot of the energy versus particle number for light nuclei with up to
$16$ nucleons. \ The line segments show the convexity lower bounds in the
SU$(4)$ limit which hold for any two-body potential whose Fourier transform is
negative definite \cite{Lee:2007eu}.}%
\label{su4_all}%
\end{center}
\end{figure}

SU(4) symmetry arises naturally in the limit of large number of colors
\cite{Kaplan:1995yg,Kaplan:1996rk}, and the fact that both the spin-singlet
and spin-triplet nucleon scattering lengths are unusually large suggests that
the physics of low-energy nucleons is close to the Wigner limit
\cite{Mehen:1999qs,Epelbaum:2001fm}. \ In Ref.~\cite{Lee:2007eu} a general
theorem on path integral positivity was derived for interactions governed by
an SU$(2N)$-invariant two-body potential whose Fourier transform is negative
definite. \ It was also shown that as a consequence of the path integral
positivity, the particle spectrum must satisfy a number of convexity lower
bounds with respect to particle number. \ In Fig.~(\ref{su4_all}) we draw all
SU(4) convexity bounds applied to the spectrum of light nuclei with up to $16$
nucleons \cite{Lee:2007eu}. \ We note that each of the lower bound constraints
are satisfied. \ While these results do not imply that Monte Carlo simulations
of nucleons using chiral effective theory can be performed without sign or
phase oscillations, they do suggest that the simulations are possible with
only relatively mild cancellations given the approximate SU(4) symmetry and
attractive interactions at low-energies.%
%TCIMACRO{\FRAME{ftbpFU}{3.6236in}{3.4714in}{0pt}{\Qcb{Plot of the energy
%versus particle number for light nuclei with up to $16$ nucleons. \ The line
%segments show the convexity lower bounds in the SU$(4)$ limit which hold for
%any two-body potential whose Fourier transform is negative definite
%\cite{Lee:2007eu}.}}{\Qlb{su4_all}}{su4_all.eps}%
%width 3.6236in;  height 3.4714in;  depth 0pt;  original-width 7.99in;
%original-height 7.6519in;  cropleft "0";  croptop "1";  cropright "1";
%cropbottom "0";  filename 'su4_all.eps';file-properties "XNPEU";}} }%
%BeginExpansion
%EndExpansion

\section{Some recent results}

\subsection{Ground state energy at unitarity}

At zero temperature there are no dimensionful parameters in the unitarity
limit other than particle density. \ For $N_{\uparrow}$ up spins and
$N_{\downarrow}$ down spins in a given volume we denote the energy of the
unitarity-limit ground state as $E_{N_{\uparrow},N_{\downarrow}}^{0}$. \ For
the same volume we call the energy of the free non-interacting ground state
$E_{N_{\uparrow},N_{\downarrow}}^{0\text{,free}}$ and define the dimensionless
ratio%
\begin{equation}
\xi_{N_{\uparrow},N_{\downarrow}}=E_{N_{\uparrow},N_{\downarrow}}%
^{0}/E_{N_{\uparrow},N_{\downarrow}}^{0\text{,free}}.
\end{equation}
The parameter $\xi$ is defined as the thermodynamic limit for the
spin-unpolarized system,%
\begin{equation}
\xi=\lim_{N\rightarrow\infty}\xi_{N,N}.
\end{equation}

Several experiments have measured $\xi$ using density profiles of $^{6}$Li and
$^{40}$K expanding upon release from a harmonic trap. \ Some recent measured
values for $\xi$ are $0.51(4)$ \cite{Kinast:2005}, $0.46_{-05}^{+12}$
\cite{Stewart:2006}, and $0.32_{-13}^{+10}$ \cite{Bartenstein:2004}. \ There
is some disagreement among these recent measurements as well as with larger
values for $\xi$ were reported in earlier experiments
\cite{O'Hara:2002,Bourdel:2003,Gehm:2003}.

There are a number of analytic calculations for $\xi$ using techniques such as
BCS saddle point and variational approximations, Pad\'{e} approximations, mean
field theory, density functional theory, exact renormalization group,
dimensional $\epsilon$-expansions, and large-$N$ expansions
\cite{Engelbrecht:1997,Baker:1999dg,Heiselberg:1999,Perali:2004,Schafer:2005kg,Papenbrock:2005,Nishida:2006a,Nishida:2006b,JChen:2006,Krippa:2007A,Arnold:2007,Nikolic:2007,Veillette:2006}%
. \ The values for $\xi$ range from $0.2$ to $0.6$. \ Fixed-node Green's
function Monte Carlo simulations for a periodic cube find $\xi_{N,N}$ to be
$0.44(1)$ for $5\leq N\leq21$ \cite{Carlson:2003z} and $0.42(1)$ for larger
$N$ \cite{Astrakharchik:2004,Carlson:2005xy}. \ A restricted path integral
Monte Carlo calculation finds similar results \cite{Akkineni:2006A}, and a
mean-field projection lattice calculation yields $0.449(9)$
\cite{Juillet:2007a}.

There have also been simulations of two-component fermions on the lattice in
the unitarity limit at nonzero temperature. \ When data are extrapolated to
zero temperature the results of \cite{Bulgac:2005a,Bulgac:2008b} produce a
value for $\xi$ similar to the fixed-node results. \ The same is true for
\cite{Burovski:2006a,Burovski:2006b}, though with significant error bars.
\ The extrapolated zero temperature lattice results from
\cite{Lee:2005is,Lee:2005it} established a bound, $0.07\leq\xi\leq0.42$.

\begin{figure}
[ptb]
\begin{center}
\includegraphics[
height=2.2961in,
width=3.2768in
]%
{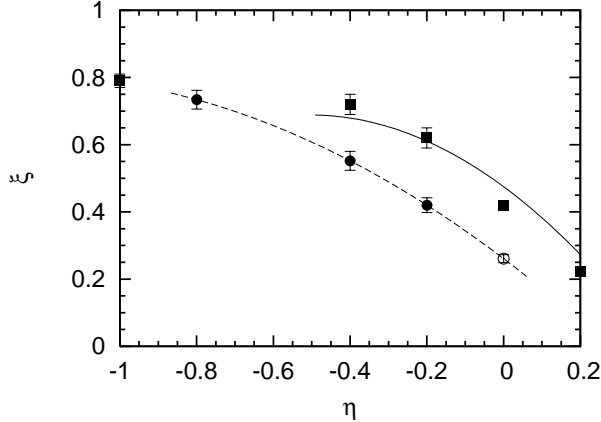}%
\caption{Plot of $\xi$ as a function of $\eta=k_{F}^{-1}a_{\text{scatt}}^{-1}%
$. \ \ The circles show the lattice results of \cite{Abe:2007ff}. \ The
squares are fixed-node Green's function Monte Carlo results
\cite{Astrakharchik:2004}, and the solid line corresponds with epsilon
expansion results \cite{Chen:2006A}.}%
\label{Seki}%
\end{center}
\end{figure}

Recent lattice calculations in the grand canonical ensemble yield a value for
$\xi=0.261(12)$ \cite{Abe:2007fe,Abe:2007ff}. \ These calculations used
lattice volumes of $4^{3}$, $6^{3}$, $8^{3}$, $10^{3}$ and also probed the
behavior at finite scattering length. $\ $In Fig.$~$(\ref{Seki}) we show $\xi$
as a function of $\eta=k_{F}^{-1}a_{\text{scatt}}^{-1}$ \cite{Abe:2007ff}.
\ The circles show the lattice results of \cite{Abe:2007ff}, and the dotted
line shows a quadratic fit through the points. \ The squares are fixed-node
Green's function Monte Carlo results \cite{Astrakharchik:2004}, and the solid
line corresponds with results calculated using the epsilon expansion
\cite{Chen:2006A}.%
%TCIMACRO{\FRAME{ftbpFU}{3.2768in}{2.2961in}{0pt}{\Qcb{Plot of $\xi$ as a
%function of $\eta=k_{F}^{-1}a_{\text{scatt}}^{-1}$. \ \ The circles show the
%lattice results of \cite{Abe:2007ff}. \ The squares are fixed-node Green's
%function Monte Carlo results \cite{Astrakharchik:2004}, and the solid line
%corresponds with epsilon expansion results \cite{Chen:2006A}.}}{\Qlb{Seki}%
%}{Seki.eps}{\special{ language "Scientific Word";  type "GRAPHIC";
%width 3.2768in;  height 2.2961in;  depth 0pt;  original-width 2.4993in;
%original-height 1.7452in;  cropleft "0";  croptop "1";  cropright "1";
%cropbottom "0";  filename 'xi_unitary.eps';file-properties "XNPEU";}} }%
%BeginExpansion
%EndExpansion

\begin{figure}
[ptb]
\begin{center}
\includegraphics[
height=3.1652in,
width=3.2076in
]%
{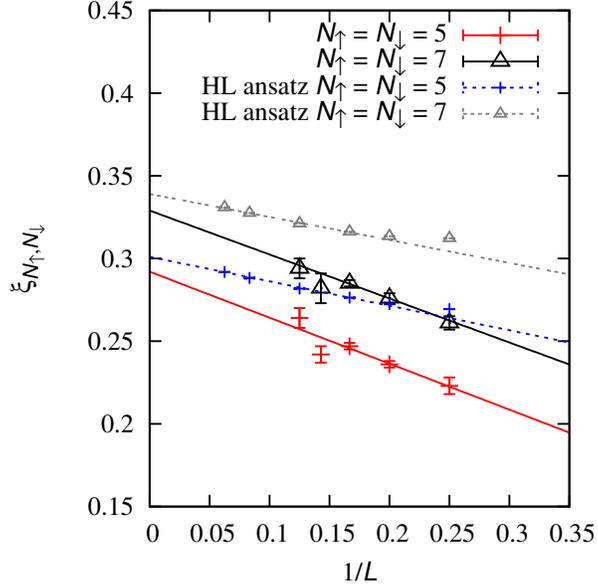}%
\caption{Results for $\xi_{5,5}$ and $\xi_{7,7}$ at finite $L$ and the
corresponding continuum limit extrapolations \cite{Lee:2008xs}. \ For
comparison we also show Hamiltonian lattice results using the symmetric
heavy-light ansatz in the lowest filling approximation \cite{Lee:2007A}.}%
\label{ldependence_hl}%
\end{center}
\end{figure}

In Ref.~\cite{Lee:2005fk}\ $\xi_{N,N}$ was calculated on the lattice using
Euclidean time projection in small volumes where it was estimated that
$\xi=0.25(3)$. \ More recent results using a technique called the symmetric
heavy-light ansatz found similar values for $\xi_{N,N}$ at the same lattice
volumes and estimated $\xi=0.31(1)$ in the continuum and thermodynamic limits
\cite{Lee:2007A}. \ A very recent lattice calculation using Euclidean time
projection with a bounded continuous auxiliary field used lattice volumes
$4^{3}$, $5^{3}$, $6^{3}$, $7^{3}$, $8^{3}$ and extrapolated to the continuum
limit \cite{Lee:2008xs}. \ The results found were%
\begin{equation}
\xi_{5,5}=0.292(12), \label{xsi_55}%
\end{equation}%
\begin{equation}
\xi_{7,7}=0.329(5). \label{xsi_77}%
\end{equation}
In Fig.~\ref{ldependence_hl} we show results for $\xi_{5,5}$ and $\xi_{7,7}$
at finite $L$ and the corresponding continuum limit extrapolations
\cite{Lee:2008xs}. \ For comparison we also show Hamiltonian lattice results
using the symmetric heavy-light ansatz in the lowest filling approximation
\cite{Lee:2007A}. \ These lattice calculations show close agreement with each
other and disagreement with fixed-node Green's function Monte Carlo results
for the same number of particles in a periodic cube \cite{Carlson:2003z}.%
%TCIMACRO{\FRAME{ftbpFU}{3.2076in}{3.1652in}{0pt}{\Qcb{Results for $\xi_{5,5}$
%and $\xi_{7,7}$ at finite $L$ and the corresponding continuum limit
%extrapolations \cite{Lee:2008xs}. \ For comparison we also show Hamiltonian
%lattice results using the symmetric heavy-light ansatz in the lowest filling
%approximation \cite{Lee:2007A}.}}{\Qlb{ldependence_hl}}{ldependence_hl.eps}%
%width 3.2076in;  height 3.1652in;  depth 0pt;  original-width 4.542in;
%original-height 4.4832in;  cropleft "0";  croptop "1";  cropright "1";
%cropbottom "0";  filename 'Ldependence_HL.eps';file-properties "XNPEU";}} }%
%BeginExpansion
%EndExpansion

\subsection{Critical temperature at unitarity}

At unitarity the critical temperature $T_{c}$ can be written as a fraction of
the Fermi energy. \ Experimentally $T_{c}/E_{F}$ has been measured using
trapped $^{6}$Li and found to be $0.27(2)$ \cite{Kinast:2005}. \ However the
interpretation of this result is made difficult by modifications caused by the
trap\ potential and the problem of relating empirical and actual temperature
scales \cite{Perali:2004a,Bulgac:2005x,Burovski:2006a}. \ A number of
approximate theoretical calculations suggest a value for the critical
temperature above \cite{Nozieres:1985JLTP,Holland:2001,Perali:2004a} as well
as below \cite{Haussmann:1994,Ohashi:2002a,Liu:2005} the Bose-Einstein
condensation temperature $T_{\text{BEC}}=0.218E_{F}$. \ An epsilon expansion
calculation around $d=2$ yields $T_{c}/E_{F}\approx0.15,$ while the epsilon
expansion around $d=4$ yields $T_{c}/E_{F}\approx0.25$
\cite{Nishida:2006a,Nishida:2006b,Nishida:2007}. \ Omitting terms at
$O(N^{-2})$, the large $N$ expansion yields $T_{c}/E_{F}\approx0.14$
\cite{Nikolic:2007}. \ A continuum-space restricted path integral Monte Carlo
calculation found $T_{c}/E_{F}\approx0.25$ \cite{Akkineni:2006A}.

\begin{figure}
[ptb]
\begin{center}
\includegraphics[
height=2.7821in,
width=3.8865in
]%
{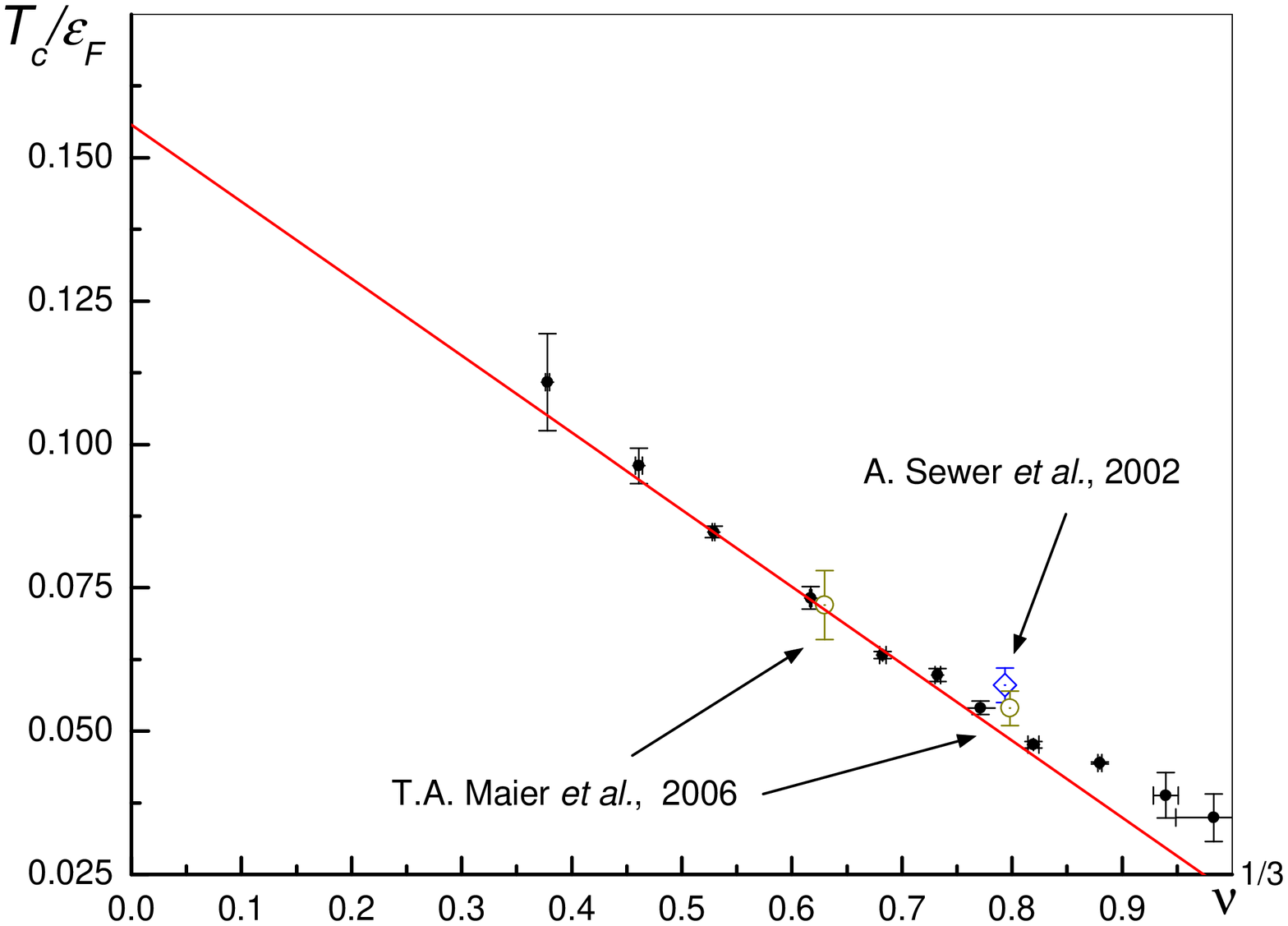}%
\caption{The critical temperature $T_{c}/E_{F}$ versus $v^{1/3}$, where $v$ is
the lattice filling fraction \cite{Burovski:2006a,Burovski:2006b}.}%
\label{umassdata}%
\end{center}
\end{figure}

Lattice simulations measuring long-range order in the pair correlation
function find values $T_{c}/E_{F}<0.14$ \cite{Lee:2005it}, $T_{c}%
/E_{F}<0.15(1)$ \cite{Bulgac:2008c}$,$ $T_{c}/E_{F}=0.152(7)$
\cite{Burovski:2006a,Burovski:2006b}$,$ and $T_{c}/E_{F}=0.183(12)$
\cite{Abe:2007ff}. \ The spread in values can likely be explained by lattice
discretization errors, which are visible in Fig.~(\ref{umassdata}) showing the
dependence of $T_{c}/E_{F}$ on $v^{1/3}$, where $v$ is the lattice filling
fraction \cite{Burovski:2006a,Burovski:2006b}. \ The simulations were done
with lattice sizes $6^{3}$, $8^{3}$, $12^{3}$. \ The point labelled A. Sewer
et al. corresponds with \cite{Sewer:2002}, while the points labelled T. A.
Maier et al. correspond with unpublished work which appears to be unavailable
in print. \ The results of \cite{Wingate:2005xy} are also consistent with a
point along this line.%

%TCIMACRO{\FRAME{ftbpFU}{3.8865in}{2.7821in}{0pt}{\Qcb{The critical temperature
%$T_{c}/E_{F}$ versus $v^{1/3}$, where $v$ is the lattice filling fraction
%\cite{Burovski:2006a,Burovski:2006b}.}}{\Qlb{umassdata}}{umassdata.eps}%
%width 3.8865in;  height 2.7821in;  depth 0pt;  original-width 8.5737in;
%original-height 6.1203in;  cropleft "0";  croptop "1";  cropright "1";
%cropbottom "0";  filename 'UMassData.eps';file-properties "XNPEU";}} }%
%BeginExpansion
%EndExpansion

\begin{figure}
[ptb]
\begin{center}
\includegraphics[
height=3.2603in,
width=3.9392in
]%
{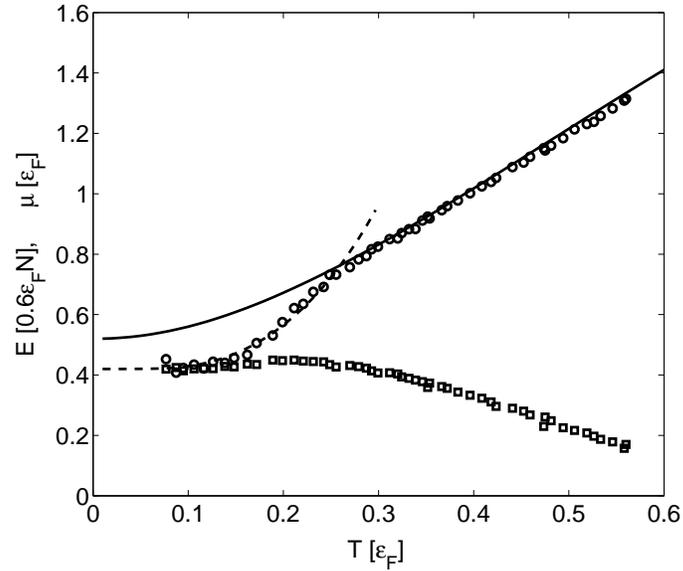}%
\caption{Plot of the average energy per particle in units of $\frac{3}{5}%
E_{F}$ versus temperature in units of $E_{F}$ \cite{Bulgac:2005a,Bulgac:2008c}%
$.$}%
\label{Bulgac}%
\end{center}
\end{figure}

While coherence measurements of the pair correlation function in
\cite{Bulgac:2008c} indicate an upper bound on the critical temperature,
$T_{c}/E_{F}<0.15(1)$, the calculation of the average energy has a peculiar
structure at $T/E_{F}=0.23(2)$ at lattice volumes $6^{3}$, $8^{3}$
\cite{Bulgac:2005a,Bulgac:2008c}. \ This data is shown in Fig.~(\ref{Bulgac}).
\ The physical significance of this effect is presently unknown. \ Meanwhile
lattice calculations of the pair correlation function using projection Monte
Carlo find low-energy string-like excitations winding around the periodic
lattice \cite{Lee:2006hr}. \ These excitations may play some role in spoiling
pair coherence at relatively low temperatures.%
%TCIMACRO{\FRAME{ftbpFU}{3.9392in}{3.2603in}{0pt}{\Qcb{Plot of the average
%energy per particle in units of $\frac{3}{5}E_{F}$ versus temperature in units
%of $E_{F}$ \cite{Bulgac:2005a,Bulgac:2008c}$.$}}{\Qlb{Bulgac}}{bulgac.eps}%
%width 3.9392in;  height 3.2603in;  depth 0pt;  original-width 7.8222in;
%original-height 6.4662in;  cropleft "0";  croptop "1";  cropright "1";
%cropbottom "0";  filename 'Bulgac.eps';file-properties "XNPEU";}} }%
%BeginExpansion
%EndExpansion

\subsection{Dilute neutron matter at NLO in chiral effective field theory}

\begin{figure}
[ptb]
\begin{center}
\includegraphics[
height=3.224in,
width=3.6608in
]%
{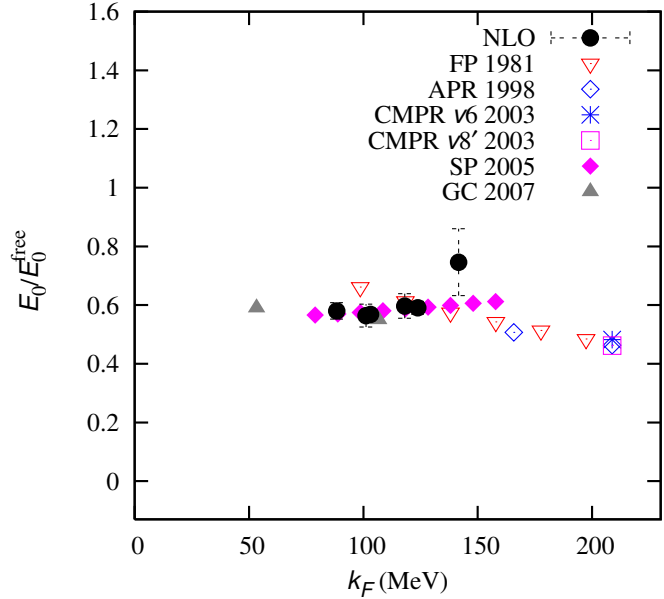}%
\caption{Results for $E_{0,\text{NLO}}/E_{\text{0}}^{\text{free}}$ versus
Fermi momentum $k_{F}$ \cite{Borasoy:2007vk}.}%
\label{xsi_literature}%
\end{center}
\end{figure}

In Ref.~\cite{Borasoy:2007vk} the ground state energy of dilute neutrons was
calculated on the lattice at next-to-leading order in chiral effective field
theory. \ The simulations used $8$ and $12$ neutrons in lattice volumes
$5^{3}$, $6^{3}$, $7^{3}$ at lattice spacings $a=(100$ MeV$)^{-1}$,
$a_{t}=(70$ MeV$)^{-1}$. \ In Fig.~\ref{xsi_literature} we show results for
the ratio of the interacting ground state energy to non-interacting ground
state energy, $E_{0,\text{NLO}}/E_{\text{0}}^{\text{free}}$, as a function of
Fermi momentum $k_{F}$. \ For comparison we show other results from the
literature: \ FP 1981 \cite{Friedman:1981qw}, APR 1998 \cite{Akmal:1998cf},
CMPR $v6$ and $v8^{\prime}$ \cite{Carlson:2003wm}, SP 2005
\cite{Schwenk:2005ka}, and GC 2007 \cite{Gezerlis:2007fs}. \ There is good
agreement between the different results near $k_{F}=120$ MeV, but there is
some disagreement on the slope. \
%TCIMACRO{\FRAME{ftbpFU}{3.6608in}{3.224in}{0pt}{\Qcb{Results for
%$E_{0,\text{NLO}}/E_{\text{0}}^{\text{free}}$ versus Fermi momentum $k_{F}$
%\cite{Borasoy:2007vk}.}}{\Qlb{xsi_literature}}{xsi_literature.eps}%
%width 3.6608in;  height 3.224in;  depth 0pt;  original-width 4.542in;
%original-height 3.9946in;  cropleft "0";  croptop "1";  cropright "1";
%cropbottom "0";  filename 'xsi_literature.eps';file-properties "XNPEU";}} }%
%BeginExpansion
%EndExpansion
The analysis in Ref.~\cite{Borasoy:2007vk} shows that much of the $P$-wave
contributions from different spin channels cancel numerically.

\subsection{Comparison with other methods and future outlook}

At nonzero temperature there are unfortunately very few ab initio calculations
that can be used to compare with results obtained using lattice effective
field theory. \ We have already mentioned a restricted path integral Monte
Carlo calculation for cold atoms at unitarity \cite{Akkineni:2006A}. \ However
the size of systematic errors due to path restriction is difficult to
estimate, and the final result for the critical temperature is in strong
disagreement with each of the lattice results presented above.

More comparisons can be made for calculations of low energy spectra. \ At
present the most accurate ab initio calculations of light nuclei binding
energies for up to twelve nucleons have been obtained using Green's Function
Monte Carlo. \ The overall accuracy of these calculations are at the $1-2\%$
level. \ Current lattice calculations are not as accurate as this, but it is
hoped that lattice simulations for light nuclei at N$^{3}$LO in the future can
reach comparable accuracies. \ At cutoff momentum $\Lambda=500$ MeV, No-Core
Shell Model calculations using the NNLO chiral potential (plus N$^{3}$LO terms
for the two-nucleon potential) give binding energies for $^{6}$Li and $^{7}$Li
at $3\%$ accuracy \cite{Nogga:2005hp,Nogga:2006ir}. \ The first lattice
effective field theory calculations at NNLO are currently in progress.
\ Preliminary results for lattice spacing $a=(100$ MeV$)^{-1}$ give an alpha
binding energy accurate at the $5\%$ level \cite{Epelbaum:2008a}. \ However
much further work remains in developing the lattice formalism at higher order
and studying larger nuclei.

For the ground state of cold atoms at unitarity and dilute neutron matter, the
quality of lattice effective field theory calculations are competitive with or
exceed other computational approaches. \ Here the relative success of lattice
simulations over other methods is probably due to the nature of the ground
state and the use of determinantal Monte Carlo. \ For cold atoms at unitarity
and dilute neutron matter the competition between attractive binding forces
and Fermi antisymmetry is somewhat evenly matched, resulting in a complicated
superfluid ground state somewhere in between BCS and BEC. \ This makes it
difficult to use techniques where nodal constraints must be approximately
guessed and easier to rely on determinantal Monte Carlo methods which
automatically incorporate Fermi antisymmetry.

In addition to probing more nucleons and higher orders in effective field
theory, future work must also probe simulations at larger lattice volumes.
\ This includes both smaller physical lattice spacings as well as larger
physical volumes. \ The transition from small lattice systems to large
production runs should be possible as more experience and data is collected on
the efficiency of various lattice algorithms and more computing resources are
devoted to important calculations. \ It is probably unlikely that lattice
effective field theory simulations can match the spatial lattice lengths
$L=30\sim40$ used in some large-scale lattice QCD simulations. \ For cold atom
simulations at unitarity the most significant computational barrier with
increasing system size is the increase in condition number for the auxiliary
field matrices $\mathbf{M}_{A}(s,t)$. \ Similar computational slowdown occurs
in unquenched lattice QCD simulations at very small sea quark masses or at
large chemical potential (in addition to the problem caused by complex phase oscillations).

For a single bound nucleus it is not necessary to probe volumes much larger
than the size of the bound state since the finite volume errors are
exponentially small. \ In this case it would be more useful to probe smaller
physical lattice spacings, as much as constraints such as sign or complex
phase oscillations will allow. \ For unbound nuclear systems the finite volume
dependence of energy levels should be more interesting. \ This data can be
used to probe nucleon-nucleus scattering or nucleus-nucleus scattering using
L\"{u}scher's finite volume scattering method.

It is difficult to predict the development of the field in the future, but one
general hope is that collaborative efforts develop with researchers not
directly involved in large-scale lattice calculations. \ One working model has
already been pioneered in the lattice QCD community, where large numbers of
gauge configurations are stored and shared for general use. \ A similar model
may be useful for lattice effective field theory calculations for systems with
significant general interest.

\section{Summary}

In this article we have reviewed the current literature on lattice simulations
for few- and many-body systems. \ We discussed methods which combine the
theoretical framework of effective field theory with computational lattice
methods. \ The lattice spacing serves as the ultraviolet cutoff for the
low-energy effective theory, and all interactions are included up to some
chosen order in power counting. \ By increasing the order, the accuracy at low
energies can be systematically improved. \ One feature of this approach is the
ability to study several different phenomena using exactly the same lattice
action. \ Another feature of the lattice effective field theory approach is
the close theoretical link with standard analytic tools used in effective
field theory calculations. \ The approach also benefits from the computational
flexibility provided by a number of efficient lattice algorithms. \ We have
discussed many of these in this article.

The idea of lattice simulations using effective field theory is relatively
new, and this review article represents a snapshot of the current progress in
the field. \ We have attempted to cover the relevant principles from effective
field theory as well as different formalisms and algorithms used in recent
lattice calculations. \ We have focused much attention on techniques which can
be applied to both cold atoms and low-energy nuclear physics as well as common
methods used in work by different collaborations.

\section*{Acknowledgements}

The author thanks a long list of collaborators and colleagues for discussing
work and topics covered in this review. \ The list includes Bugra Borasoy,
Aurel Bulgac, Shailesh Chandrasekharan, Jiunn-Wei Chen, Evgeny Epelbaum, Hans
Hammer, Hermann Krebs, Ulf-G. Mei\ss ner, Nikolay Prokof'ev, Gautam Rupak,
Thomas Sch\"{a}fer, Ryoichi Seki, Boris Svistunov, Bira van Kolck, and Matt
Wingate. \ This work is supported in part by DOE grant DE-FG02-03ER41260.

\bibliographystyle{apsrev}
\bibliography{NuclearMatter}

\end{document}